\documentclass[journal,10pt,]{IEEEtran}
\usepackage[printonlyused]{acronym}
\usepackage{adjustbox}
\usepackage[style=ieee,backend=bibtex]{biblatex}
\usepackage{array}
\renewcommand{\figurename}{Figure}
\usepackage[font=small]{caption}
\usepackage{subfigure}
\usepackage{listings}
\usepackage{amsmath}
\usepackage{graphicx}
\usepackage{amssymb}
\usepackage{multirow}
\usepackage{amsmath}
\usepackage[monochrome]{color}
\usepackage{soul}
\usepackage[justification=centering]{caption}
\bibliography{biblio}
\usepackage{array}
\newcolumntype{x}[1]{>{\centering\arraybackslash\hspace{0pt}}p{#1}}
\usepackage{footnote}
\makesavenoteenv{tabular}
\makesavenoteenv{table}
\usepackage{xcolor}
\usepackage{tikz}
\usetikzlibrary{shapes,arrows}

\begin{document}

\title{Survey on Terahertz Nanocommunication and Networking: A Top-Down Perspective}
\vspace{1mm}
\author{
\IEEEauthorblockN{Filip Lemic\IEEEauthorrefmark{1}, Sergi Abadal\IEEEauthorrefmark{2}, Wouter Tavernier\IEEEauthorrefmark{3}, Pieter Stroobant\IEEEauthorrefmark{3}, Didier Colle\IEEEauthorrefmark{3},\\Eduard Alarc\'{o}n\IEEEauthorrefmark{2}, Johann Marquez-Barja\IEEEauthorrefmark{1}, Jeroen Famaey\IEEEauthorrefmark{1}\\}
\vspace{1mm}
\IEEEauthorblockA{\IEEEauthorrefmark{1}Internet Technology and Data Science Lab (IDLab), Universiteit Antwerpen - imec, Belgium\\} 
\IEEEauthorblockA{\IEEEauthorrefmark{2}NaNoNetworking Center in Catalunya (N3Cat), Universitat Polit\`{e}cnica de Catalunya, Spain\\}
\IEEEauthorblockA{\IEEEauthorrefmark{3}Internet Technology and Data Science Lab (IDLab), Ghent University - imec, Belgium\\}
Email: filip.lemic@uantwerpen.be\vspace{-5mm}}

\maketitle

\begin{abstract}
Recent developments in nanotechnology herald nanometer-sized devices expected to bring light to a number of groundbreaking applications.
Communication with and among nanodevices will be needed for unlocking the full potential of such applications. 
As the traditional communication approaches cannot be directly applied in nanocommunication, several alternative paradigms have emerged.
Among them, electromagnetic nanocommunication in the terahertz (THz) frequency band is particularly promising, mainly due to the breakthrough of novel materials such as graphene.
For this reason, numerous research efforts are nowadays targeting THz band nanocommunication and consequently nanonetworking.
As it is expected that these trends will continue in the future, we see it beneficial to summarize the current status in these research domains.
In this survey, we therefore aim to provide an overview of the current THz nanocommunication and nanonetworking research.
Specifically, we discuss the applications envisioned to be supported by nanonetworks operating in the THz band, together with the requirements that such applications pose on the underlying nanonetworks.
Subsequently, we provide an overview of the current contributions on the different layers of the protocol stack, as well as the available channel models and experimentation tools.
As the final contribution, we identify a number of open research challenges and outline several potential future research directions.
\end{abstract}


\IEEEpeerreviewmaketitle


\acrodef{RSS}{Received Signal Strength}
\acrodef{RSSI}{Received Signal Strength Indicator}
\acrodef{KL}{Kullback-Leibler}
\acrodef{RF}{Radio-Frequency}
\acrodef{UWB}{Ultra-Wide Band}
\acrodef{EU}{European Union}
\acrodef{ToA}{Time of Arrival}
\acrodef{AoA}{Angle of Arrival}
\acrodef{OSI}{Open Systems Interconnection}
\acrodef{kNN}{k-Nearest Neighbors}
\acrodef{LQI}{Link Quality Information}
\acrodef{CSMA}{Carrier-Sense Multiple Access}
\acrodef{PRR}{Packet Reception Rate}
\acrodef{LLC}{Logical Link Control}
\acrodef{BER}{Bit-Error Ratio}
\acrodef{AP}{Access Point}
\acrodef{MT}{Mobile Terminal}
\acrodef{MAC}{Media Access Control}
\acrodef{FPS}{Fingerprinting Localization Algorithms} 
\acrodef{HT}{Hypothesis Testing} 
\acrodef{TCN}{Timing Channel for Nanonetworks}
\acrodef{TDMA}{Time Division Multiple Access}
\acrodef{PHLAME}{PHysical Layer Aware MAC protocol for Electromagnetic nanonetworks in the Terahertz Band}
\acrodef{EEWNSN-MAC}{Energy Efficient Wireless Nano Sensor Network MAC protocol}
\acrodef{RTR}{Ready-to-Receive}
\acrodef{DRIH-MAC}{Distributed Receiver-Initiated Harvesting-Aware MAC}
\acrodef{CTR}{Critical Transmission Ratio}
\acrodef{D2D}{Device-to-Device}
\acrodef{CDF}{Cumulative Distribution Function}
\acrodef{ML}{Machine Learning}
\acrodef{TS-OOK}{Time Spread ON-OFF keying}
\acrodef{GPS}{Global Positioning System}
\acrodef{BS}{Base Station}
\acrodef{SNR}{Signal-to-Noise Ratio}
\acrodef{IoT}{Internet of Things}
\acrodef{2D}{2-Dimensional}
\acrodef{NoC}{Network-on-Chip}
\acrodef{NACK}{Negative ACKnowledgment}
\acrodefplural{NoC}[NoCs]{Networks-on-Chip}
\acrodef{WNoC}{Wireless Network-on-Chip}
\acrodefplural{WNoC}[WNoCs]{Wireless Networks-on-Chip}
\acrodef{GSM}{Global System for Mobile communications}
\acrodef{3D}{3-Dimensional}
\acrodef{FM}{Frequency Modulation}
\acrodef{SDM}{Software-Defined Metamaterial}
\acrodef{SLR}{Stateless Linear-path Routing}
\acrodef{BER}{Bit-Error-Ratio}
\acrodef{WNSN}{Wireless Nano-Sensor Network}
\acrodef{SoC}{System-on-Chip}
\acrodef{SINR}{Signal to Interference Plus Noise Ratio}
\acrodef{RTS}{Ready-to-Send}
\acrodef{CTS}{Clear-to-Send}
\acrodef{UDP}{User Datagram Protocol}
\acrodef{IoNT}{Internet of Nano-Things}
\acrodef{SotA}{State-of-the-Art}
\acrodef{IoE}{Internet of Everything}
\acrodef{IoMNT}{Internet of Multimedia Nano-Things}
\acrodef{DoS}{Denial of Service}
\acrodef{LoS}{Line-of-Sight}
\acrodef{SPP}{Surface Plasmon Polariton}
\acrodef{WoS}{Web of Science}
\acrodef{WiP}{Work-in-Progress}
\acrodef{WSN}{Wireless Sensor Network}
\acrodef{ZnO}{Zinc Oxide}

\section{Introduction}

``There's Plenty of Room at the Bottom: An Invitation to Enter a New Field of Physics''~\cite{feynman1960there} was the title of a visionary lecture given by the Nobel Prize recipient Prof. Richard Feynman at the annual American Physical Society meeting at the California Institute of Technology (Caltech) in December 1959.
Prof. Feynman discussed the possibility of directly manipulating materials on an atomic scale - and he surely wasn't joking.
The concepts originally outlined by Prof. Feynman later became circumscribed under the term ``nanotechnology'', first introduced by Prof. Norio Taniguchi from Tokyo University of Science in 1974.
Due to substantial research efforts in recent years, nanotechnology is today paving the way toward sub-$\mu$m scale devices (i.e., from one to a few hundred nanometers).
Controlling materials on such a scale is expected to give rise to integrated nanodevices with simple sensing, actuation, data processing and storage, and communication capabilities, opening the horizon to a large variety of novel, even groundbreaking applications.      

Communication and coordination among the nanodevices, as well as between them and the macro-scale world, will be required to achieve the full promise of such applications. 
Thus, several alternative nanocommunication paradigms have emerged in the recent years, the most promising ones being electromagnetic, acoustic, mechanical, and molecular communication~\cite{abbasi2016nano}. 

In molecular nanocommunication, a transmitting device releases molecules into a propagation medium, with the molecules being used as the information carriers~\cite{hiyama2006molecular}. 
Acoustic nanocommunication utilizes pressure variations in the (fluid or solid) medium to carry information between the transmitter and receiver.
In mechanical (i.e., touch-based) nanocommunication, nanorobots are used as carriers for information exchange~\cite{chen2015touch}.
Finally, electromagnetic nanocommunication uses the properties of electromagnetic waves (e.g., amplitude, phase, delay) as the information carriers~\cite{akyildiz2010electromagnetic}.
Among the above paradigms, molecular and electromagnetic nanocommunication show the greatest promise for enabling communication between nanodevices~\cite{akyildiz2010propagation}.
This work focuses on \textit{electromagnetic nanocommunication}.
This is due to the fact that it is among the two most promising nanocommunication paradigms, as well as due to its suitability to different propagation mediums (e.g., in-body, free space, on-chip) and potential for meeting the applications' requirements (more details in Section~\ref{sec:applications}).

Classical electromagnetic communication and networking paradigms are not directly applicable to the majority of nanoscale communication and networking scenarios, predominantly due to the small sizes and limited capabilities of nanodevices~\cite{jornet2010graphene,rikhtegar2013brief}. 
That it to say, microwave or even millimeter-wave (mmWave) frequencies impose relatively large antennas (i.e., mm-scale and larger) that do not fit into nanodevices.
To meet the size requirements of nanodevices, a classical metallic antenna would be required to use very high radiation frequencies. 
For example, a one-micrometer-long dipole antenna would resonate at approximately 150 terahertz (THz)~\cite{jornet2013fundamentals}. 
Although the communication bandwidth increases with the increase in the antenna's resonance frequency, the same happens with the propagation loss. 
Due to the highly constrained power of nanodevices~\cite{wang2008towards}, the feasibility of nanonetworks would be compromised if this approach would be adopted. 
In addition, nanoscale material properties of common metals are unknown, hence the common assumptions from antenna theory might not be correct~\cite{hanson2005fundamental}. 
Finally, it is currently also technologically infeasible to develop miniature transceiver that could operate at such high frequencies~\cite{jornet2013fundamentals}.

As an alternative approach, graphene has attracted substantial research attention, mainly due to its unique electrical and optical properties~\cite{llatser2012graphene}. The interaction of electromagnetic radiation with graphene and its derivatives (i.e., carbon nanotubes (CNT)~\cite{da2009carbon} - rolled graphene and graphene nanoribbons (GNRs) - thin strips of graphene), differs from that of the conventional metals.
Specifically, it has been demonstrated that graphene supports the propagation of \ac{SPP} waves in the THz frequency band~\cite{vakil2011transformation,horng2011drude,horng2011drude,mikhailov2011theory}.
Due to that, graphene-enabled electromagnetic nanocommunication in the THz frequencies is able to deliver miniaturization~\cite{zhou2014miniaturized,perruisseau2013graphene}, i.e., a prerequisite for nanocommunication. In addition, the possibility to operate at lower frequencies relaxes the energy and power requirements for the nanodevices~\cite{jornet2013fundamentals, zhou2014miniaturized}. {\color{red}{Moreover, graphene shows unique tunability properties, which allow to steer the beam or tune the resonance frequency by just changing a bias voltage \cite{perruisseau2013graphene,Correas2017}. For these reasons and given its experimentally demonstrated compatibility with CMOS manufacturing processes} \cite{wu2013graphene, han2014graphene, neumaier2019integrating}, graphene allows versatile antennas to be directly integrated in nanocommunication devices as envisaged in multiple works} \cite{akyildiz2010electromagnetic,Jornet2014TRANSCEIVER,Correas2017}.
Consequently, graphene-enabled THz nanocommunication attracted substantial attention from a broad scientific community.

As the efforts targeting THz band nanocommunication yielded highly encouraging results, the research focus soon spread from the communication to the networking community, giving birth to nanonetworking in the THz band.
Suffice to say, the results of these efforts are equally encouraging and promising, at this point arguably also abundant.
Hence, we believe a summary of the research efforts and current \ac{SotA} on THz nanocommunication and nanonetworking would be beneficial to the community, which provides the main motivation for this survey.

In this survey, we first discuss several application domains that could be enabled by nanonetworks operating in the THz band.
These include software-defined metamaterials, wireless robotic materials, body-centric communication, and \acp{WNoC}. 
Moreover, we derive a set rule-of-thumb requirements that each of the application domains posits on the supporting nanonetworks.   
Then, we utilize a top-down approach in discussing the \ac{SotA} of different layers of the nanonetworks' protocol stack.
We believe a top-down approach to be more natural than the bottom-up alternative, as it provides the reader with a straightforward mapping between the application requirements on the one side, and the underlying protocols and their design goals on the other.
In addition, we provide an overview of existing channel models and simulation and experimentation tools available for THz nanocommunication and nanonetworking research.
At the end of each section, we discuss the corresponding research ``gaps'' and  open challenges. 
Moreover, we summarize several additional open challenges related to THz nanocommunication and nanonetworking in general, but not directly pertaining to specific layers of the protocol stack, channel models, or experimentation tools. 
We aim to provide a straightforward introduction to THz nanocommunication and nanonetworking research, as well at as to identify the ``missing pieces'' in the current results and suggest potential future research directions.

\vspace{-1.5mm}
{\color{red}{\subsection{Methodology}
This survey has been developed following the Systematic Literature Review (SLR) methodology~\cite{keele2007guidelines,cruz2016systematic}.
We aimed to select all the relevant works in the context of THz nanocommunication and nanonetworking, in turn resulting in the detection of the open challenges and missing ``pieces’' in the existing research literature.
We have considered the following inclusion criteria: i) an article proposes a full novel solution rather than conceptual, informative, or ongoing efforts (i.e., position, poster, demo, and \ac{WiP} articles were not considered), ii) an article went through a peer-review process and is either publicly available (e.g., in ArXiv) or included in the \ac{WoS}, Scopus, IEEEXplore, ACM, and/or Springer databases, and iii) an article is written in English.
In terms of the step-by-step methodology, we have used the Google Scholar database in the initial review phase, during which the articles were included or excluded based on their titles and abstracts.
In the consequent step, the full texts were assessed along the above-mentioned inclusion criteria.
As a commonly accepted practice, we have placed extra emphasis on articles published since 2015, highly cited ones, and the ones published in the most important venues, e.g., IEEE Transactions on Nanotechnology, ELSEVIER’s Nano Communication Networks, IEEE Nanotechnology Magazine, and ACM International Conference on Nanoscale Computing and Communication (ACM NANOCOM).

\vspace{-1.5mm}
\subsection{Structure}}}
The rest of this paper is structured as follows. 
In Section~\ref{sec:related_work}, we provide an overview of the related surveys in the existing literature.  
The application domains that could potentially be supported by nanonetworks operating in the THz band are, together with the requirements such applications pose on the underlying nanonetworks, discussed in Section~\ref{sec:applications}.
The current research efforts on the network, link, and physical layers of the protocol stack are summarized in Sections~\ref{sec:network},~\ref{sec:link}, and~\ref{sec:phy}, respectively.
Moreover, in Section~\ref{sec:channel} we discuss the available channel models for THz nanocommunication, while the existing experimentation and simulation tools are outlined in Section~\ref{sec:testbeds}.
In Section~\ref{sec:challenges}, we discuss several more general open challenges, i.e., the ones not directly related to the content of the previous sections, yet still relevant to THz nanocommunication and nanonetworking research.
Finally, we conclude the survey in Section~\ref{sec:conclusion}.   

\vspace{3mm}

\section{Related Work}
\label{sec:related_work}

There are several contributions generically targeting THz band communication~\cite{akyildiz2014teranets,akyildiz2014terahertz,petrov2016terahertz}. 
They provide useful insights into the paradigm from the device's perspective, as well as from the communication point of view.
In terms of communication, advances and challenges pertain to channel modeling, development of communication protocols, and establishment of supporting experimentation tools and facilities. 

There are also several specialized survey papers that discuss different aspects of THz band nanocommunication. 
The pioneering work targeting this topic was published by Akyildiz and Jornet~\cite{akyildiz2010electromagnetic}, providing an in-depth view on nanotechnology and discussing several options for enabling communication among nanonodes.
In~\cite{akyildiz2010electromagnetic}, the authors take the nanodevice standpoint and first outline the envisioned architecture of the nanodevice, consisting of units for sensing, actuating, powering, data processing and storage, and communication.
The authors follow by outlining several application groups for \acp{WNSN}: biomedical, environmental, industrial and consumer goods, and military and defense.
Moreover, they briefly discuss the architecture of a \ac{WNSN} consisting of nanonodes, nanorouters, nano-micro interfaces, and gateways.
In addition, several open research questions and promising directions for future research are outlined. 
It is worth mentioning that the survey is dated to 2010.

\begin{table*}[!ht]
\begin{center}
\caption{Surveys on THz band communication and networking}
\label{tab:surveys}
\begin{tabular}{l l l m{9.2cm}} 
\hline
\textbf{Name} & \textbf{Year} & \textbf{Type} & \textbf{Summary of covered topics} \\ 
\hline
Akyildiz~\emph{et al.}~\cite{akyildiz2010internet} & 2010 & \acl{IoNT} &  \ac{SotA} in THz nanocommunication for the IoNT; research challenges (channel modeling, information encoding, and protocols for the IoNT). \\ \hline 
Akyildiz~\emph{et al.}~\cite{akyildiz2010electromagnetic} & 2010 & THz nanocommunication & \ac{SotA} in nanodevice technology; summaries of \ac{WNSN} applications and architectures; overview of nanocommunication and networking challenges. \\ \hline
Jornet~\emph{et al.}\cite{jornet2012internet} & 2012 & \acl{IoNT} & \ac{SotA}, open challenges, and research directions in the THz nanocommunication and \acf{IoMNT}. \\ \hline
Rikhtegar~\emph{et al.}~\cite{rikhtegar2013brief} & 2013 & THz nanocommunication &  Molecular and electromagnetic (THz) communication for nanoscale applications; summary of nanoscale communication paradigms and potential applications. \\ \hline
Balasubramaniam~\emph{et al.}~\cite{balasubramaniam2013realizing} & 2013 & \acl{IoNT} & Challenges in realizing the IoNT (data collection and routing, bridging the outside world and nanonetworks, etc.); possible IoNT applications.
\\ \hline
Akyildiz~\emph{et al.}~\cite{akyildiz2014terahertz} & 2014 & THz communication & THz band applications at macro and nanoscale; \ac{SotA} in THz band transceivers and antennas; open challenges from communication and networking perspectives; simulation and experimentation tools for THz band communication. \\ \hline
Akyildiz~\emph{et al.}~\cite{akyildiz2014teranets} & 2014 & THz communication &  \ac{SotA} in THz band transceivers and antennas; open challenges from communication and networking perspectives.  \\ \hline
Akyildiz~\emph{et al.}~\cite{akyildiz2015internet} & 2015 & \acl{IoNT} & Introduction to the Internet of Bio-Nano Things (IoBNT); bridging the outside world and the IoBNT; open challenges in the IoBNT. \\  \hline
Miraz~\emph{et al.}~\cite{miraz2015review} & 2015 & \acl{IoNT} & Short overview and future research directions pertaining to the \ac{IoT}, \ac{IoE}, and \ac{IoNT}. \\ \hline
Dressler~\emph{et al.}~\cite{dressler2015connecting} & 2015 & \acl{IoNT} & Bridging the outside world and in-body nanonodes for health-care applications; IoNT network architectures; simulation tools for in-body nanonetworks. \\ \hline
Petrov~\emph{et al.}~\cite{petrov2016terahertz} & 2016 & THz communication & \ac{SotA} in THz band communication; engineering trade-offs in typical applications; open challenges and research directions in THz band communication. \\ \hline
Alsheikh~\emph{et al.}~\cite{alsheikh2016mac} & 2016 & THz nanocommunication & Existing MAC protocols for \ac{WNSN}; performance analysis and design guidelines for WNSN MAC protocols. \\\hline
Rizwan~\emph{et al.}~\cite{rizwan2018review} & 2018 & \begin{tabular}{@{}l@{}}Nanocommunication in \\healthcare\end{tabular} & Nanosensors and nanonetworks for healthcare systems; big data analytics (data  sources, preprocessing, feature extraction, visualization, predictive modelling)\\ \hline
Ghafoor~\emph{et al.}~\cite{ghafoor2020mac} & 2019 & THz communication & Features of the THz band; THz macro and nanoscale applications; design requirements for THz MAC protocols; classification of existing MAC protocols; open challenges for MAC protocols. \\ \hline
\hline
This survey & 2021 & THz nanocommunication & THz nanoscale applications; THz nanocommunication and nanonetworking protocols; THz nanoscale channel models; simulation and experimentation tools  \\ \hline
\end{tabular}
\end{center}
\vspace{-4mm}
\end{table*}

It is also worth noting the work in~\cite{rikhtegar2013brief}, which provides an overview of molecular and electromagnetic nanocommunication.
In addition,~\cite{rizwan2018review} briefly discusses electromagnetic communication from a stand-point of future healthcare systems, with the main focus of the work being big data analytics for healthcare data.
In terms of electromagnetic nanocommunication utilizing THz frequencies,~\cite{rikhtegar2013brief} provides similar insights as~\cite{akyildiz2010electromagnetic}. 
Moreover, in~\cite{alsheikh2016mac}, the authors provide a comparative survey of different \ac{MAC} protocols for \ac{WNSN}. 
In~\cite{ghafoor2020mac}, the authors provide a survey of \ac{MAC} protocols for THz band communication in general.  
In addition, performance analyses of these protocols have been carried out in terms of consumed energy, transmission distance, and probability of collisions.

Several contributions target different aspects of the \ac{IoNT}, most notably~\cite{akyildiz2010internet,jornet2012internet,balasubramaniam2013realizing,dressler2015connecting,miraz2015review,akyildiz2015internet}.
In these works, the authors tend to agree on the \ac{IoNT} architecture, consisting of a nanonetwork connected to the macro-world through a nano-macro gateway.
In addition, several application domains have been outlined in these works, which can roughly be grouped into health-care, environmental and agricultural monitoring, multimedia, and military and defense.
Moreover, the works converge toward the idea that there are two encouraging nanocommunication options, namely molecular and electromagnetic utilizing THz frequencies. 
The former is predominantly feasible for in-body nanocommunication. 
Among the above-mentioned contributions targeting IoNT, arguably the most relevant ones for this survey are~\cite{dressler2015connecting} and~\cite{jornet2012internet}.
The work in~\cite{dressler2015connecting} is interesting as it summarizes several requirements for nanocommunication for enabling in-body health-care applications. 
These pertain to legal (e.g., legislation on duration of the body contact, invasiveness or implantability of nanodevices), functional (i.e., the purpose of communication between nanonetwork and macro-world), and technical (e.g., reliability, safety, privacy, real-time capabilities) requirements. 
The work is also relevant because it discusses several challenges for in-body nanocommunication, as well as useful simulation tools for the problem at hand. 
Jornet \emph{et al.}~\cite{jornet2012internet} provide a relevant read due to the fact that it explicitly outlines nanocommunication challenges in the multimedia-focused \ac{IoNT}.
These challenges pertain to data compression and signal processing, THz channel modeling, and challenges on different layers of the protocol stack. 

In order to position our work in the context of the above-outlined contributions (summarized in Table~\ref{tab:surveys}), it is first relevant to note that these contributions, apart from~\cite{ghafoor2020mac,rizwan2018review}, date from 2016 or before. 
Several novel proposals have been made on various aspects of THz nanocommunication and nanonetworking in the recent years, ranging from novel protocols in different layers, to THz channel modeling and experimentation tools. 
We believe that at this point in time a critical summary of these contributions and their positioning in regard to the older ones would benefit the scientific community.
Second, some of the current surveys focus either on THz communication in general or on nanocommunication, resulting in neglecting numerous aspects relevant to THz band electromagnetic nanocommunication.
Adversely, other contributions are substantially more focused, targeting for example only the \ac{IoNT} or \ac{MAC} layer protocols for THz band (nano)communication.
Third, due to its very recent roll-out with the most important works being only a few years old, THz band nanonetworking has not received sufficient treatment in the existing surveys. 
We aim at filling the above-stated gaps by providing a full view on the problem, ranging from applications and their requirements, different layers of the protocols stack, channel models, and experimentation tools, to challenges and open research questions, but pertaining exclusively to electromagnetic nanocommunication and nanonetworking in the THz band.   

Finally, it is worth emphasizing that this work does not cover THz macro-scale applications, protocols, nor channel models. Interested readers are directed to several recent  papers surveying these topics, e.g.,~\cite{huq2018thz,mumtaz2017terahertz,chen2019survey,huq2019terahertz,ma2020intelligent}. In addition, this survey deals with electromagnetic nanocommunication only. Hence, molecular and other potential paradigms for nanocommunication and networking are out of scope of this paper. Interested readers can find more details in recent surveys from the literature~\cite{kuscu2019transmitter,akan2016fundamentals,wang2017diffusion,felicetti2016applications,chahibi2017molecular}.
Similarly, THz nanocommunication hardware-related aspects are out of scope of this work.
Although arguably not abundant, existing survey papers discuss aspects such as THz sources~\cite{lewis2014review,lallas2019key}, graphene antennas~\cite{Correas2017,elayan2016graphene}, and circuits/devices~\cite{naghavi2019filling,sengupta2018terahertz}.


\section{Application Domains Enabled by THz Nanonetworks}
\label{sec:applications}

In this section, we provide an overview of the most prominent application domains with the potential of being enabled by THz-operating nanonetworks. 
For each of the domains, we discuss its requirements, which are at the end of the section summarized in Table~\ref{tab:applications_requirements}.
Note that in the table we include the most stringent requirements for each application domain, although these can potentially be more lenient for some applications in the domain. 
Also note that we differentiate the application domains based on their requirements.
For example, we distinguish wireless robotic materials and body-centric communication, although both domains contain a variety of sensing-only applications.
Due to that, some works from the literature (e.g.,~\cite{akyildiz2012nanonetworks,akyildiz2015internet}) specify \acp{WNSN} or \ac{IoNT} as applications that could potentially be enabled by THz band nanonetworks, which is in our case ``embedded'' in some of the specified application domains.

\subsection{Software-Defined Metamaterials}
Metamaterials (and metasurfaces, their two-dimensional counterparts~\cite{walia2015flexible}) are manufactured structures that enable powerful control of electromagnetic waves. As such, metamaterials can be used to realize devices with engineered and even unnatural properties related to the reflection, absorption, or transmission of electromagnetic radiation. For instance, metamaterials have been proposed for the electromagnetic cloaking of objects~\cite{cai2007optical}, noise~\cite{wang2017reduction} cancellation, holography \cite{poddubny2013hyperbolic}, anomalous reflection \cite{smith2004metamaterials}, and focusing of energy with unprecedented accuracy~\cite{zhai2011electromagnetic}. Such unprecedented control is achieved through the careful design of a periodic array of subwavelength elements typically called \emph{unit cells}. The main issue with current metamaterials, however, is that the unit cells are ``hard-coded'' for a single application and operational condition (e.g., to work for a single angle of incidence) and cannot be reused across applications nor reprogrammed for different operations. To alleviate this issue, Liaskos \emph{et al.}~\cite{liaskos2015design} proposed \acp{SDM}, a new paradigm of programmable metamaterials where the unit cells can be reconfigured through a software interface with a set of well-defined instructions. 

By virtue of the unprecedented and real-time control of the absorption, reflection, and transmission characteristics of metasurfaces, the \ac{SDM} approach is expected to enable a plethora of applications in areas such as sensing, communications, or imaging~\cite{abadal2020programmable}. In wireless communications, \acp{SDM} have been proposed as a basic building block of \emph{intelligent reflecting surfaces} enabling the revolutionary concept of \emph{programmable wireless environments} \cite{Liaskos2018,gong2019towards}. By controlling the reflection and scattering profiles at selected spots, both propagation loss and multipath effects can be mitigated in any wireless channel. This has proven to be a true paradigm shift in wireless communications, as the recent explosion of works can attest \cite{liaskos2019novel,wu2019intelligent,huang2019reconfigurable,ozdogan2019intelligent}, because the channel has been traditionally an inescapable limiting factor. Another application related to wireless communications is the development of simplified architectures for wireless communication transmitters \cite{tang2019programmable}. There, the metasurfaces are utilized to directly modulate the carrier wave in multiple channels using the baseband signal without the need of core components of the RF chain such as mixers, filters, amplifiers, which can become hard to realize at high frequencies. Finally, we also highlight the applicability of \acp{SDM} in the area of holographic imaging, where real-time programmability opens the door to dynamic holograms with outstanding spatial resolution given by the subwavelength granularity of the unit cells \cite{Li2017b}.

\begin{figure}[!t]
\centering
\includegraphics[width=\columnwidth]{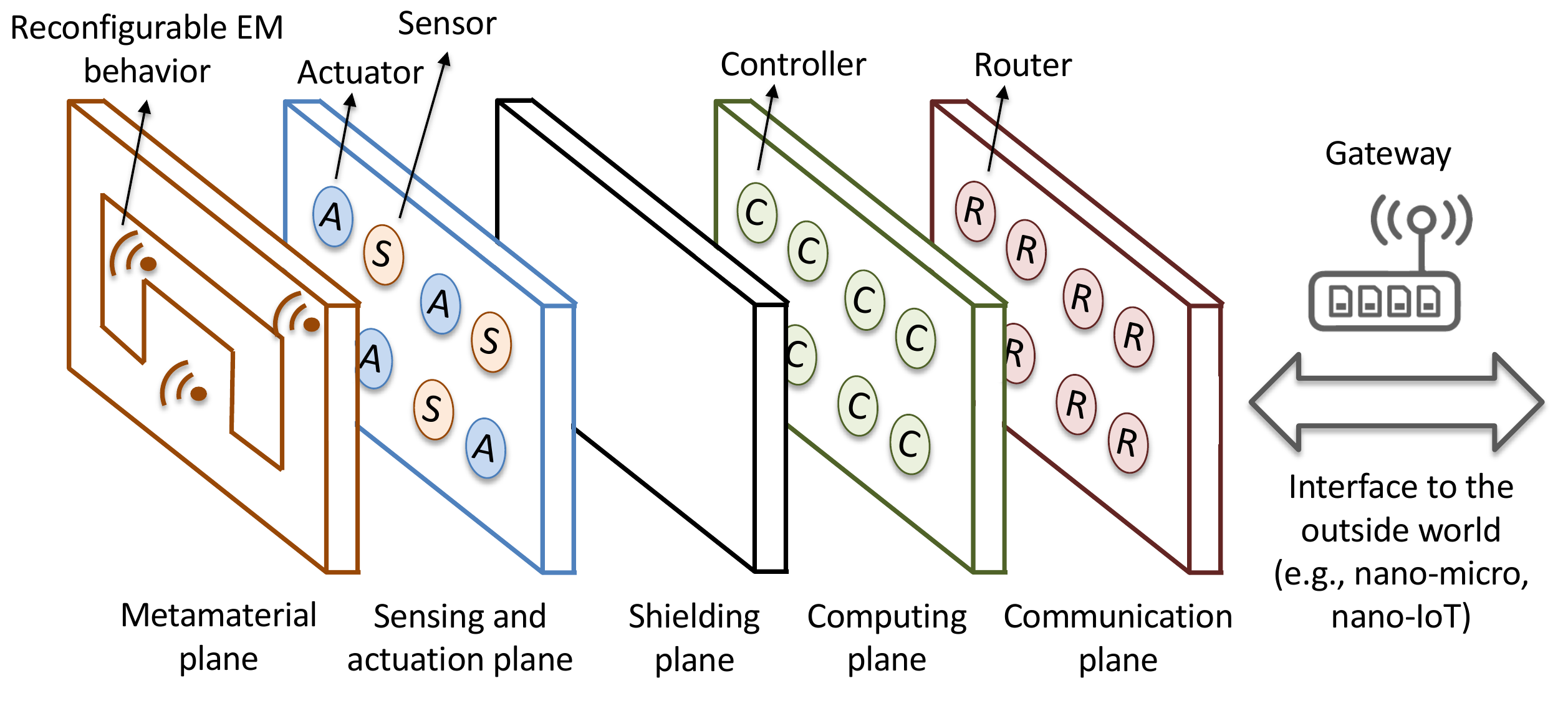}
\caption{Envisioned high-level architecture for enabling software-defined metamaterials~\cite{lemic2020idling}}
\label{fig:metamaterials}
\vspace{-3mm}
\end{figure}

Fundamentally, the behavior of \acp{SDM} hinges on the collective response of each of its subwavelength unit cells. To enable the reprogramming without compromising the autonomy of the metamaterial, the \ac{SDM} paradigm proposes to embed a communication network of controllers within the metamaterial, as shown in Figure~\ref{fig:metamaterials}. In such a scenario, each controller interacts locally with its associated unit cells to adjust its properties and communicates with other unit cells to obtain or distribute the desired behavior. However, the subwavelength scale the SDM unit cells poses a frequency-dependent form-factor limit on the internal network of controllers. In this context, to enable the SDM applications in a wide range of frequencies, nanocommunications in the THz band becomes a desired paradigm to implement the communication between the unit cell controllers within an SDM~\cite{abadal2017computing}.
Depending on the actual application, the number of unit cells can range from thousands to millions~\cite{abadal2017computing, bjornson2020power}, translating to a similar number of controllers in the THz nanonetwork. Their exact number will depend on the physical sizes of the SDMs and the final application, which in some predictions could cover the walls of an entire office for the programmable wireless environment case~\cite{liaskos2019novel}. Moreover, the controllers will have to be interconnected for better adaptation and operative range purposes. Due to the small form-factor and a huge number of envisioned controllable metamaterials comprising an intra-SDM network, the energy consumption of each metamaterial and consequently of the intra-SDM network will have to be low. Owing to practical reasons, these metamaterials will only have energy harvesting capabilities, with capacitor-based storage instead of batteries. 

Currently, a first wave of SDM designs is under development to showcase the capabilities of this new technology~\cite{abadal2020programmable}. The latency requirements are expected to be relaxed in this first stage, i.e., between a few milliseconds and a few seconds~\cite{abadal2017computing}. Moreover, very simple controllers and intra-SDM network infrastructure will be deployed in first prototypes. The network traffic is expected to be downlink mostly, predominantly used for controlling the behavior of metamaterials. Moreover, the initial phase envisions no mobility, i.e., the metamaterials and network nodes are expected to be static. Low reliability of data delivery will presumably suffice and the security is not expected to play an important role~\cite{abadal2017computing}. This is mainly because, in the current stage, the aim is to deliver a proof of concept of the SDM paradigm.

To the best of our knowledge, derivations in regard to the performance requirements for SDMs in their roll-out phase are scarce in the current literature \cite{Saeed2019,tataria2020real} and we make a brief analysis here based on the discussions within \cite{abadal2020programmable}. In essence, the \ac{SDM} is a programmable metasurface with a control sub-system that modifies the response of the metasurface at a given maximum rate that depends on the application. 
In holographic displays, \acp{SDM} should be programmable at the typical refresh rates of displays, this is, around 60 times per second. 
In wireless programmable environments, \acp{SDM} need to be fast enough to adapt to changes in the channel due to mobility or other phenomena, which typically occur at the 1-100 milliseconds scale. 
Finally, in \acp{SDM} used as wireless communication transmitters, the required refresh rate of the metasurface is thus the modulation rate of the transmitter.

The figures above allow us to make an estimation of the performance requirements of intra-SDM networks. 
On the one hand, the intra-SDM communication latency needs to be a fraction of the required refresh time: in the millisecond range for holographic displays and programmable wireless environments and below for metasurface-based wireless transmitters. 
With respect to the throughput, we can assume that an \ac{SDM} module will consist of 10000 unit cells and will cover an area of 100 cm\textsuperscript{2} based on recent estimates \cite{liaskos2019novel,tang2019programmable}. 
If the range of an individual intra-SDM controller operating in THz frequencies is 1~cm, roughly 10x10 interconnected controllers are needed for controlling the whole metamaterial. 
Say that every 100~ms the metamaterial elements have to be updated and that this update is performed in a flood-like multi-hop fashion using a byte-long command, starting from a controller positioned in one corner of the metamaterial surface.
Note that a byte-long command has been selected as most EM functionalities can be achieved with 8-16 states~\cite{liu2019intelligent}.
Under these assumptions, the number of hops until all controllers are reached equals $2\times(10-1)\times10=180$.
Note that each transmission between controllers can potentially also be utilized by instrumenting the metamaterials controlled by the controller, hence we arrive to the required minimal data throughput of 1800 transmissions per sec = 14.4~kbps. 
Arguably, the signaling overhead in such scenario is relatively low and, to further simplify the analysis, we can assume that the network throughput equals the data throughput.
Note that, due to a variety of simplifications made in our derivations, the derived value should be used only as a very rough indication of the minimum required network throughput.
With that in mind, we can then provide a rule of thumb estimate of the required network throughput in the order of 1 to 50~kbps (Table~\ref{tab:applications_requirements}).

As the SDM technology evolves, new prototypes will arise that explore its full potential. Demonstrations of mission-critical applications with stronger timing requirements on the order of microseconds are expected~\cite{abadal2017computing,abadal2020programmable}. Moreover, \acp{SDM} are also envisioned to become wearable, thus having the ability to bend, stretch, and roll~\cite{walia2015flexible}. For the supporting network, this will represent an additional requirement in terms of shape resiliency and operation in high mobility scenarios. For enabling mission-critical applications and although \acp{SDM} show certain resilience to faults \cite{taghvaee2020error}, the reliability of communication will have to be high, while some guarantees for security will also have to be in place. For supporting a variety of potential applications in this domain, addressing will be required on a level of an individual controller, or even on the level of an individual metamaterial element. For similar reasons, the communication links will have to be bidirectional enabling communication from the user to the SDM unit cells and vice versa, as well as among SDM unit cells to implement distributed sensing and intelligence within the device \cite{abadal2020programmable}. In terms of throughput and using the same approach for analysis as before and using 10~ms and 10~$\mu$s as the metamaterial unit update period, we arrive to a minimum required network throughputs of 144~kbps and 144~Mbps, respectively. Hence, roughly speaking the required network throughput for SDMs of the second generation will be in the range of 50~kbps to 500~Mbps. Note that in the derivation we assumed that each transmission between controllers is simultaneously used as a command for controlling the corresponding metamaterials. For more detailed traffic analyses, we refer the reader to \cite{Saeed2019}.

\subsection{Wireless Robotic Materials}

\begin{figure}[!t]
\vspace{1mm}
\centering
\includegraphics[width=\columnwidth]{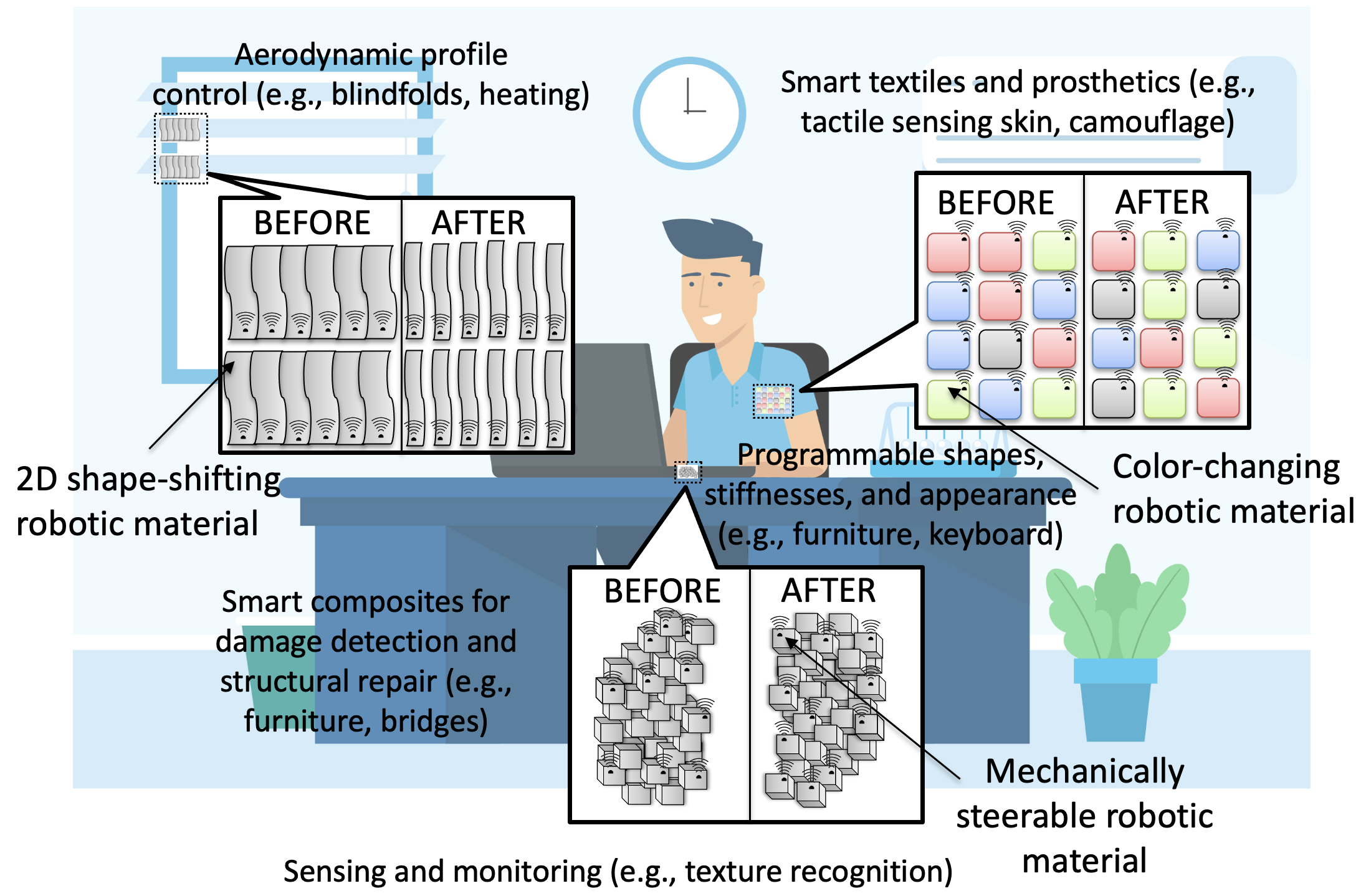}
\caption{Example applications of wireless robotic materials}
\label{fig:robotic_materials}
\vspace{-3mm}
\end{figure}

In contrast to \acp{SDM} that are envisioned to control electromagnetic waves, wireless robotic materials are expected to enable smart composites that autonomously change their shape, stiffness, or physical appearance in a fully programmable way~\cite{mcevoy2015materials,mcevoy2016shape}.
The term wireless robotic material has been coined in~\cite{mcevoy2015materials,Correll:2017:WRM:3131672.3131702}. 
They define the robotic materials as \textit{multi-functional composites that tightly integrate sensing, actuation, computation, and communication to create smart composites that can sense their environment and change their physical properties in an arbitrary programmable manner.} 
The applications that the authors in~\cite{mcevoy2015materials} suggest are \textit{airfoils that change their aerodynamic profile, vehicles with camouflage abilities, bridges that detect and repair damage, or robotic skins and prosthetics with a realistic sense of touch.} 
Similarly, the authors in~\cite{Correll:2017:WRM:3131672.3131702} envision applications such as tactile sensing skin, robots (i.e., nanodevices) that can reproduce patterns projected onto them for camouflage, and a dress that can localize the direction of incoming sound and display it to its wearer. 
Several promising applications envisioned to be enabled by wireless robotic materials are depicted in Figure~\ref{fig:robotic_materials}. 

As argued in~\cite{Correll:2017:WRM:3131672.3131702}, the envisioned applications will require very large swarms of nanodevices tightly integrated into fabric, on skin, etc.  
This poses limitations in terms of the size of the elements of robotic materials, yielding the THz band as one of the most promising communication paradigms for controlling these elements. 
Moreover, the network size and density will be largely influenced by the application that the network is envisioned to support.  
We believe that the application of enabling camouflage abilities will require the largest (i.e., covering an entire vehicle or human body) and most dense networks. 
Nonetheless, the requirements for network size and node density are expected to be less pronounced than for the \acp{SDM}, primarily due to the expected difference in sizes of the robotic materials (a few millimeters) and metamaterials (potentially much smaller than 1~mm).

In terms of network traffic, the authors in~\cite{Correll:2017:WRM:3131672.3131702} argue that the envisioned sensors and actuators embedded in wireless robotic materials could generate information ranging from binary (e.g., for enabling distributed gesture recognition~\cite{hosseinmardi2015distributed}) to a few-hundreds-of-hertz-bandwidth signals (e.g., localized texture recognition by robotic skin~\cite{hughes2015texture}). 
Let us provide a simple calculation for deriving for supported traffic load by the network of wireless robotic materials.
Similar to the previous calculations, we make a vague assumption that a robotic material patch will consist of 100 elements that have to be updated every 20~ms, which is the assumption taken from the Tactile Internet use-cases~\cite{fettweis2014tactile} in which the network latency has to be comparable to the human observational abilities. 
Furthermore, we assume that this update is performed in a flood-like multi-hop fashion, starting/ending at the source/sink node positioned in the corner of the wireless robotic material patch.
Under these assumptions and following the same approach as before, the network traffic needed for controlling the wireless robotic material patch then equals 450~kbps and 3.6~Mbps for control signals carrying 1 and 8 bits of information, respectively.
Utilizing the numbers, we estimate that the network throughput of roughly between 100~kbps and 10~Mbps will be needed for enabling the wireless robotic materials-related applications.
Note that, intuitively, the network will have to support bidirectional traffic, primarily for enabling the vision of sensing and actuating networks~\cite{van1993sensor}.  

As summarized in Table~\ref{tab:applications_requirements}, the reliability of data delivery in the networks enabling wireless robotic materials will have to be high for some applications, e.g., for detecting and repairing damage on bridges, ceilings, and other ``critical'' structures.
Taking the same application as an example, the security of data transmission will have to be high for the above-mentioned applications.
The applications also involve wearable electronics, ``smart'' dresses, and artificial skin, all being carried by a person.
Hence, the mobility is expected to be very high for some of the applications.
In some cases, there will be a need to localize the nodes of the network under such mobility conditions, e.g., for localized texture recognition by robotic skin~\cite{hughes2015texture}.  
Furthermore, some applications will tolerate cluster-based addressing of nodes (e.g., bridge repairs), while some others may require individual addressing (e.g., camouflage).
Finally, the energy consumption of the devices and consequently in the networks will in some cases have to be low, e.g., when the devices are expected to have long lifetime such as in construction monitoring scenarios.
However, since these devices are expected to be larger than the metamaterials discussed above, their energy efficiency and power profile requirements are not expected to be as stringent as for the metamaterials.
Although the larger robotic materials could be powered by smaller batteries, for the smaller robotic materials and dense networks the authors in~\cite{Correll:2017:WRM:3131672.3131702} suggest to use energy scavenging and harvesting~\cite{jornet2012joint2}. 

Note that in the literature, researchers often make a distinction between wireless robotic materials and wireless nanosensor networks. 
We find such a distinction unnecessary, given that we separate different applications based on the requirements they pose on the supporting nanonetwork. 
Wireless nanosensor networks have been proposed in~\cite{akyildiz2012nanonetworks} and envision applications such as high-resolution environmental monitoring~\cite{afsharinejad2016performance}, wearables~\cite{petrov2015feasibility}, nanocameras-based extreme spatial resolution recordings~\cite{simonjan2018nano}, nanoscale imaging~\cite{knap2017thz}, and the Internet of Multimedia Nano-Things~\cite{jornet2012internet}.
Nonetheless, these applications can be viewed as a sensing-only subset of applications enabled by the wireless robotic materials, hence we do not group them into a separate category.

\subsection{In-body Communication}

Mobile medical nanodevices are a promising technology for \textit{in-situ} and \textit{in-vivo} applications~\cite{sitti2015biomedical,agoulmine2012enabling}.
These nanodevices will be access small regions of the human body (e.g., gastrointestinal, brain, spinal cord, blood capillaries, inside the eye), while essentially being non-invasive. 
The authors in~\cite{sitti2015biomedical} argue that mobile medical nanodevices \textit{could even enable access to unprecedented sub-millimeter size regions inside the human body, which have not been possible to access currently with any medical device technology}. 
{\color{red}{Similarly, the authors in~\cite{piro2015terahertz} envision: ``\textit{therapeutic nanomachines able to operate either in inter- and intra-cellular areas of the human body}``.
By doing so, pioneering applications, such as immune system support, bio-hybrid implant, drug delivering system, health monitoring, and genetic engineering, will be enabled, argue the authors.}
The authors in~\cite{abbasi2016nano} envision applications such as human physical movement monitoring, early diagnosis and treatment of malicious agents (e.g., viruses, bacteria, cancer cells), bone-growth monitoring in diabetes patients, and organ, nervous track, or tissue replacements (i.e., bio-hybrid implants). 
Finally, the authors in~\cite{ali2015internet} outline several example applications that could be supported by THz nanonetworks, such as smart drug administration, nanoscale surgeries, and epidemic spread detection and management.}

\begin{figure}[!t]
\vspace{-1mm}
\centering
\includegraphics[width=\columnwidth]{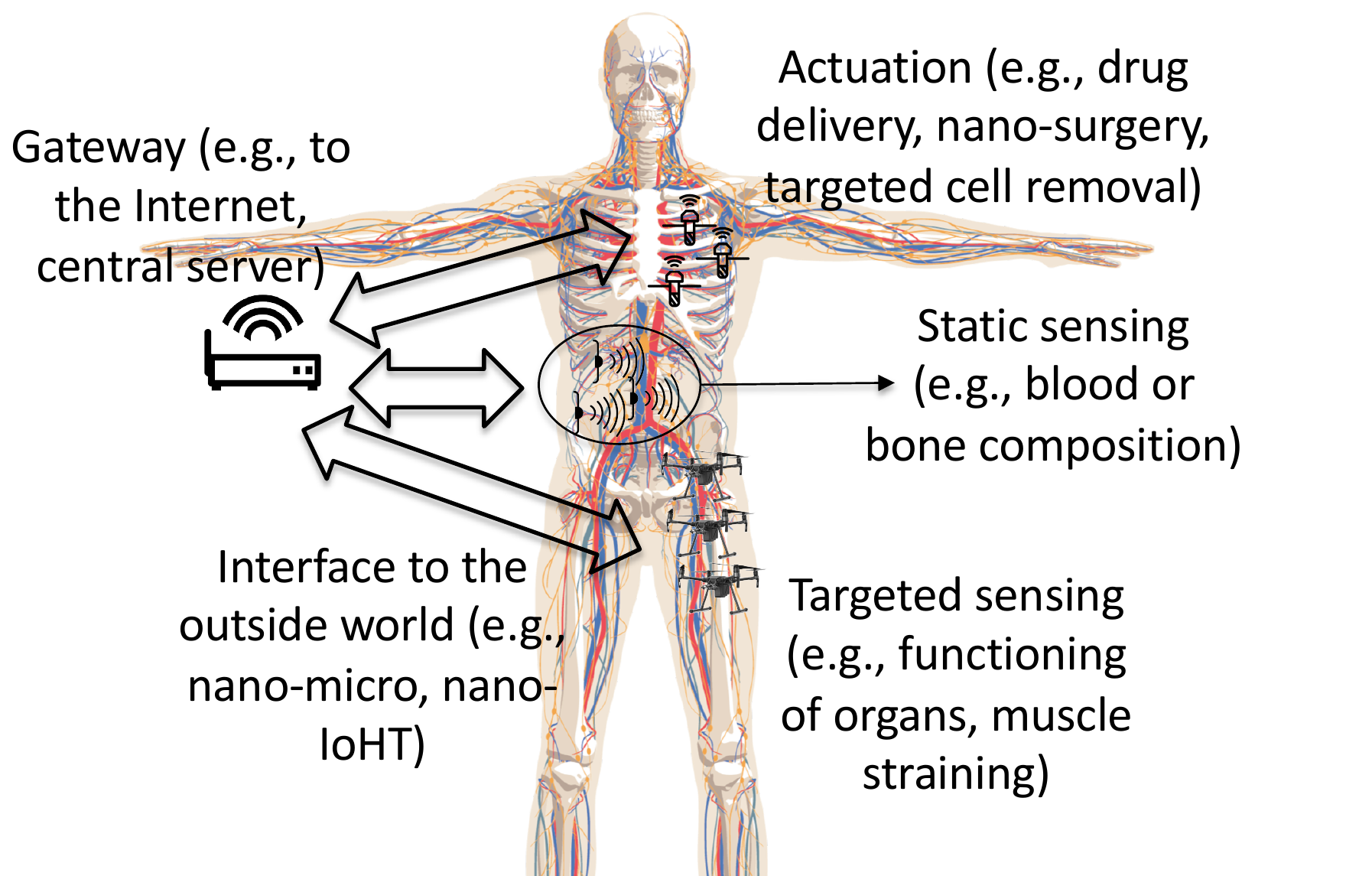}
\caption{Envisioned high-level architecture for enabling body-centric applications}
\label{fig:healthcare}
\vspace{-5mm}
\end{figure}

As indicated in Figure~\ref{fig:healthcare}, the nanodevices would be able to perform sensing (e.g., blood composition or functioning of specific organs) and actuation (e.g., targeted drug delivery), all while reporting or being controlled from the outside world. 
Obviously, the form-factor of these nanodevices will be of prime importance, again yielding THz band communication as one of the most suitable communication paradigms.     
The number of mobile medical nanodevices is expected to be very large (up to a billion according to some estimations~\cite{lewis1992behavioral}) for some applications (e.g., for tissue engineering or detecting bacteria via swarms of sub-millimeter-scale nanodevices).

The amount of traffic that the network of mobile medical nanodevices will have to support will largely depend on the application.
Given that the aim of the nanodevice is to enter and sense/influence something in sub-millimeter regions inside the human body, the main goals of the network will be to deliver this information to the outside world and to support the control of the nanodevice.
The network \textit{per-se} will be formed by only a set of devices relaying the information to/from the outside world. 
However, given that the aim of a swarm of medical nanodevices is to sense a variety of events inside of a body (e.g., the presence of a bacteria) or to form a tissue, the supporting network will arguably be mesh-like. 
The nanodevices will in this case also need to create control loops with the outside world, hence the network will have to operate under real-time constraints.
Let us assume one such scenario, i.e., a swarm of mobile nanodevices is sensing a human brain and potentially creating an artificial tissue if there is damage in the brain.  
Roughly speaking, the network is distributed across the area of 1000~cm\textsuperscript{3} and consists of 10\textsuperscript{6} mobile nanodevices (a relatively conservative assumption), each one sending or receiving 8 bits of information every 20~msec.
We assume flood-based distribution of traffic from or toward the outside world.
Utilizing similar calculation approach as before, we come to the staggering number of transmissions which equals roughly 3$\times$10\textsuperscript{6} transmissions/s, resulting in the network throughput of 24~Mbps.
Along the above derivation, the required network throughput for enabling body-centric applications will - as a rule of thumb - be in the range between 1 and 50~Mbps.

The body-centric communication related network requirements are arguably the most challenging among the applications potentially supported by THz band communication, as summarized in Table~\ref{tab:applications_requirements}.
In addition to large network sizes, high throughput requirements, and in-body propagation of signals, the energy consumption of the nanodevices and consequently the networks supporting their operation have to be very low, primarily due to the required small form-factor of the devices. 
Hence, these nanodevices are presumably going to use energy harvesters are their only energy source. 
Moreover, the reliability of communication will have to be very high (e.g., for controlling medical nanodevices in a brain), while the end-to-end latencies will have to be very low for enabling real-time control of the nanodevices.
Obviously, the security of communication will have to be very high, especially for the ones envisioned to stay in a person's body for a longer period (e.g., for monitoring purposes).
This is to avoid the nanodevices being ``hijacked'' by the attackers, while the patient is not being in a controlled environment shielded from the potential attackers.
In cases when the nanonetworks are not extremely large, the addressing will have to be individual in order to e.g., control the movements of a particular nanodevice~\cite{sitti2015biomedical}.
Moreover, the nanodevices are envisioned to be localizable and traceable~\cite{sitti2015biomedical} for enabling localized sensing and movement control of the nanodevices.
Finally, due to both blood streams in a person's body and potential movement of the person (primarily the relative movement of person's body parts respective to one another), the nodes are expected to be highly mobile, which poses an additional challenge for the supporting nanonetworks~\cite{sitti2015biomedical}.

{\color{red} Note that the above example, as well as the requirements that the applications are expected to pose on the supporting nanonetworks, have been derived for the most challenging applications. 
In contrast to the other promising application domains discussed in this work, for in-body communication there are several recent efforts targeting the derivation of network requirements for only certain applications in the domain.
In this direction, the interested reader is referred to~\cite{asorey2020analytical,asorey2020throughput}, where the authors derive the achievable throughput as a function of various network parameters such as transmission rate and nanonodes' available energy.
The work was carried out assuming flow-guided nanonetworks (i.e., the ones operating in the blood circulatory system).
Moreover, the work was extended in~\cite{asorey2020flow} to demonstrate the feasibility of flow-guided nanonetworks in enabling low throughput applications such as viral load
monitoring, restenosis, sepsis and bacterial blood infections, and heart attacks.
Moreover, the authors in~\cite{ali2015internet} qualitatively distinguish nanonetwork requirements based on the high-level function of the applications. 
They envision the monitoring, detection, and therapy-related functionalities, however without providing quantitative characterizations.}

\subsection{On-chip Communication}
\label{sec:chipCom}
Virtually all processors nowadays are based on multi-core architectures where a single chip contains multiple independent processor cores and a given amount of on-chip memory. The different processors compute in parallel and use the memory to share data and synchronize their executions. In this context, the current trend to increase performance is to integrate more cores within the same chip \cite{Bohnenstiehl2017}. This, however, places an increased burden to the on-chip interconnect, which is used to send control signals and move shared data across the chip, to the point of turning intra-chip communication into the key determinant of the processor's computational performance and energy efficiency \cite{Abadal2018a}. Hence, substantial research efforts focused on the on-chip interconnects, with initial bus-based interconnects soon giving way to more efficient and resilient \acp{NoC}. Initially, \acp{NoC} were wired solutions, which posed challenges in terms of delay, power requirements, and area overhead {\color{red}as more cores were integrated within a single chip~\cite{Marculescu2009,Nychis2012,Bertozzi2014}. This has prompted the proposal of multiple alternative interconnect technologies \cite{Kim2012Survey,Bertozzi2014}, among which we find wireless on-chip communications.}


\begin{table*}[!ht]
\vspace{1mm}
\begin{center}
\caption{Summary of requirements in different application domains}
\label{tab:applications_requirements}
\begin{tabular}{l  cx{1.9cm} cx{3.3cm} cx{3.6cm} cx{3.9cm} cx{3.6cm}} 
\hline
\textbf{Requirements} & \multicolumn{2}{c}{\textbf{Software-defined metamaterials}} & \textbf{Wireless robotic materials}  & \textbf{In-body communication} & \textbf{On-chip communication} \\ 
 & \textbf{Gen.~1} & \textbf{Gen.~2} &  &  &  \\ \hline
Network size        & $10^3$ to $10^6$    & $10^9$     & 10 to 10$^6$              & 10$^3$ to 10$^{9}$          & Up to 10$^3$ \\
Node density        & \multicolumn{2}{c}{100 to 10000 nodes per cm$^2$}     & 1 to 100 nodes per cm$^2$ & $>$10$^3$ nodes per cm$^3$ & 10-100 per mm$^2$ \\
Latency             & ms to s & $\mu$s              & ms                        & ms to s                    & 10-100~ns \\
Throughput          & 1-50~kbps & 50~kbps to 500~Mbps            & 100~kbps-10~Mbps      & 1-50~Mbps              & 10-100~Gbps \\
Traffic type        & downlink & bidirectional      & bidirectional 	        & bidirectional              & bidirectional \\ 
Reliability  	    & low & medium                    & high          	        & very high                  & very high \\ 
Energy consumption  & very low & very low   & low           	        & very low                   & low \\
Mobility            & none & medium to high         & high          	        & high                       & none \\
Addressing          & none to cluster & individual  & cluster to individual     & individual                 & individual \\
Security            & none & low to medium          & high                      & very high                  & medium \\
Additional features &             &                 & localization              & in-body communication      & \\
					&             &                 &                           & localization \& tracking   & \\
\hline
\end{tabular}
\end{center}
\vspace{-4.5mm}
\end{table*}

\begin{figure}[!t]
\centering
\includegraphics[width=0.87\columnwidth]{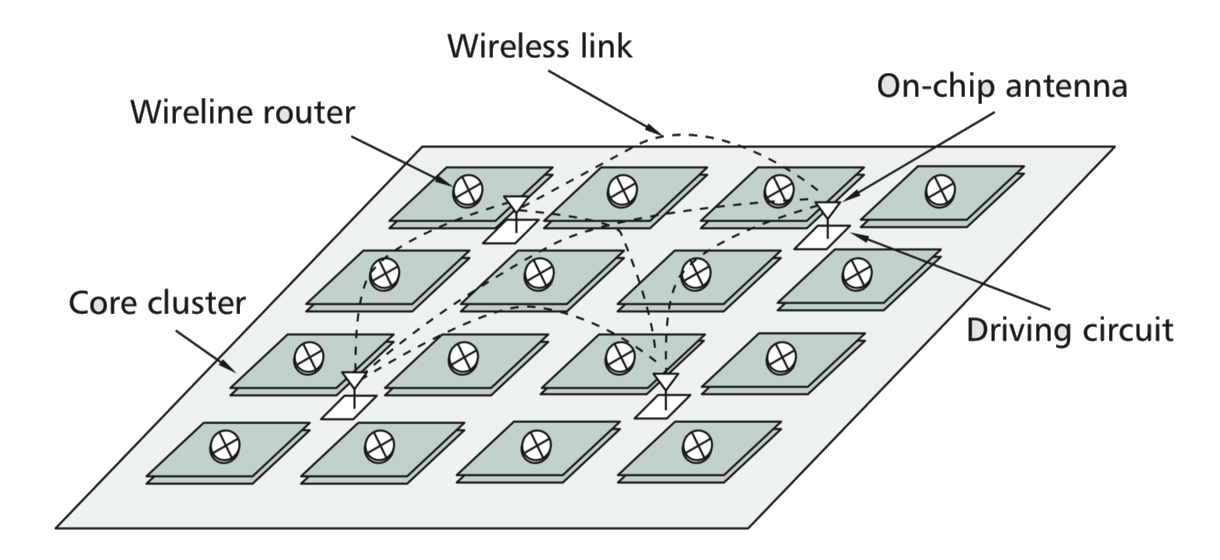}
\caption{Envisioned high-level architecture for enabling on-chip communication~\cite{abadal2013graphene}}
\label{fig:on_chip}
\vspace{-4mm}
\end{figure}

The advantages of employing wireless communication for intra-chip networks include reduced propagation delay, reconfigurability, and improved scalability in terms of latency, throughput, and energy consumption~\cite{ganguly2011scalable, wang2011wireless}.
Nevertheless, as shown in Figure~\ref{fig:on_chip}, current \acp{WNoC} are predominantly utilized for long-range point-to-point links for decreasing the average hop count of traditional NoC solutions.
In other words, WNoC are currently deployed for enhancing the wired NoC, mostly due to the relatively large sizes of the metallic antennas required for enabling wireless communication in the mmWave band, which has generally been assumed in this context. 
Recently, Abadal \emph{et al.}~\cite{abadal2013graphene} proposed the employment of nanoscale \acp{WNoC} by means of graphene nanoantennas. 
Graphene-based nanoantennas with sizes of only a few micrometers, i.e., two orders of magnitude below the dimensions of metallic antennas, could provide inter-core communication utilizing the THz frequencies. Moreover, the antennas are inherently tunable, providing new ways to reconfigure the network. 
This novel approach is expected to fulfill the stringent requirements of the area-constrained, latency-bound, and throughput-intensive on-chip communication. This concept has recently been further developed~\cite{saxena2017folded, saxena2020scalable}.

In the envisioned multicore on-chip communication, on-chip traffic typically consists of a mixture of short control messages employed for cache coherence, data consistency, and synchronization purposes, together with larger data transfers. The communication is clearly bidirectional, while the addressing in the network has to be individual, as reflected in Table~\ref{tab:applications_requirements}. Depending on the memory access patterns exhibited by the application, communication will have varying degrees of unicast, multicast, and broadcast transmissions. Note, for instance, that some applications have strong all-to-all communication patterns. 
Although latency is the primary concern in on-chip communications, as delays in the serving of packets essentially delay the whole computation, throughput is an important metric to avoid throttling of the computing cores. The recent literature reports the on-chip network throughput requirement in the range of 10-100~Gbps~\cite{Abadal2018a,Mestres2016}, which can be further pushed to the Tbps barrier in communication-intensive architectures such as accelerators \cite{Kwon2017}.


As mentioned above, the WNoC scenario has very challenging end-to-end latency requirements in the range of sub-$\mu$s~\cite{Abadal2018a} with high reliability (typical \acp{BER} lower than $10^{-15}$~\cite{postman2012survey,Abadal2018a} to compete with that of on-chip wires). As the chips are generally shielded, security issues related to on-chip communication are not of paramount importance. 
In the unlikely event of hardware trojans being present within the chip, several lightweight measures can be taken to avoid spoofing, eavesdropping, or \ac{DoS} attacks \cite{Lebiednik2018a}.
Moreover, although the chips are expected to be mobile, the relative locations of the nodes on a chip are not going to change. 
In that sense, there is no mobility of network nodes that should be anticipated.  
Due to very small sizes of the chips, the energy efficiency of the network should be high.
This is to avoid overheating of the chip due to high energy dissipation~\cite{yalgashev2014performance}.  $\blacksquare$

A summary of the requirements that different application domains pose on the supporting communication networks is given in Table~\ref{tab:applications_requirements}. 
Compared to other application domains, software-defined metamaterials posit the least stringent constrains for the majority of requirements. 
This pertains to the throughput and security requirements, as well as the traffic type and mobility support. 
On-chip communication can be considered as a relatively unique application domain as the topology of a network and propagation characteristics can be considered as static and known upfront~\cite{abadal2015broadcast}.
Nonetheless, compared to other domains, this domain poses the most stringent constraints in terms of node density, delivery latency, achievable throughput, and reliability of communication. 
We have mentioned before that electromagnetic nanocommunication can be utilized in different propagation mediums, in contrast to molecular nanocommunication.
In addition, we've stated that electromagnetic communication in THz frequencies, primarily due to graphene, can support device minimization and antenna tunability, in contrast to mmWave, microwave, and other electromagnetic frequency bands.
Due to that and in contrast to other paradigms, electromagnetic nanocommunication in THz frequencies is a promising candidate for supporting \textit{all} of the discussed application domains.

In theory, the THz band can support very large bit-rates, up to several Tbps. 
However, it is clear from the discussion above and Table~\ref{tab:applications_requirements} that such throughputs will not be required for enabling the envisioned applications domains.
Nonetheless, a very large bandwidth enables new simple communication mechanisms suited for the expected limited capabilities of nanodevices, primarily in terms of their energy levels~\cite{jornet2012phlame}.
In addition, a large bandwidth enables the development of efficient medium sharing schemes~\cite{wang2013energy}, which can both enable low energy performance and scale to the required numbers and densities of the nanonodes.
In addition, it is worth emphasizing again that the THz frequency band is ``the last piece of RF spectrum puzzle''~\cite{elayan2019terahertz}.
Hence, the interferences from other communication sources in the same frequencies are currently virtually non-existent.
In the future, they are expected to be very low, due to the low utilization, high attenuation, and large bandwidth of the THz frequency band~\cite{elayan2019terahertz}.
This suggests that highly reliable nanocommunication, which is required by several of the outlined application domains, can be enabled by utilizing the THz band.

The above considerations suggesting the potential for low power, highly scalable, and reliable communication make a strong argument in favor of using electromagnetic nanocommunication in the THz frequency band for supporting the desired application domains. 
Certainly, there are unresolved challenges to be addressed, and one aim of this survey is to detect, summarize, and discuss them.


\section{Network Layer}
\label{sec:network}

The network layer functionality is responsible for enabling data communication between connected THz nanonodes at arbitrary distance from each other.  
To enable such communication, nanonodes might rely on intermediate nanonodes (hops) to forward information. 
Forwarding functionality might be available for all or a subset of nanonodes. 
Routing functionality ensures that (collective) forwarding behaviour results into successful paths between data sources and destinations. 
The nanonodes therefore might rely on mechanisms to be identified and/or addressed, either individually or as a group (e.g., based on their physical location). 
Due to hardware limitations in transceivers, traditional routing, forwarding, and addressing schemes are often not applicable. 
Traditional routing protocols, as used in e.g., the Internet, rely on the exchange of control or meta messages to learn and distribute information about the network topology or reachability. 
However, memory, channel, and energy restrictions in nanonetworks impose stringent requirements on the relay, forwarding, and routing functionalities.
These constraints are similar to those encountered in the domain of \acp{WSN}. Some THz routing protocols expand upon conventional mechanisms by incorporating THz-specific link models when deciding between various paths (e.g., by incorporating molecular absorption loss).
However, an additional challenge of routing in THz nanonetworks is the scale, since the majority of WSN routing protocols are optimized for up to 200 motes~\cite{al2004routing}, while some of the application domains enabled by THz nanonetworks require significantly larger number of nanonodes, as summarized in Table~\ref{tab:applications_requirements}.
In addition, in many applications nanonodes are more prone to failures than sensor motes, as their only powering option is to harvest energy from the available environmental sources.
Therefore, other research efforts take application-specific information into account for settings which are outside the scope of WSNs.

\begin{figure}[!t]
\centering
\scalebox{.4}{
\tikzstyle{block} = [rectangle, draw, fill=blue!10, 
 text width=13.5em, text centered, rounded corners, minimum height=4em]
\tikzstyle{line} = [draw, -latex']
\makebox[\textwidth][c]{
\begin{tikzpicture}[node distance = 2.5cm, auto]
 \node [block] (main) {\Large \textbf{THz Network Layer Protocols}};
 \node [block, below of=main, xshift=-110, yshift=-60] (relaying) {\Large
     \textbf{Relaying \& Forwarding\\(Sect. \ref{ssec:relaying})} \\
     Xia \emph{et al.}~\cite{xia2017cross} \\
     Rong \emph{et al.}~\cite{rong2017relay} \\
     Yu \emph{et al.}~\cite{yu2015forwarding} \\
     PESAWNSN~\cite{yen2017energy} \\
     Pierobon \emph{et al.}~\cite{pierobon2014routing} \\
      {\color{red}C\'{a}novas-Carrasco \emph{et al.}~\cite{canovas2018nanoscale} \\
     C\'{a}novas-Carrasco \emph{et al.}~\cite{canovas2019optimal}} \\
 };
 \node [block, below of=main, xshift=115, yshift=15] (routing) {\Large \textbf{Routing}};
 \node [block, below of=routing, xshift=-80,  yshift=-40] (flooding) {\Large 
     \textbf{Flooding-based routing\\(Sect. \ref{ssec:flooding})} \\
     Liaskos \emph{et al.}~\cite{liaskos2015promise} \\
     Tsioliaridou \emph{et al.}~\cite{tsioliaridou2016lightweight} \\
     Afsana \emph{et al.}~\cite{afsana2015outage} \\
     LENWB~\cite{stelzner2018body} \\
     Buther \emph{et al.}~\cite{buther2018hop} \\
     E\textsuperscript{3}A~\cite{al2017cognitive} \\
 };
 \node [block, below of=routing, xshift=150, yshift=25] (pathfinding) {\Large
     \textbf{Pathfinding} \\
 };
 \node [block, below of=pathfinding, left of=pathfinding, xshift=-10, yshift=-30] (regular) {\Large
     \textbf{Regular topologies\\(Sect. \ref{ssec:regular_topologies})} \\
     Radetzki  \emph{et al.}~\cite{radetzki2013methods} \\
     Maze routing~\cite{fattah2015low} \\
     Fukushima \emph{et al.}~\cite{fukushima2009fault} \\
     Jovanovic \emph{et al.}~\cite{jovanovic2009new} \\
     Ebrahimi \emph{et al.}~\cite{ebrahimi2013md} \\
     Saeed \emph{et al.}~\cite{saeed2018fault} \\
 };
 \node [block, below of=pathfinding, right of=pathfinding, xshift=10] (irregular) {\Large
     \textbf{Irregular topologies\\(Sect. \ref{ssec:irregular_topologies})} \\
     Tsioliaridou \emph{et al.}~\cite{tsioliaridou2015corona} \\
     Tsioliaridou \emph{et al.}~\cite{tsioliaridou2017packet} \\
 };
 \path [line] (main) -| (relaying);
 \path [line] (main) -| (routing);
 \path [line] (routing) -| (flooding);
 \path [line] (routing) -| (pathfinding);
 \path [line] (pathfinding) -| (regular);
 \path [line] (pathfinding) -| (irregular);
\end{tikzpicture}
}
}
\caption{Classification of network layer protocol for THz nanocommunication}
\label{fig:network_layer_classification}
\vspace{-4mm}
\end{figure}
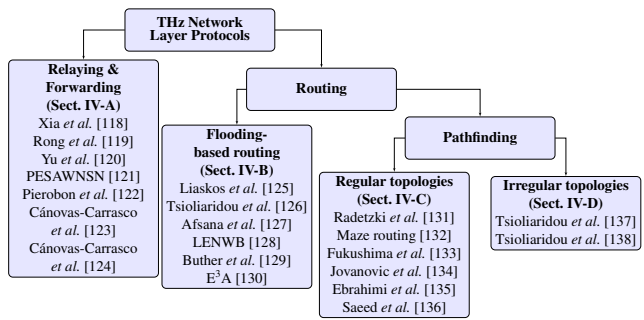

Below we provide an overview on existing research categorized according to the core mechanism they rely on.
Figure~\ref{fig:network_layer_classification} depicts an overview of our approach. 
Subsection~\ref{ssec:relaying} discusses forwarding and relaying methods. These works focus on whether to transmit a packet directly to a hop which is further away, or whether to relay the information via one or more intermediary nanonodes. 
The other subsections focus on algorithms  aiming at finding the nanonodes that should send and/or forward a packet in order to reach one or more destinations (i.e., routing). 
Subsection~\ref{ssec:flooding} focuses on flooding-based mechanisms, which are used for one-to-many routing. 
Subsections~\ref{ssec:regular_topologies} and~\ref{ssec:irregular_topologies} discuss one-to-one routing in regular and irregular topologies, respectively.


\subsection{\vspace{-1mm}Relaying and Forwarding for THz Nanonetworks} \label{ssec:relaying}
Enabling multi-hop communication in THz nanonetworks introduces design restrictions at the lower layers. 
Xia and  Jornet~\cite{xia2017cross} mathematically characterize relaying strategies maximizing network throughput with respect to transmission distance, transmission power energy and packet generation rate in a THz band network consisting of nanonodes with directional antennas. The unique characteristics of THz communication are considered by modeling the distance-dependent and frequency-selective characteristics of molecular absorption.
To reduce the \ac{BER} of cooperative relaying strategies in \acp{WNSN}, amplify-and-forward and decode-and-forward relaying nanonodes are evaluated by Rong \emph{et al.}~\cite{rong2017relay}, whilst taking into account spreading and molecular absorption losses. Using the former forwarding technique, relaying nanonodes amplify the received signal as it is, whilst using the latter technique, relaying nanonodes demodulate the signal before it is forwarded. The bit error rate of both schemes is compared to that of direct transmission, and a relaying gain of approximately 2.2 dB was observed for amplify-and-forward and 5 dB for decode-and-forward.
As molecular absorption and other frequency-selective features seriously hamper multi-hop throughput in the THz band, Yu \emph{et al.}~\cite{yu2015forwarding} propose channel-aware forwarding schemes in \acp{WNSN}, ensuring that data will not be forwarded to: i) a relaying nanonode in a region which is adversely affected by molecular absorption, or ii) to a short-distant nanonode, which will result in unnecessarily large hop count. Compared to direct transmission, the authors report an increase of end-to-end capacity by a factor 12-13, whilst limiting the delay increase to less than 0.02\%. Similarly accounting for molecular absorption, Yen~\emph{et al.}~\cite{yen2017energy} combine signal quality-aware forwarding with data aggregation in \acp{WNSN}, while also taking free path space loss into account. Data aggregation (e.g., taking the minimum of received values) in intermediary nanonodes is often possible, depending on the particular application (e.g., if only the lowest value is needed). The resulting design problem is formulated as an optimization of routing decisions, for which an efficient heuristic has been proposed for calculating minimum cost spanning trees considering transmission power and signal quality. 

Energy efficiency is crucial in THz band nanonetworks. The throughput attainable in a these environments is therefore strongly related to the associated nanoscale energy harvesting processes. Pierobon \emph{et al.}~\cite{pierobon2014routing} present a hierarchical cluster-based architecture for \acp{WNSN}, extending their work on MAC protocols~\cite{wang2013energy}. Each cluster is assumed to have a nanocontroller, which is a nanonode with more advanced capabilities and which can transmit directly to all nanonodes (when given sufficient transmission power). The nanocontroller is in charge of optimizing the trade-off between multi-hop forwarding among individual nanonodes vs. communication via the nanocontroller from the perspective of throughput and lifetime of the network and its associated connectivity. Based on the probability of energy saving through multi-hop transmission, the nanocontroller instructs nanonodes what transmission power to use for optimal hop distance and throughput, as well as selects the next hop on the basis of their energy and current load. Both total path loss and molecular absorption loss are considered in modeling the channel. For nanonodes at a distance of 1 to 10 mm, the authors find that the total energy spent per bit halves using forwarding as compared to direct transmission, whilst the delay improves by a factor 3-4. 

 {\color{red}Another similar hierarchical architecture with a focus on energy management is proposed by C\'{a}novas-Carrasco \emph{et al.} (\cite{canovas2018nanoscale} and \cite{canovas2019optimal}). Concretely, a body area network is proposed, compromising a large number of nanonodes injected into the bloodstream and thus always moving. These nanonodes communicate with one or more nanorouters, which are more powerful devices implanted under the skin. The nanorouters in turn connect to an Internet gateway.
The first of these works \cite{canovas2018nanoscale} focuses on a scenario in which a nanorouter is implanted in a human hand. The work discusses a molecular absorption loss model which takes the different layers of tissues inside the hand into account. Also, it proposes concrete power supply components which harvest energy from the bloodstream itself and from an external ultrasound source (the ultrasound-based power source also allows nanonodes to detect whether they are in vicinity of the nanorouter). The authors estimate that each nanonode is able to send a 40 bit packet every 52 minutes.
\cite{canovas2019optimal} considers a scenario in which the nanonodes report events with different levels of importance. A Markov Decision Process (MDP) for the transmission policy of a nanonode is proposed, the states of which take the type of event detected by the nanonodee into account, as well as the battery level and the distance between the nanonode and the last detected nanorouter. 
The authors compare their approach to policies in which all events are transmitted, only high priority events are transmitted or a decision whether to report an event is made based on an estimate of the nanonode's power level when reaching the next nanorouter. The throughput of the proposed approach always exceeds the throughput of any other policy by at least a couple of percents. Moreover, as events become more frequent, the advantage of the MDP over the other approaches becomes more pronounced.}

\subsection{\vspace{-1mm}Flooding-based Routing} \label{ssec:flooding}
Baseline THz nanonetworking mechanisms are in essence direct extensions of MAC protocols which rely on flooding to deliver data to their intended destinations.  Flooding involves unconditional packet retransmission by all involved nanonodes. Advantages of flooding include its simplicity, its reliability through redundant transmissions and a lack of (topology-dependent) initialization, making it a good choice for applications with mobile nanonodes. Unfortunately, unmodified flooding involves a high number of redundant transmissions.

Therefore, several efforts have been undertaken to mitigate the number of redundant transmissions by limiting the involvement of particular nanonodes or through selectively forwarding in a probabilistic manner.  Liaskos \emph{et al.}~\cite{liaskos2015promise} adopt an adaptive flooding scheme where wireless nanonodes in a static environment can deactivate themselves based on their perceived signal to interference ratio and resource levels and thus create a dynamic flooding infrastructure. Compared to fine-tuned probabilistic flooding, the proposed scheme achieves an improvement of approximately 10\% in packet rate to achieve full network coverage and a similar latency. It also manages to fully remove interference effects. Tsioliaridou \emph{et al.}~\cite{tsioliaridou2016lightweight} follow a similar mechanism relying on the Misra-Gries algorithm to determine if each nanonode should act as user or retransmitter. Concretely, each incoming packet is classified as (i) suffering from a parity error, (ii) a duplicate of an earlier received packet, (iii) a correctly decoded packet which is received for the first time. This results in a stream of events, with the Misra-Gries algorithm is employed to estimate the frequency of each event, thus resulting in statistics which are used to decide if a nanonode should act as a retransmitter. Molecular absorption, fading effects and electromagnetic scattering (simulated using a ray tracing engine) are taken into account in simulating the physical layer. 
In the context of body area networks, Afsana \emph{et al.}~\cite{afsana2015outage} propose a cluster-based forwarding scheme in which cluster controllers are elected as a function of their residual energy. Network communication is then optimized for intra- and inter-cluster communication considering energy consumption at the link level. The resulting performance has been evaluated in terms of \ac{SINR} (for an SNR up to 20 dB) and outage probability up to 0.5, considering the impact of shadowing, molecular absorption, and spreading loss. In all cases, the proposed scheme results in a data-rate improvement of 3-15\% over the baseline.

Various flooding mechanisms are discussed by Stelzner \emph{et al.}~\cite{stelzner2018body} in the context of in-body nanonetworks. The baseline forwarding mechanism floods messages based on a fixed, predetermined probability. A more advanced scheme, `probA' involves adaptively changing that probability according to the estimated density of transmitters close to a given nanonode. Both mechanisms are evaluated within an environment modeling an aorta and artery (a cylindrical environment with a wide or narrow radius). Probabilistic flooding requires approximately 60\% of nanonodes to participate in the routing to achieve full network coverage in both cases. ProbA performs well in the aorta and requires participation of only 40\% of nanonodes, but fails to achieve full network coverage in the more restricted artery.

As an alternative to probabilistic flooding, Stelzner \emph{et al.}~\cite{stelzner2018body} also propose a scheme (i.e., \emph{LENWB}) 
requiring  nanonodes to store information up to 2-hop distant neighbors. In the aorta, LENWB achieves full network coverage when 50\% of nanonodes participate, and in the artery LENWB performs even better, achieving full network coverage with only 20\% of nanonodes forwarding the messages. However, we do note that this high delivery reliability and coverage comes at a significant cost in terms of memory requirements.
Buther \emph{et al.}~\cite{buther2018hop} go a step further, and propose individual nanonodes to learn and store their hop distance to a micro-scale gateway. The hop count is then used as a direction indicator to determine if the packet needs to be flooded or not. They consider a naive version of the algorithm, in which nanonodes do not remember if a message has been forwarded, and propose an optimized approach by invalidating the hop count of nanonodes once they have acted as a forwarding nanonode. This results in decreasing the total massage count from exponential to quadratic behaviour. A similar approach is taken by Al-Turjman \emph{et al.}~\cite{al2017cognitive} in the context of IoNT, where the next hop towards a gateway node is not only determined by its hop count, but is also restricted to the first next hop which satisfies some energy-related constraints (e.g., at least 50\% battery remaining). The path loss model takes both molecular absorption and spread path loss into account. When compared to a shortest path data collection strategy, the proposed approach avoids overusing nanonodes which are close to a gateway, thus resulting in a larger network lifetime (6 times or more). This however comes at the cost of a higher latency due to the selection of longer paths (2-3 times higher).

Each of the discussed algorithms is (to a varying degree) successful in reducing the number of redundant transmissions, and may be considered for implementing one-to-many communication. However, for when one-to-one communication is required, flooding is inefficient by nature, since all nanonodes in the network receive each message. (\cite{buther2018hop} is an exception, but even there all nanonodes will receive a message if the target is far enough from the gateway.) Furthermore, the reduction in the number of transmitted messages is typically at the cost of some advantage of flooding: probabilistic flooding may fail to deliver messages if not tuned carefully, and other mechanisms collect topology-dependent information, making them less useful if the nanonodes are mobile.

\begin{table*}[!ht]
\vspace{-1mm}
\begin{center}
\caption{Summary of network layer protocols}
\label{tab:network_layer_protocols}
\begin{tabular}{l l l l l} 
\textbf{Protocol} & \textbf{Distinct Features} & \textbf{Potential Applications} & \textbf{Evaluation Topology} & \textbf{Evaluation Metrics} \\ 
\hline

Xia \emph{et al.}~\cite{xia2017cross} &
\begin{tabular}{@{}l@{}}- directional antennas \\ - directivity-sensitive buffer \\ \quad and queuing \end{tabular} &
- wireless robotic materials &
\begin{tabular}{@{}l@{}}- 2D Poisson distributed nodes \\ - grid of relay nodes \end{tabular} &
\begin{tabular}{@{}l@{}}- throughput vs. packet \\ \quad generation rate and \\ \quad transmission distance \end{tabular}  \\
\hline

Rong \emph{et al.}~\cite{rong2017relay} &
\begin{tabular}{@{}l@{}} - amplify-and-forward \\ \quad vs. decode-and-forward \\  \end{tabular} &
\begin{tabular}{@{}l@{}} - wireless robotic materials \\ - body-centric communication \end{tabular} & - 3 node network &
- bit-error rate  \\
\hline

Yu \emph{et al.}~\cite{yu2015forwarding} & - channel-aware forwarding  &
\begin{tabular}{@{}l@{}} - wireless robotic materials \\ - body-centric communication \end{tabular} &
\begin{tabular}{@{}@{}l@{}} - 1D (string) network \\ - uniform or random nodes  \\ - 5 to 50 nodes \end{tabular} &
\begin{tabular}{@{}l@{}} - end-to-end capacity \\ - average latency\end{tabular}  \\
\hline

PESAWNSN~\cite{yen2017energy} &
\begin{tabular}{@{}l@{}}- power and signal quality \\ \quad-aware arc weight\\ - data aggregation \end{tabular} &
\begin{tabular}{@{}@{}l@{}} - software-defined met. (gen. I) \\- wireless robotic materials \\- body-centric communication\end{tabular} &
\begin{tabular}{@{}l@{}} - 2D random nodes \\  - 10000 nodes \end{tabular} &
\begin{tabular}{@{}l@{}} - power consumption\\ \quad vs. transmission range\end{tabular}  \\
\hline

Pierobon \emph{et al.}~\cite{pierobon2014routing} &
\begin{tabular}{@{}l@{}}- hierarchical cluster-based  \\ - distributed probabilistic \\ \quad energy-based forwarding  \end{tabular} &
\begin{tabular}{@{}l@{}} - wireless robotic materials \\ - body-centric communication \end{tabular} &
\begin{tabular}{@{}l@{}} - 2D Poisson distributed nodes \\ - 100 nodes \end{tabular} &
\begin{tabular}{@{}l@{}}- average latency \\- capacity per node \\- energy efficiency\end{tabular}  \\
\hline

 {\color{red}\begin{tabular}{@{}l@{}}C\'{a}novas-Carrasco  \\ \quad \emph{et al.}~\cite{canovas2018nanoscale}  \end{tabular}}  &
 {\color{red}\begin{tabular}{@{}l@{}}- hierarchical cluster-based  \\ - circulating nanonodes \\ - energy harvesting  \end{tabular}} &
 {\color{red}\begin{tabular}{@{}l@{}} - body-centric communication \end{tabular}} &
 {\color{red}\begin{tabular}{@{}l@{}} - 1 nanorouter implanted in \\ \quad the skin and 1 nanonode \\ \quad in the center of a vein \end{tabular}} &
 {\color{red}\begin{tabular}{@{}l@{}} - energy efficiency\end{tabular}} \\
\hline

 {\color{red}\begin{tabular}{@{}l@{}}C\'{a}novas-Carrasco  \\ \quad \emph{et al.}~\cite{canovas2019optimal}  \end{tabular}}  &
 {\color{red}\begin{tabular}{@{}l@{}}- hierarchical cluster-based  \\ - circulating nanonodes \\ - MDP-guided  event reporting  \end{tabular}} &
 {\color{red}\begin{tabular}{@{}l@{}} - body-centric communication \end{tabular}} &
 {\color{red}\begin{tabular}{@{}l@{}} - cylindrical volume \\ - random nodes \\ - 6000 nodes \end{tabular}} &
 {\color{red}\begin{tabular}{@{}l@{}} - energy efficiency \\ - effect of nanorouter \\ \quad position on \\ \quad throughput\end{tabular}} \\
\hline

Liaskos \emph{et al.}~\cite{liaskos2015promise} &
\begin{tabular}{@{}l@{}} - adaptive flooding \\ - nodes may deactivate \end{tabular} &
\begin{tabular}{@{}@{}@{}l@{}} - software-defined met. (gen. I) \\ - wireless robotic materials \\ - body-centric communication \\ - on-chip communication\end{tabular} &
\begin{tabular}{@{}l@{}} - 2D uniform grid \\ - 625 to 4000 nodes \end{tabular} &
\begin{tabular}{@{}l@{}} - achieved coverage \\ - mean service time \\ - energy efficiency\end{tabular}  \\
\hline

Tsioliaridou \emph{et al.}~\cite{tsioliaridou2016lightweight} &
\begin{tabular}{@{}@{}l@{}} - adaptive flooding \\ - based on reception statistics \\ - Misra-Gries algorithm \end{tabular} &
\begin{tabular}{@{}l@{}} - software-defined metamaterials \\ - on-chip communication\end{tabular} &
\begin{tabular}{@{}l@{}} - 2D and 3D uniform grids \\ - 1000 to 8000 nodes \end{tabular} &
\begin{tabular}{@{}l@{}} - achieved coverage \\ - mean service time \\ - energy efficiency\end{tabular}
\\
\hline

Afsana \emph{et al.}~\cite{afsana2015outage} &
\begin{tabular}{@{}l@{}}- cluster-based forwarding \\ - centers selected based\\ \quad on residual energy\end{tabular} &
- body-centric communication & - unspecified &
\begin{tabular}{@{}l@{}}- throughput vs. SNR\end{tabular}  \\
\hline

LENWB~\cite{stelzner2018body} &
\begin{tabular}{@{}l@{}}- probabilistic flooding vs. \\ \quad (locally) adaptive flooding \end{tabular} &
- body-centric communication &
\begin{tabular}{@{}@{}l@{}} - cylindrical volumes of \\ \quad various sizes \\ - random nodes \\ - 10 to 2000 nodes \end{tabular} &
\begin{tabular}{@{}l@{}} - coverage \\ - memory usage \end{tabular}  \\
\hline

Buther \emph{et al.}~\cite{buther2018hop} &
\begin{tabular}{@{}l@{}}- flooding based comm. \\ - using hop-distance\\- destructive mode to\\ \quad avoid broadcast storms  \end{tabular} &
- body-centric communication &
\begin{tabular}{@{}@{}l@{}} - cylindrical volume \\  - random and uniform nodes \\ - up to 60 nodes \end{tabular} &
\begin{tabular}{@{}l@{}}- power consumption \\ \quad vs. network size \end{tabular}  \\
\hline

E\textsuperscript{3}A~\cite{al2017cognitive} &
\begin{tabular}{@{}l@{}}- sensor-to-gateway comm.\\ - forwarding based on hop-\\ \quad distance and energy level\end{tabular} &
\begin{tabular}{@{}l@{}} - software-defined met. (gen. I) \\ - wireless robotic materials \end{tabular} &
\begin{tabular}{@{}l@{}} - 2D uniform grid \\ - 1500 nodes \end{tabular} &
\begin{tabular}{@{}l@{}}- average latency \\- failure rate \\ - energy efficiency\end{tabular}  \\
\hline

Tsioliaridou \emph{et al.}~\cite{tsioliaridou2015corona} &
\begin{tabular}{@{}l@{}}- geographic routing using \\ \quad hop-distance coordinates \\- integer calculations only \end{tabular} & - software-defined met. (gen I.)  &
\begin{tabular}{@{}l@{}} - 2D uniform grid \\ \quad and random nodes \\ - 10000 nodes \end{tabular} &
\begin{tabular}{@{}l@{}} - failure rate \\ - energy efficiency\end{tabular}  \\
\hline

Tsioliaridou \emph{et al.}~\cite{tsioliaridou2017packet} &
\begin{tabular}{@{}l@{}}- geographic routing using \\ \quad hop-distance coordinates \\- integer calculations only \\ - coordinates selection alg.\end{tabular} &
\begin{tabular}{@{}l@{}}- software-defined metamaterials \\ - wireless robotic materials \end{tabular} &
\begin{tabular}{@{}l@{}} - 3D uniform grid \\ \quad and random nodes \\ - 5000 nodes \end{tabular} &
\begin{tabular}{@{}l@{}} - failure rate \\ - energy efficiency\end{tabular}  \\
\hline


\end{tabular}
\end{center}
\vspace{-5mm}
\end{table*}

\subsection{Routing in (Semi-)Regular Topologies} \label{ssec:regular_topologies}
Some applications require to transmit messages to a specific nanonode, in which case flooding (which transmits a message to all nanonodes in the network) is an inefficient solution. To achieve such unicast communication, some strong form of addressing is needed (i.e., nanonodes needing an identifier, physical or logical location).  When network topologies follow regular patterns, relatively simple addressing and routing mechanisms become possible. This is the case for NoC applications, where nanonodes are often static and laid out according to a regular grid pattern, enabling hard-coded coordinates for routing nanonodes among the two main axes. Researchers designing wireless NoC protocols often envision cores communicating using THz frequencies because carbon nanotube-based antennas have the potential for higher transmission rates and lower power and area overhead than e.g., ultra-wide band communication \cite{wang2014wireless}.

A simple XY routing mechanism lets routers forward messages along the X-axis (horizontal) first, until a router is found with the same X coordinate, and subsequently the routers forward along the Y-axis until the destination is reached. Greedy forwarding mechanisms rely on a distance metric which can be calculated between coordinates and aim to choose a neighbor which minimizes the distance to the destination. 
In the presence of faults or in non-planar topologies, greedy forwarding might result in a message being routed to a non-destination nanonode with no distance decreasing neighbors. Thus, fault tolerant routing techniques have been developed, an overview of which is provided by Radetzki \emph{et al.}~\cite{radetzki2013methods}.
Since some of the core ideas for fault-tolerance were originally proposed for (wired or wireless) NoC settings, we provide an overview of these papers. Since these efforts do not specifically target THz nanocommunication, they are omitted in Tables~\ref{tab:network_layer_protocols} and~\ref{tab:mapping_to_requirements}.
One important technique in fault tolerant routing is face routing, which defines rules to route around spots with faulty nanonodes. Maze routing \cite{fattah2015low} adds extra fields to a message, allowing to route around fault regions without requiring nanonodes to store any information other than their coordinates. Maze routing is however limited to planar topologies, and the route may be far from optimal.

Fukushima \emph{et al.}~\cite{fukushima2009fault} propose to collect local information to group faulty nanonodes in rectangular non-overlapping fault regions. The restriction that the fault regions must be non-overlapping may require to expand a fault region to include some operational nanonodes. Such healthy nanonodes inside a fault region cannot receive packets and thus must be switched off. More complex shaped fault regions can be realized by expanding the information collection region \cite{jovanovic2009new}. 

Using fault-tolerant routing to reroute packets around faulty regions will increase the packet latency and create congestion around the faulty region. Ebrahimi \emph{et al.} \cite{ebrahimi2013md} augments the XY algorithm with local nanonode information to route along failures reducing congestion through the incorporation of local queue and buffer information. Similarly, Saeed \emph{et al.}~\cite{saeed2018fault} propose two fault-adaptive XY routing mechanisms (one avoiding loops, and another maximizing success delivery probability) enabling communication between a network of controllers in the context of hyper-surfaces.
These fault tolerant algorithms typically trade off the information required to take a routing decision (and the corresponding algorithmic) complexity, with flexibility to deal with more complex fault scenarios. An exception to this is maze routing, but that approach is restricted to planar topologies and may result in suboptimal paths.

\vspace{-1mm}
\subsection{Irregular Topologies} \label{ssec:irregular_topologies}
When network topologies get even more irregular, nanonodes need a mechanism to determine their own coordinates. Tsioliaridou \emph{et al.}~\cite{tsioliaridou2015corona} was one of the first approaches to use a number of fixed anchor points in a 2D or 3D space (as in software-defined meta-materials) to flood announcements of their existence across the network. Based on triangulation, nanonodes could determine their (non-unique) coordinate. Once the initialization phase is over, nanonodes can participate in a stateless manner in the packet forwarding process which consists of selective flooding towards the destination. The authors compare their approach with probabilistic flooding and a dynamic infrastructure flooding approach. Since the flooding approaches deliver each packet to all nanonodes in the network, they both require many more nanonodes to unnecessarily spend energy on decoding the massage (5 times more in the paper, but this depends on the network size). 
This approach was further refined in Tsioliaridou \emph{et al.}~\cite{tsioliaridou2017packet} by proposing a routing approach which further minimizes required retransmits. 
A mechanism is proposed relying only on integer calculations and nanonode-local information enabling each nanonode to deduce whether it is located on the linear segment connecting the sender to the recipient nanonode. The energy efficiency of the scheme can be further optimized by tuning the width of the linear path. This path width allows to trade-off reliability and energy efficiency: a larger path width allows the algorithm to deal with more irregular scenarios, but it also results in an increased amount of transmitted messages. THz-band specific features such as molecular absorption and shadow fading are also taken into account when evaluating the algorithm.

Both of these schemes offer efficient one-to-one routing solutions, but they require a static environment, as nanonode mobility would result in frequent invalidation of the coordinate systems. Another open challenge is that these techniques may enable nanonodes to determine their own coordinates, but these works do not address how they might obtain the coordinates or addresses of other nnanonodes with whom communication is needed.  

The particular addressing needs of medical application scenarios are conceptually investigated in Stelzner \emph{et al.}~\cite{stelzner2017function}. The authors distinguish addressing from guidance concepts. 
The latter refer to alternate solutions to reach a target, without requiring an explicit address in communication, e.g., through kinds of wiring, electromagnetic fields, or bio-circuits. 
Function-centric nanonetworking refers to a scheme were location and functional capability of groups of nanonodes are addressed rather than the communication (individual) endpoint(s). 
The location can refer to an area defined in the human body, the function refers to a type or category of nanonodes rather than an individual one (e.g., blood pressure sensor).
$\blacksquare$

Table \ref{tab:network_layer_protocols} lists publications focusing on one or more network layer protocols. For each publication, the main novel features are given, as well some applications proposed by the authors themselves. We also mention the specific topology in which the routing schemes were evaluated and the evaluation metrics considered by the respective authors.

Across different applications and topologies, there is a large focus on minimizing \emph{power consumption} and \emph{energy efficiency}.  However, these metrics are often exclusively evaluated based on the number of sent (and received) messages. This provides an incomplete picture, since the power per message also depends on the transmission distance and the message length.

\emph{Resiliency} has also been studied in various settings: papers focusing on flooding typically discuss network coverage (the percentage of nanonodes which receive a one-to-all broadcast). Another example is Tsioliaridou \emph{et al.}~\cite{tsioliaridou2017packet}, where a path redundancy parameter is introduced, which allows to tune the number of nanonodes which participate in the transmission of a point-to-point message.

Most applications require large numbers of participating nanonodes. Nonetheless, the \emph{scalability} of routing protocols is often not evaluated. Table \ref{tab:network_layer_protocols} shows that almost all works focus on networks with a set number of nanonodes, typically corresponding to a small fraction of the larger network. However, papers which do look into the effect of network size, such as Liaskos \emph{et al.}~\cite{liaskos2015promise} and Stelzner \emph{et al.}~\cite{stelzner2018body} find that parameters such as network coverage and power consumption depend non-linearly on both the area/volume covered by the network, and the density of the nanonodes. Future work should evaluate the sensitivity in function of this metric, as limiting the evaluation to a set network size may result in conclusions which do not generalize well to larger or smaller networks. 

Whilst collecting these results,the effect of \emph{nanonode mobility} {\color{red}remains often overlooked}. This is permitted in some applications, where the topology is indeed static, or network changes occur very infrequently. Such scenarios include network-on-chip, the monitoring of mission critical materials and some software-defined metamaterials and IoNT/WSN settings. However, in-body network applications, the network is intrinsically mobile: when nanonodes operate within a bloodstream they move due to the blood flow, if they are attached to tissue, the body in which they are situated may move and change their relative positioning. From the existing research which targets in-body network applications,  {\color{red}only Buther \emph{et al.}~\cite{buther2018hop} and C\'{a}novas-Carrasco \emph{et al.}~\cite{canovas2018nanoscale} \cite{canovas2019optimal} mention nanonode mobility. The work of \cite{canovas2018nanoscale} and \cite{canovas2019optimal} focuses on nanonodes reporting events rather than nanonodes communicating directly with each other. Since an event message can be transmitted via any nanorouter, the main issue in this setting is a) knowing if, and b) estimating how long it will be until, a particular nanonode is in reach of a nanorouter. The first issue is addressed in \cite{canovas2018nanoscale} via an ultrasound energy source. The second issue is approached in \cite{canovas2019optimal} using MDPs.} In \cite{buther2018hop}, the mobility issue is addressed by periodically invalidating and reinitializing all routing information. Whilst such an approach is (in theory) always possible, we note that this may place a considerable burden on the network load and is costly in terms of energy consumption. Also, even in a slow moving environment, the frequency at which an individual connection change between any pair of nanonodes may be high (due to there being a large number of nanonodes), and thus, routing protocols which depend on global network information may require frequent updates. In future work, the effect of mobility on a network protocol should be carefully evaluated and simulated in an environment in which nanonodes are continuously moving. Ideas from research regarding mobile ad-hoc networks could provide a valuable starting point.

Finally, due to the large variety of application scenarios, evaluation topologies and metrics, it is difficult to \emph{compare} the results of various authors. There are various approaches to remedy this problem: open-sourcing the network topologies may allow researchers to easily evaluate their algorithms in a standardized variety of settings (without needing the in-depth knowledge to generate the specific topologies). However, there are many other modeling and evaluation parameters which may influence the results and complicate any comparison. In view of these observations, contributions focusing on the numerical comparison of various existing algorithms in a variety of settings, such as Stelzner \emph{et al.}~\cite{stelzner2018body} are highly valuable tools to provide a comparison of existing work. Such work is unfortunately rare, and should be further encouraged.


\section{Link Layer}
\label{sec:link}

In THz band nanonetworks, link layer protocols are used for enabling direct communication between a pair of nanonodes or between a nanonode and a more powerful device (e.g,  nanorouter, nanocontroller, gateway, nano-macro interface). 
The primary functions of link layer protocols are channel access coordination, which is traditionally performed on the \ac{MAC} sub-layer, and recovery from bit transmission errors, usually performed on the \ac{LLC} sub-layer.  
Classical MAC and LLC protocols cannot be directly applied in THz band nanocommunication for the following reasons~\cite{jornet2012phlame,alsheikh2016mac}. 
First, the existing link layer protocols have been predominantly designed for band-limited channels. 
However, the THz frequency band provides almost a 10 THz wide bandwidth window, which constitutes the main difference between graphene-enabled THz nanocommunication and classical link layer protocols. 
Second, existing link layer protocols are mostly carrier-sensing techniques and are therefore too complex for THz nanocommunication in a number of application domains and scenarios. 
Finally, nanodevices feature highly limited energy, in various scenarios requiring the usage of energy-harvesting systems~\cite{wang2008towards,xu2010self,mohrehkesh2014optimizing}. 
This changes the availability of the nanodevices' communication systems over time, which is not a constraint posited to the classical link layer protocols~\cite{lemic2020idling}. 
This has been recognized in the research community and several link layer (often MAC sub-layer only) protocols for THz nanocommunication have been proposed.

\begin{figure}[!t]
\centering
\scalebox{.49}{
\tikzstyle{block} = [rectangle, draw, fill=blue!10, 
 text width=13.5em, text centered, rounded corners, minimum height=4em]
\tikzstyle{line} = [draw, -latex']
\makebox[\textwidth][c]{
\begin{tikzpicture}[node distance = 2.5cm, auto]
 \node [block] (main) {\Large \textbf{THz Link Layer Protocols}};
 \node [block, below of=main, xshift=-150, yshift=-40] (distributed) {\Large
     \textbf{Distributed\\(Sect.~\ref{ssec:distributed})} \\
     Akkari \emph{et al.}~\cite{akkari2016distributed} \\
     Alsheikh \emph{et al.}~\cite{alsheikh2016grid} \\
     PHLAME~\cite{jornet2012phlame} \\
     DRIH-MAC~\cite{mohrehkesh2015drih} \\
     TCN~\cite{d2015timing} \\
     Xia \emph{et al.}~\cite{xia2015link} \\
     Smart-MAC~\cite{piro2013nano} \\
     APIS~\cite{yu2017pulse} \\
 };
 \node [block, below of=main, xshift=0,  yshift=-20] (hierarchical) {\Large 
     \textbf{Hierarchical\\(Sect.~\ref{ssec:hierarchical})} \\
		Wang \emph{et al.}~\cite{wang2013energy} \\
		EEWNSN-MAC~\cite{rikhtegar2017eewnsn} \\
		EESR-MAC~\cite{srikanth2012energy} \\
		CSMA-MAC~\cite{lee2018slotted} \\
 };
 \node [block, below of=main, xshift=150, yshift=-20] (chip) {\Large
     \textbf{On-Chip Communication\\(Sect.~\ref{ssec:on_chip})} \\
		Mansoor \emph{et al.}~\cite{Mansoor2016} \\
		BRS-MAC~\cite{Mestres2016} \\
		Dynamic \ac{MAC}~\cite{Mansoor2015} \\
 };
 \path [line] (main) -| (distributed);
 \path [line] (main)  -- (hierarchical);
 \path [line] (main) -| (chip);
\end{tikzpicture}
};
}
\caption{Classification of link layer protocols for THz nanocommunication}
\label{fig:link_layer_classification}
\vspace{-3mm}
\end{figure}
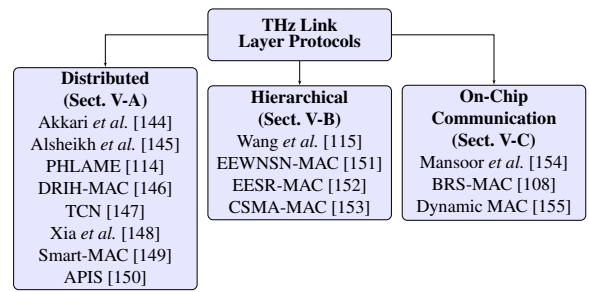

\begin{figure}[!t]
\centering
\includegraphics[width=0.8\columnwidth]{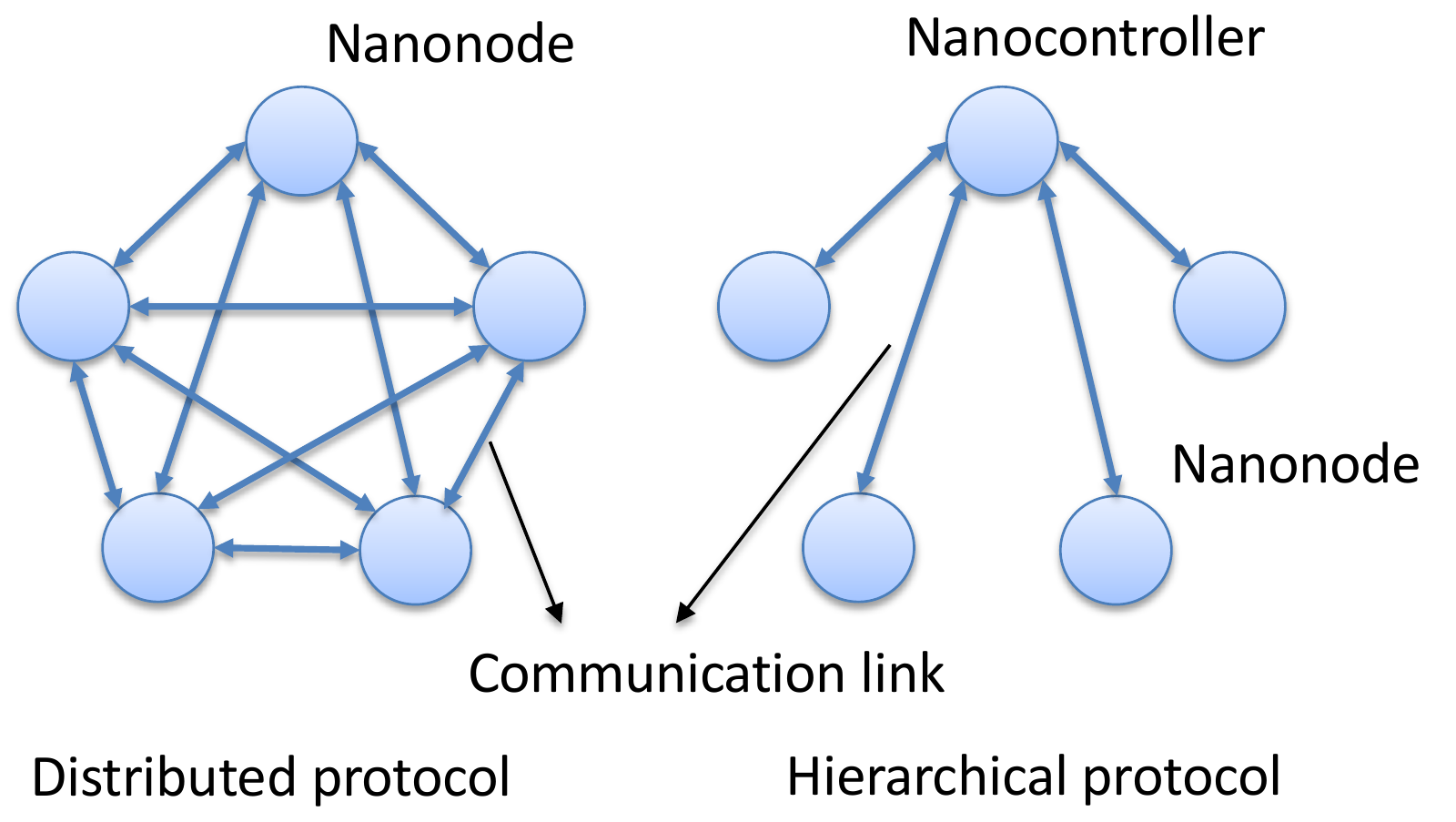}
\caption{Distributed vs. hierarchical MAC protocols}
\label{fig:distributed_vs_hierarchical}
\vspace{-3mm}
\end{figure}

As shown in Figure~\ref{fig:link_layer_classification}, they can be grouped into hierarchical protocols that assume the availability of more powerful nanocontrollers, and distributed, in which all nanonodes are considered as equal (Figure~\ref{fig:distributed_vs_hierarchical}).
In addition, protocols specifically designed for on-chip nanocommunication can be separately grouped, due to the uniqueness of communication features, as in more details discussed below.  
Compared to the distributed protocols, the main advantages of the hierarchical link-layer protocols include reduced energy consumption, increased network scalability, and increased reliability due to interference and collision probability~\cite{al2017ah}.
Their main disadvantages are their comparably higher complexity and delivery latency.  


\subsection{Distributed Protocols}
\label{ssec:distributed}

Akkari \emph{et al.}~\cite{akkari2016distributed} reason that there is a need for novel link layer protocols for nanonetworks due to the fact that nanodevices have strict power, memory, energy, and computation constraints. 
Thus, the authors argue, the nanonodes may only be able to store one packet, requiring packets to be delivered before certain hard deadlines. 
Motivated by this claim, they propose a distributed and computation-light \ac{MAC} protocol for nanonetworks. 
The protocol determines the optimal transmission times for nanonodes so that the largest set of traffic rates can be supported, while ensuring delivery within a hard deadline.
Optimal transmission decisions are made locally using \ac{CSMA} Markovian chain and Lyapunov optimization, which are based on nanonode's incoming traffic rate and queue length.
The protocol achieves continuous communication by simultaneously considering its energy consumption and harvesting rate.

Motivated by the fact that nanonodes communicating in the THz band are capable of achieving very high transmission bit-rates at very short distances, the authors in~\cite{alsheikh2016grid} also argue that classical \ac{MAC} protocols cannot be directly applied in THz band nanocommunication. 
Therefore, they propose an energy-aware \ac{MAC} protocol for synchronizing the communication between nanodevices. 
The proposed protocol is based on a \ac{TS-OOK} scheme (more details in Section~\ref{sec:phy}), which is a pulse-based communication scheme for THz band nanonetworks, and assumes grid-like distribution of the nanonodes.
In the protocol, active nanonodes transmit their packets in an interleaved way to the receivers based on the calculated \ac{CTR}, i.e., the maximum ratio between the transmission time and the energy harvesting time needed for continuous operation. 

Jornet \emph{et al.}~\cite{jornet2012phlame} propose the \ac{PHLAME}, which is built on top of modified TS-OOK. 
In PHLAME, the transmitting and receiving nanonodes jointly select the transmission symbol rate (the ratio between time between symbols and pulse duration) and channel coding scheme.
They do that by performing a handshaking process initiated by a nanonode that has information to transmit and enough energy to complete the transmission. 
Using the common coding scheme, the transmitter generates a packet containing the synchronization trailer, transmitter ID, receiver ID, packet ID, randomly selected transmitting data symbol rate, and error detecting code.
The handshaking acknowledgment is issued by the receiver upon the reception of the packet.
If the handshake is accepted, the receiver issues a packet containing the synchronization trailer, transmitter ID, receiver ID, packet ID, transmitting data coding scheme, and error detecting code.
The transmitter then transmits a data packet using the symbol rate specified by the transmitter, and encoded with the weight and repetition order specified by the receiver during the handshake process.
The authors show that the proposed protocol is able to support densely populated nanonetworks in terms of energy consumption per useful bit of information, average packet delay, and achievable throughput.

Another distributed \ac{MAC} protocol is \ac{DRIH-MAC}~\cite{mohrehkesh2015drih}.
Simply stated, in DRIH-MAC (RIH-MAC in its preliminary version~\cite{mohrehkesh2014rih}) the communication is initiated by the receiver by transmitting a \ac{RTR} packet to one or multiple transmitters. 
The recipients of the RTR packet transmit a data packet to the receiver.
Scheduling in DRIH-MAC is based on a probabilistic scheme-based on the edge-coloring problem.
The idea of edge-coloring is to color the edges of a graph such that two edges incident to the same node are of different colors.
Obviously, in DRIH-MAC different edge colors translate to transmission sequences.
The authors claim DRIH-MAC to be scalable and light-weight, with the minimized probability of collisions and maximized utilization of harvested energy.

The authors in~\cite{d2015timing} propose \ac{TCN}, a link layer THz band nanocommunication protocol that exploits ``timing channels''. 
They define the timing channels as logical channels in which the information is encoded in the silence duration between two consequent transmissions.
The authors argue that, by using timing channel-based communications, \ac{TCN} enables energy efficient low data-rate communication.
Moreover, by introducing acknowledgment-based collision detection, \ac{TCN} enables recovery from transmission errors.
Retransmissions are then envisioned for improving the reliability of communication.

A link-layer synchronization and \ac{MAC} protocol for wireless communication in the THz band is presented in~\cite{xia2015link}. 
The link-layer synchronization capability is achieved through a receiver-initiated (i.e., one-way) handshake procedure. 
The core idea of the handshake is to prevent data transmissions when the receiver does not have enough energy for reception.
Additionally, the protocol aims at maximizing the channel utilization and minimizing the packet discard probability. 
It does that by making use of a sliding flow-control window at the link-layer, i.e., the receiver specifies the amount of data that can be received for its current energy level.

The authors in~\cite{piro2013nano} propose Smart-MAC, a MAC protocol delivered as a part of the NanoSim simulator and often used as a baseline for benchmarking of other MAC protocols.
Smart-MAC uses a handshake procedure for discovering nanonodes within its transmission range.
If at least one nanonode is discovered, the packet is transmitted. 
In case the nanonode does not have any neighbors, Smart-MAC uses a random back-off delay prior to restarting the handshake procedure. 
In case of multiple packets in the queue, a new transmission is scheduled after a specific time computed through the same back-off mechanism, which reduces the probability of collisions.

A distributed scheduling mechanism named Adaptive Pulse Interval Scheduling (APIS) is proposed~\cite{yu2017pulse} for the interconnection between nanonetworks and an IoT gateway, targeting software-defined metamaterial and wireless robotic material applications. 
In APIS, the sink distribute a transmission schedule, followed by the nanonodes transmitting data based on the channel sensing results and traffic patterns. 
The proposed protocol was evaluated in terms of reliability of data delivery, achievable throughput, and energy consumption. 
Due to the scheduled transmissions, high time precision and continuous operation of the nanonodes are required for scheduling, which could be infeasible under limited computational capacity and stored energy.
Moreover, the communication between the nanonodes is abstracted and not considered in this work.

\subsection{Hierarchical Protocols}
\label{ssec:hierarchical}

The authors in~\cite{wang2013energy} propose an energy and spectrum-aware \ac{MAC} protocol for THz band nanocommunication. 
First, they propose to utilize the hierarchical network architecture characteristic to \acp{WNSN} for shifting the protocol complexity to more resourceful nanocontrollers.
Therefore, the nanocontroller regulates the channel access on behalf of the nanonodes of its cluster.
The nanocontroller does that by utilizing \ac{TDMA} and based on the nanonodes' data requirements and energy constraints. 
The proposed \ac{MAC} protocol utilizes \ac{CTR}, i.e., the maximum allowable ratio between the transmission and energy harvesting times below which the nanonode consumes less energy than harvested.
Based on this and assuming TS-OOK as the physical layer communication scheme, a symbol-compression scheduling protocol is proposed for assigning each nanonode with different sets of transmission slots in such a way that the overall nanonetwork achieves optimal throughput, while maintaining transmission ratios below the CTR for achieving energy balancing. 
Note that the protocol utilizes the TS-OOK's elasticity in the inter-symbol spacing, allowing multiple nanonodes to transmit their packets in parallel without inducing collisions.

Rikhtegar \emph{et al.}~\cite{rikhtegar2017eewnsn} present the Energy Efficient Wireless Nano Sensor Network MAC (EEWNSN-MAC), a MAC protocol for mobile multi-hop THz band nanonetworks. 
They assume a network comprised of nanonodes moving randomly at a constant speed, as well as static nanorouters and a nano-micro interface. 
EEWNSN-MAC is divided into three steps: i) handshaking-based selection of a cluster head (i.e., nanorouter); ii) TDMA-based scheduling phase in which a nanorouter schedules the transmission times for the nanonodes in its cluster; iii) the data transmission phase in which the nanonodes send their packets to the nanorouters, followed by their aggregation and forwarding to the nano-micro interface.

The authors in~\cite{srikanth2012energy} propose EESR-MAC, an energy efficient, scalable, and reliable MAC protocol for THz nanonetworks. 
In the protocol, the cluster head is first derived as the one that is equidistant from other nanonodes in the cluster.  
Then, the classic \ac{TDMA} approach is used for inter and intra-cluster communication, where the master node is assumed to transmit the schedule to the other nanonodes.
The approach is presented here for completeness purposes, as it has not been evaluated nor benchmarked against other approaches (and is therefore omitted in Table~\ref{tab:mapping_to_requirements}).

A slotted CSMA/CA based MAC protocol (CSMA-MAC) is proposed for energy harvesting nanonodes in~\cite{lee2018slotted}. The protocol assumes a coordinator node periodically transmitting beacon packets containing the super-frame structure. A nanonode that wants to transmit receives the beacon, synchronizes itself to the super-frame structure, followed by transmitting its data using slotted CSMA/CA channel access mechanism. The protocol has been benchmarked in terms of achievable data-rate against a simple round robin approach.

\begin{table*}[!ht]
\begin{center}
\caption{Summary of link layer protocols}
\label{tab:link_layer_protocols}
\begin{tabular}{l l l l l} 
\hline
\textbf{Protocol} & \textbf{Sub-layer} & \textbf{Distinct Features} & \textbf{Potential Applications}  & \textbf{Evaluation Metrics} \\ 
\hline

Akkari \emph{et al.}~\cite{akkari2016distributed} & \ac{MAC} & \begin{tabular}{@{}@{}l@{}} - \ac{CSMA}-based with hard deadline \\ - continuous communication\end{tabular} & \begin{tabular}{@{}l@{}} - software-defined metamaterials \\- wireless robotic materials\end{tabular} & - timely-delivery ratio\footnotemark  \\ 
\hline

Alsheikh \emph{et al.}~\cite{alsheikh2016grid}    &  \ac{MAC}  & \begin{tabular}{@{}@{}l@{}} - blind transmission \\ - continuous communication \\ - grid constellation\end{tabular} & \begin{tabular}{@{}@{}l@{}} - software-defined metamaterials (gen. I) \\- wireless robotic materials \end{tabular} & \begin{tabular}{@{}@{}@{}l@{}}- consumed energy \\ - collision probability \\ - transmission delay \\ - network throughput\end{tabular} \\
\hline

PHLAME~\cite{jornet2012phlame} &  \ac{MAC}, \ac{LLC} & \begin{tabular}{@{}@{}l@{}}- handshake-based \\ - interference minimization \\ - retransmissions possible\end{tabular} & \begin{tabular}{@{}@{}l@{}} - software-defined metamaterials (gen. I) \\- wireless robotic materials \\- body-centric communication\end{tabular} & \begin{tabular}{@{}@{}l@{}}- consumed energy \\ - collision probability \\ - network throughput\end{tabular} \\
\hline

DRIH-MAC~\cite{mohrehkesh2015drih}  & \ac{MAC} & \begin{tabular}{@{}l@{}} - receiver-initiated communication \\- schedule-based transmissions\end{tabular} & - wireless robotic materials & \begin{tabular}{@{}@{}l@{}}- transmission delay \\- capacity utilization \\- energy utilization\end{tabular} \\
\hline

TCN~\cite{d2015timing} &  \ac{MAC}, \ac{LLC} & 
\begin{tabular}{@{}@{}l@{}} - scheduled transmissions (TDMA) \\ - acknowledgement-based \\ - retransmissions possible\end{tabular} & - body-centric communication & \begin{tabular}{@{}@{}l@{}}- energy per bit \\- collision probability \\- energy consumption \end{tabular} \\
\hline

Xia \emph{et al.}~\cite{xia2015link}  &  \ac{MAC} & \begin{tabular}{@{}@{}l@{}} - handshake-based \\ - receiver-initiated synchronization \\- sliding window flow control\end{tabular}
  & - wireless robotic materials  & \begin{tabular}{@{}l@{}} - packet discard probability \\- network throughput \end{tabular} \\
\hline

Smart-MAC~\cite{piro2013nano}   &  \ac{MAC} & \begin{tabular}{@{}l@{}} - default MAC in NanoSim \\- handshake and backoff-based\end{tabular}  & - wireless robotic materials  & \begin{tabular}{@{}l@{}} - packet-loss ratio \\- scalability \end{tabular} \\
\hline

APIS~\cite{yu2017pulse} & \ac{MAC} & \begin{tabular}{@{}l@{}} - designed for gateway to sink links \\ - based on traffic and channel sensing \end{tabular}  & \begin{tabular}{@{}l@{}} - software-defined metamaterials \\- wireless robotic materials \end{tabular} & \begin{tabular}{@{}@{}l@{}} - achievable throughput \\- energy efficiency \\- delivery reliability  \end{tabular} \\ \hline

Wang \emph{et al.}~\cite{wang2013energy}  &  \ac{MAC} & \begin{tabular}{@{}@{}l@{}}- \ac{TDMA}-based \\ - continuous communication \\- fairness-oriented\end{tabular}  &  \begin{tabular}{@{}l@{}} - software-defined metamaterials (gen. I) \\- wireless robotic materials \end{tabular} & \begin{tabular}{@{}l@{}} - single-user throughput \\- achievable data-rate\end{tabular} \\
\hline

EEWNSN-MAC~\cite{rikhtegar2017eewnsn} &  \ac{MAC} & \begin{tabular}{@{}l@{}} - TDMA with clustering \\- mobility and multi-hopping \end{tabular}  & \begin{tabular}{@{}l@{}} - software-defined metamaterials (gen. I) \\- wireless robotic materials \end{tabular}  & \begin{tabular}{@{}@{}l@{}} - energy consumption \\- scalability \\- packet-loss ratio\end{tabular} \\
\hline

EESR-MAC~\cite{srikanth2012energy} & \ac{MAC} & - TDMA with clustering  &   \begin{tabular}{@{}l@{}} - software-defined metamaterials (gen. I) \\- wireless robotic materials \end{tabular}  & - none \\
\hline

CSMA-MAC~\cite{lee2018slotted} & \ac{MAC} & \begin{tabular}{@{}l@{}} - beaconing-based \\-CSMA/CA channel access \end{tabular}  & \begin{tabular}{@{}l@{}} - body-centric communication \\- wireless robotic materials \end{tabular}  & - achievable throughput \\
\hline

Mansoor \emph{et al.}~\cite{Mansoor2016} &  \ac{MAC}, \ac{LLC} & \begin{tabular}{@{}l@{}} - token-passing-based \\- based on traffic estimates \end{tabular}  & - on-chip communication & \begin{tabular}{@{}@{}l@{}} - average data-rate \\- energy efficiency \\- transmission delay\end{tabular} \\
\hline

BRS-MAC~\cite{Mestres2016} &  \ac{MAC}, \ac{LLC} & \begin{tabular}{@{}l@{}} - CSMA and NACK-based \\ - preamble-based collision detection \end{tabular}  & - on-chip communication  & \begin{tabular}{@{}l@{}} - achievable throughput \\- transmission delay \end{tabular} \\
\hline

Dynamic \ac{MAC}~\cite{Mansoor2015} & \ac{MAC}, \ac{LLC} & \begin{tabular}{@{}l@{}} - combines \ac{CSMA} and token-passing \\ - based on expected traffic loads \end{tabular}  & - on-chip communication  & \begin{tabular}{@{}@{}l@{}} - achievable throughput \\- energy efficiency \\- protocol overhead\end{tabular} \\

\hline
\end{tabular}
\end{center}
\vspace{-6.5mm}
\end{table*}

\subsection{Protocols for On-Chip Communication}
\label{ssec:on_chip}

The authors in~\cite{Abadal2018a} provide a context analysis of \ac{MAC} protocols for on-chip communication.
They argue that, from the link layer perspective, on-chip communication represents a unique scenario with respect to traditional wireless communication.
This is predominantly due to the fact that the topology of a network, chip layout, and characteristics of the building materials are static and known in advance~\cite{abadal2015broadcast}.
Therefore, the on-chip wireless channel can be accurately characterized as quasi-deterministic from the link layer perspective.

As discussed before, the on-chip applications require very low and deterministic latency, high reliability, and very high throughput. 
As the secondary requirement, the energy consumption should be constrained to limit the heat dissipation on the chip.
This has to an extent been recognized in the existing literature.
Note that these works do not consider the usage of THz frequencies for communication, but instead focus on the sub-THz (i.e., 30-300~GHz) band.
Nonetheless, we will outline them here due to the intrinsic similarities between the THz and mmWave bands in the context on on-chip nanocommunication. 

Two flavors of a token-passing (i.e., dynamic TDMA-based) \ac{MAC} protocol for on-chip wireless communication are proposed in~\cite{Mansoor2016}.
In token-passing, the channel access is based on the possession of a token which circulates between nodes in a round-robin fashion to ensure fairness. 
The duration of token possession is in~\cite{Mansoor2016} determined on predicted estimates of communication demands for different nanonodes.
The protocol is evaluated in terms of average data-rate, energy efficiency, and latency of packet delivery.

Mestres \emph{et al.}~\cite{Mestres2016} propose the \{Broadcast, Reliability, Sensing\} protocol (BRS-MAC). 
BRS-MAC combines the CSMA/CA and CSMA/CD mechanisms by utilizing preamble-based collision detection.
Specifically, as the first envisioned step and only if the channel is sensed idle, the sender transmits a preamble (otherwise the node backs off).  
Nanonodes that correctly received it remain silent for the rest of the transmission, while the ones that detected a collision respond with a \ac{NACK}.
If the NACK was received, the original sender cancels the transmission and backs off, while the other nodes discard the preamble.
Adversely, the sender transmits the rest of the packet.
The protocol has been evaluated in terms of achievable throughput and delivery latency.

The authors in~\cite{Mansoor2015} argue that there is a need for reconfigurable wireless links for optimizing the utilization of the on-chip channel bandwidth.
Grounded on that observation, they propose a dynamic \ac{MAC} protocol by integrating the \ac{CSMA} and token-based mechanisms.
The proposed protocol utilizes the token-passing mechanism in case of high traffic loads.
When the traffic loads are low, token passing becomes energy inefficient.
Therefore, in such cases the \textit{dynamic MAC unit} of the protocol switches the operation to a \ac{CSMA}-based mechanism in which the consumed energy is due to valid transmission only.
The protocol is evaluated in terms of achievable throughput, energy efficiency, and protocol overhead (i.e., power, area, and delay characteristics). $\blacksquare$

Table~\ref{tab:link_layer_protocols} provides a summary of the above-discussed link layer protocols for THz band nanonetworks. 
As visible in the table, there are several open challenges pertaining to the link layer protocols for nanonetworks.
First, the majority of the current protocols deal with the \ac{MAC} sub-layer, with only the PHLAME and TCN protocols additionally proposing a \ac{LLC} sub-layer mechanism. 
This implies that the majority of current protocols are not tailor-made for applications that require high communication reliability (e.g., body-centric and on-chip communication).  
One of the open research questions is to improve the reliability of THz band nanocommunication on the link layer.
As an example, PHLAME supports packet repetitions as means for improving the nanonetwork reliability (i.e., \ac{PRR}). 
However, when and how many repetitions should be utilized is still unclear. 
Such a decision could potentially be based on the current energy levels of the nanonodes, their distances, and/or the amounts of traffic, as in more detail discussed in~\cite{lemic2019assessing}. 

Second, the majority of the existing protocols do not optimize for the latency of data delivery.
The only protocol that goes in this direction is Akkari \emph{et al.}~\cite{akkari2016distributed}, where a hard deadline on data delivery is imposed, however the minimization of delivery latency is not attempted.
Optimization of link layer protocols in terms of latency is required for several of the envisioned applications (e.g., the second generation of software-defined metamaterials, on-chip communication).
In addition and to the best of our knowledge, protocols aiming at jointly optimizing throughput and latency, which is required for on-chip nanocommunication, are currently missing.

Third, apart for EEWNSN-MAC, none of the protocols explicitly accounts for the fact that nanonodes can be mobile. 
Even the EEWNSN-MAC protocol, given that it hierarchical, requires the selection of cluster heads, which is known to yields unsatisfactory performance in scenarios with high mobility. 
In mobility scenarios, are handshake and clustering-based hierarchical protocols are generally expected to yield poor performance.
This is further accentuated by the fact that, due to harvesting, a certain amount of time will usually have to pass between the handshake and data transmission.   
In high mobility scenarios, the optimal strategy for data transmission could be to just send data when there is data to send and enough energy for transmission. 
Nonetheless, such strategies have yet to be investigated.

\footnotetext{Percentage of packets successfully delivered before the deadline.}

In Table~\ref{tab:link_layer_protocols}, we have listed the application domains that could potentially be supported by a given protocol.
There are seemingly no link layer protocols explicitly targeting on-chip THz band communication.
We have outlined the most promising candidates from the mmWave band, which is indeed similar to the signal propagation in THz frequencies. 
Yet, propagation is not the same for these two bands, mostly due to the fact that THz signals attenuate faster and resonate with water, in contrast to mmWave signals resonating with oxygen molecules.
Hence, the applicability of the listed protocols for on-chip THz nanocommunication is yet to be evaluated.   

Finally, all of the existing protocols have been evaluated either analytically or by means of simulation. 
Their performance results are potentially not accurate, as they have not been derived with a very high level of realism.
For example, the energy consumption of a nanonode's communication system is in the evaluations of all protocols attributed to either transmission of reception, while idling  energy has been fully neglected.
Evaluation results with higher levels of realism are certainly needed.
In addition, the metrics used in the evaluations are non-standardized and non-exhaustive, as shown in the table.
Hence, various performance insights are currently lacking.
For example, one of the primary requirements for many of the envisioned applications is scalability. 
Nonetheless, the scalability of link layer protocols has been evaluated only in EEWNSN-MAC and even there the conclusion is that ``\textit{EEWNSN-MAC is potentially a scalable protocol}''. 
Due to the fact that the evaluation metrics are currently non-standardized and non-exhaustive, comprehensive comparison of protocols for different application scenarios is at the moment infeasible.  
Given that THz nanocommunication is in many aspects challenging, even minor improvements in the protocol design could yield high benefits.
Thus, comprehensive protocol benchmarking is certainly a promising research direction of high priority.


\section{Physical Layer}
\label{sec:phy}

The physical layer defines the means of transmitting raw bits over a physical link interconnecting two nodes. 
In the specific case of wireless communication, the physical layer is concerned with the modulation, the coding, error control, and other methods that determine the data rate and error rate of the solution, as well as its area and power.

The scenario of THz nanocommunication has a unique blend of constraints and requirements that greatly impacts the physical layer and prevents the use of well-established techniques. The nanoscale dimension imposes very stringent restrictions on the available resources (i.e., area, energy, memory) that, despite being dependent on the particular application context, suggest the use of simple and ultra-efficient modulations and coding. This is especially limiting in intermittent computing applications where devices are powered via energy harvesting. This is evidently impacting the devices' available energy, but also causing intermittency in devices' operation, posing an additional challenge in regard to their reliability.

The THz dimension of the scenario affects the physical layer of the design as well. The main reason is technological, as mature THz circuits and systems for communication are yet to come, although the community is making significant leaps forward~\cite{Wu2017c, Lee2019ISSCC, Aggrawal2016, han2019filling, Correas2017}. Moreover, the THz channel introduces the effect of molecular absorption which, for increasing distances, becomes another impairment for communication.

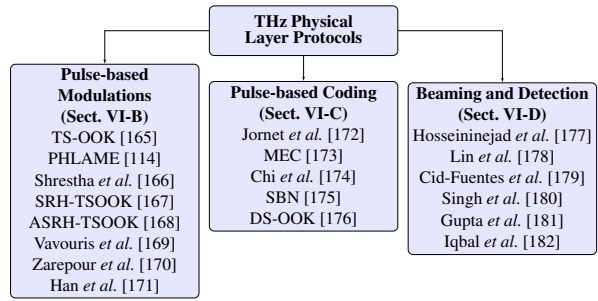
\begin{figure}
\vspace{-2mm}
\centering
\scalebox{.47}{
\tikzstyle{block} = [rectangle, draw, fill=blue!10, 
 text width=14.5em, text centered, rounded corners, minimum height=4em]
\tikzstyle{line} = [draw, -latex']
\makebox[\textwidth][c]{
\begin{tikzpicture}[node distance = 2.5cm, auto]
 \node [block] (main) {\Large \textbf{THz Physical Layer Protocols}};
 \node [block, below of=main, xshift=-160, yshift=-50] (modulation) {\Large
     \textbf{Pulse-based Modulations\\(Sect.~\ref{sec:modulations})} \\
     TS-OOK~\cite{jornet2014femtosecond} \\
     PHLAME~\cite{jornet2012phlame} \\
     Shrestha \emph{et al.}~\cite{shrestha2016enhanced} \\
     SRH-TSOOK~\cite{mabed2017enhanced} \\
     ASRH-TSOOK~\cite{mabed2018flexible}  \\
     Vavouris \emph{et al.}~\cite{vavouris2018energy} \\
     Zarepour \emph{et al.}~\cite{zarepour2014frequency} \\
     Han \emph{et al.}~\cite{han2015multi} \\
 };
 \node [block, below of=main, xshift=0,  yshift=-30] (coding) {\Large 
     \textbf{Pulse-based Coding\\(Sect.~\ref{sec:coding})} \\
     Jornet \emph{et al.}~\cite{jornet2014low}  \\
     MEC~\cite{kocaoglu2013minimum}  \\
     Chi \emph{et al.}~\cite{chi2014energy} \\ 
     SBN~\cite{zainuddin2016sbn} \\
     DS-OOK~\cite{singh2020ds} \\
 };
 \node [block, below of=main, xshift=160, yshift=-40] (detection) {\Large
     \textbf{Beaming and Detection\\(Sect.~\ref{sec:detection})} \\
     Hosseininejad \emph{et al.}~\cite{Hosseininejad2018WCNC} \\
     Lin \emph{et al.}~\cite{Lin2018} \\
     Cid-Fuentes \emph{et al.}~\cite{Cid-fuentes2012} \\
     Singh \emph{et al.}~\cite{Singh2019} \\
     Gupta \emph{et al.}~\cite{gupta2015joint} \\
     Iqbal \emph{et al.}~\cite{iqbal2018modulation} \\
 };
 \path [line] (main) -| (modulation);
 \path [line] (main)  -- (coding);
 \path [line] (main) -| (detection);
\end{tikzpicture}
};
}
\caption{Classification of physical layer protocol for THz nanocommunication}
\label{fig:phy_layer_classification}
\vspace{-3mm}
\end{figure}

Overall, the existing proposals for THz nanocommunication have embraced simplicity as one of the main design drivers. As we discuss next in Section \ref{sec:CWvsIR}, on-chip communication works mostly advocate for fast Continuous-Wave (CW) On-Off Keying (OOK) to avoid power-hungry circuits and minimize signal processing delay, whereas other applications need to simplify the physical layer further via Impulse Radio (IR)-like techniques. In the latter case, modulations rely on the transmission of femtosecond-long pulses and, not surprisingly, adopt cross-layer strategies in an attempt to further simplify the protocol stack. As shown in Figure~\ref{fig:phy_layer_classification}, we outline the main alternatives of such pulse-based modulations and coding techniques in Sections \ref{sec:modulations} and \ref{sec:coding}, respectively. Finally, we analyze recent proposals for simplified beaming, detection, and synchronization in Section~\ref{sec:detection}.


\subsection{Continuous-Wave (CW) vs. Impulse Radio (IR)}
\label{sec:CWvsIR}
Wireless communication networks have been established and have grown dominated by CW technologies. This means that the modulation is based on the manipulation of the amplitude, phase, or other characteristics of a continuous carrier wave at the desired frequency. Technology scaling has allowed to increase the carrier frequency in the quest for higher bandwidths and device miniaturization. Higher bandwidths are naturally attained as the carrier frequency increases. Miniaturization is achieved because the antenna and other passive elements, which are typically the largest components within RF transceivers, scale with the inverse of the carrier frequency. As a simple but illustrative example, the resonance frequency $f_{R}$ of a dipole antenna of length $L$ surrounded by air is given by $f_{R} = \tfrac{c_{0}}{2L}$ where $c_{0}$ is the speed of light.

\begin{figure}[!t]
    \centering
    \includegraphics[width=0.95\columnwidth]{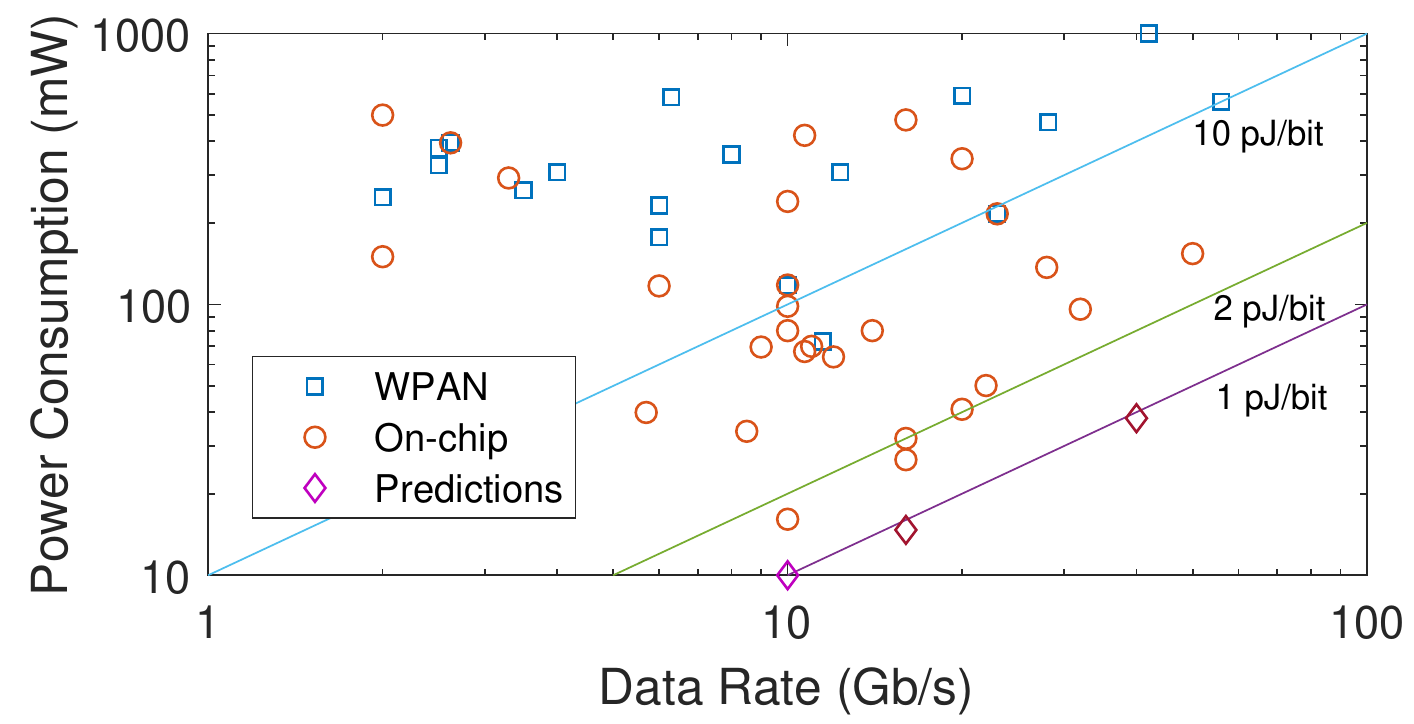}
    \caption{Energy efficiency of sub-THz and THz transceivers (from 0.06 to 0.43 THz) for short-range high-rate wireless applications. Each data point indicates the power and data rate of a single transceiver prototype from the literature for Wireless Personal Area Networks (WPAN, blue squares) and on-chip communication applications (red circles). Purple diamonds correspond to theoretical predictions made in the literature for future transceivers. Finally, straight lines represent the frontiers of energy efficiency of 1, 2, and 10 pJ per transmitted bit, so that transceivers located above (below) each line are less (more) efficient than indicated by the frontier. Data extracted from \cite{Tasolamprou2019}.}
    \label{fig:powerTHz}
    \vspace{-3mm}
\end{figure}

The energy efficiency of CW transceivers in the THz band depends on several factors such as the carrier frequency, the choice of modulation, or the transmission range. Based on an analysis of recent transceivers from 0.06 to 0.43 THz \cite{Yu2014, Lee2019ISSCC, hu2012sige, Tasolamprou2019}, Figure \ref{fig:powerTHz} shows how CW transceivers are reaching 10+ Gb/s speeds with around 1 pJ/bit for high-rate applications at the on-chip, off-chip, and indoor scales. Such trend is expected to continue in the THz band, where significant efforts are devoted to filling the so-called \emph{THz gap} \cite{Wu2017c, Lee2019ISSCC, Aggrawal2016, han2019filling, Correas2017}.

The existing scaling tendencies in CW transceivers are good news for on-chip communication applications as multiprocessors demand very high transmission speeds. Since the energy supply in computer systems is typically sustained, classical CW modulation techniques can be used. 
Nevertheless, as described in Section \ref{sec:chipCom}, the scenario demands ultra-low latency and low power consumption. Due to these factors and to the relatively immature state of THz technology, high-order modulations or techniques requiring significant signal processing are discouraged. Instead, most proposals advocate for simple modulations such as OOK and non-coherent (i.e., amplitude) detection \cite{Yu2014, Laha2015}. The OOK modulation consists in transmitting silence when the symbol is '0' and the carrier wave when the symbol is '1'. Such modulation can be achieved by simply connecting the stream of bits to the circuit that generates the carrier wave. This avoids the use of power-hungry circuits such as those needed for coherent (i.e., phase) modulation and detection. It has been thus shown that OOK can be 1.5X and 2.5X more energy efficient than BPSK and QPSK in on-chip environments \cite{Yu2014}. The downside of using low-order modulations is that, to scale the transmission rates, one may need to resort to multiple carriers to combat dispersion (i.e., a single carrier modulated at ultra-high speeds leads to an ultra-high bandwidth signal very sensitive to multipath and delay spread). Notably, Han \emph{et al.} proposes a multi-carrier modulation that adapts to the frequency-selective molecular absorption effects of the THz band \cite{Han2014a}.


In applications where energy availability is intermittent and not guaranteed, CW techniques cannot be used due to the cost of generating and using a carrier signal. Therefore, IR-like modulations where the information is encoded in short pulses instead of a carrier wave have been proposed instead. In the work by Zarepour \emph{et al.} \cite{zarepour2015performance}, carrier-less pulse-based OOK, Binary Phase Shift Keying (BPSK), Pulse Amplitude and Position Modulations (PAM and PPM, respectively) were compared in order to assess their fitness for IoNT applications. Using analytical models, it was concluded that although BPSK is relatively more complex in terms of decoding logic, it is the most efficient and reliable among all the contenders. OOK and PPM are simpler, but less reliable and efficient than BPSK. The analysis discouraged the use of PAM due to its low performance and efficiency. 


Zarepour~\emph{et al.} revisited a widely regarded trade-off between complexity and performance. BPSK and, by extension, other signaling schemes such as Transmitted Reference (TR)~\cite{Witrisal2009}, are preferable over OOK but may not be affordable in extreme scenarios. In such cases, in fact, even conventional OOK may be prohibitive. As a result and as we see next, the physical layer research has continued to push the efficiency and simplicity boundaries of pulse-based modulations.

\subsection{Pulse-based Modulations}
\label{sec:modulations}
One of the first works to discuss modulations suitable for nanoscale wireless communication was \cite{jornet2014femtosecond}. The proposed scheme is named Time-Spread On-Off Keying (TS-OOK) and is a pulse-based modulation. The main characteristics are that (i) pulses are around 100-fs long, thus leading to bandwidths in the THz range, and that (ii) the separation between pulses is much larger than the duration of the pulse. This scheme retains the simplicity of conventional OOK and, by having such a large separation between pulses, it is compatible with applications where energy is very limited or needs to be harvested. Moreover, by knowing the time between pulses, synchronization is only needed at the preamble and can be kept throughout the communication. The work in \cite{jornet2014femtosecond}, besides proposing the modulation, confirms that the achievable capacity is in the order of Tbps and also opens the door to simple multi-user approaches that exploit the long time between pulses to interleave other communication. The authors provide an interference model that, in subsequent works, have been validated experimentally~\cite{Hossain2019}.

\begin{figure}[!t]
    \centering
    \includegraphics[width=0.98\columnwidth]{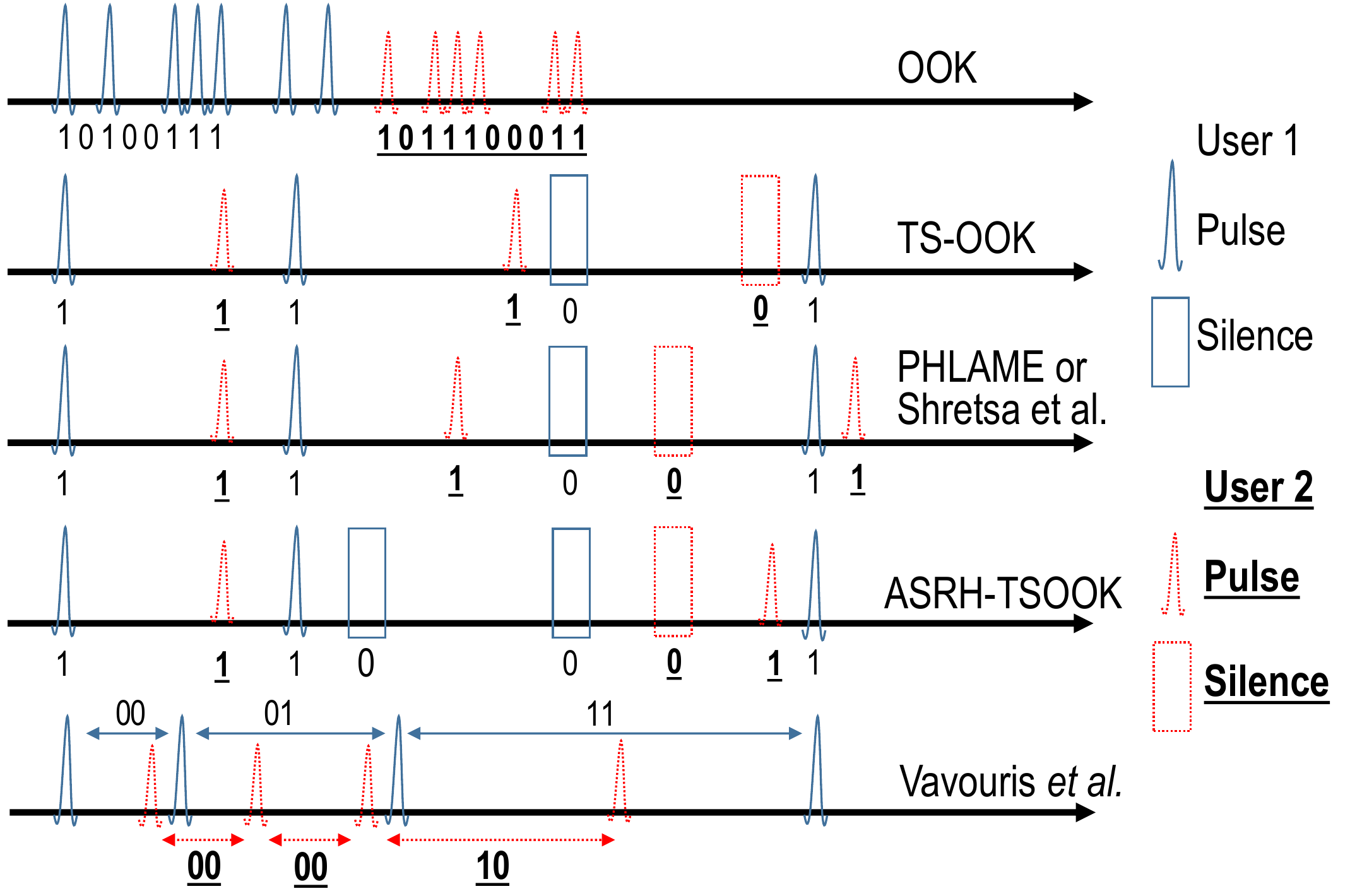}
    \caption{Comparison between the different variants of pulse-based OOK found in the literature \cite{jornet2014femtosecond, jornet2012phlame, shrestha2016enhanced, mabed2018flexible, vavouris2018energy}.}
    \label{fig:tsook}
    \vspace{-3mm}
\end{figure}

The seminal work by Jornet \emph{et al.} has been followed by several variants that optimize or particularize TS-OOK for different scenarios as graphically summarized in Figure \ref{fig:tsook}. For instance, in~\cite{jornet2012phlame,shrestha2016enhanced} the authors tackle one of the weaknesses of TS-OOK: if multiple users transmit with the same rate and collide in one pulse, they are bound to collide in all pulses. In the multi-user scheme proposed in~\cite{jornet2012phlame}, referred to as Rate Division Multiple Access (RDMA), users are assigned co-prime transmission rates during handshake to minimize interference at a reduced cost. In~\cite{shrestha2016enhanced}, the RDMA scheme is generalized for both ad hoc and infrastructure-based networks and the choice of prime numbers is further justified. Later, Mabed \emph{et al.} argued that RDMA leads to rate imbalance as users are assigned different effective rates. To overcome this issue, they proposed to employ pseudo-random time-hopping sequences to determine the time between pulses that, on average, would yield similar rate for all users~\cite{mabed2017enhanced} or capable of adjusting the rate to the user needs~\cite{mabed2018flexible}. Finally, Singh \emph{et al.} propose a completely different approach to accommodate multiple users, which is to combine OOK with sender-distinguishing direct sequence codes \cite{singh2020ds}. In this case, energy efficiency is sacrificed to achieve higher performance.

Another TS-OOK variant in the literature is~\cite{vavouris2018energy}. In this case, the authors aim to maximize energy efficiency and, to that end, propose to combine TS-OOK with PPM. The approach consists of the modulation of a symbol as the time between pulses, which is at all times much larger than the pulse duration. It is demonstrated that when increasing the symbol order, multiple bits can be encoded as a silence between pulses, therefore improving the energy efficiency at the cost of a degradation of the data rate. The PPM variant has also been combined with time-hopping in~\cite{singh2018th}, where a thorough evaluation is carried out assuming non-coherent detection and multiple modulation orders. 

Finally, it is worth mentioning proposals that also adapt to the particularities and new features of wireless communication in the THz band. On the one hand, Zarepour \emph{et al.} propose the use of frequency-hopping as a means of overcoming the problem of dynamic molecular absorption in composition-varying channels~\cite{zarepour2014frequency}. The \emph{blind} use of frequency-hopping eliminates the need for channel state observation, simplifying the modulation, while still ensuring that the transmission will succeed with a given probability. This is opposed to~\cite{han2015multi}, which assumes static channel composition and proposes to estimate distance between transmitter and receiver to select the most appropriate waveforms or frequencies for transmission. On the other hand, we also highlight the work of Lin \emph{et al.} which hinges on the use of graphene-based directional agile antennas in the THz band. More specifically, they propose to use beam hopping to switch among spatial channels during transmission~\cite{Lin2018}, a technique that can further reduce interference among users in TS-OOK scenarios.

\subsection{Pulse-based Coding}
\label{sec:coding}
Coding to reduce power consumption and interference without increasing the transceiver complexity has been another hot topic in nanocommunication research. Jornet \emph{et al.} first proposed the use of low-weight coding together with TS-OOK~\cite{jornet2011low, jornet2014low}. Rather than utilizing channel codes to detect and correct transmission errors, this simple mechanism exploits silences to save power and mitigate interference without reducing the transmission rate of each individual user.

It was later observed in~\cite{chi2013optimal} that minimizing only the average weight does not ensure minimum energy. Following this argument, the authors derive optimal codebooks that minimize the energy of transmissions for arbitrary codeword lengths. Further, Kocaoglu \emph{et al.} extend the discussion to account for arbitrary input probability distributions and keeping the codeword length unconstrained, arguing that minimum energy coding with high reliability is achieved in all cases~\cite{kocaoglu2013minimum}. Later, the authors in~\cite{chi2014energy} add the property of prefix freedom and the constraint of maximum average codeword length to the problem of minimum energy coding. Prefix-free codes ensure that no codeword is contained within any other codeword, allowing instantaneous decoding of information. The concept of simple block nanocodes is applied in~\cite{zainuddin2016sbn} to add reliability with very small cost in nanonetworks. The same authors provide a comprehensive comparison between the different proposed coding schemes in \cite{zainuddin2016low}, evaluating energy efficiency, bandwidth expansion, robustness, and interference.  

\begin{table*}[!ht]
\begin{center}
\caption{Summary of physical layer protocols}
\label{tab:phy_layer_protocols}
\begin{tabular}{l l l l l} 
\hline
\textbf{Protocol} & \textbf{Functionality} & \textbf{Distinct Features} & \textbf{Potential Applications}  & \textbf{Evaluation Metrics} \\ 
\hline

TS-OOK~\cite{jornet2014femtosecond} & Modulation & \begin{tabular}{@{}l@{}}- pulse-based \\
- time spread 100-fs pulses \\
- sync only in preamble \end{tabular} & \begin{tabular}{@{}l@{}} - body-centric communication \\ - wireless robotic materials \end{tabular} & \begin{tabular}{@{}@{}@{}l@{}} - energy consumption \\ - user inf. rate \\ - aggregated inf. rate \end{tabular} \\
\hline

PHLAME~\cite{jornet2012phlame}  &  Modulation  & \begin{tabular}{@{}l@{}}- TS-OOK with co-prime rates \\ - minimizes collisions \end{tabular} & \begin{tabular}{@{}@{}l@{}} - body-centric communication \\ - wireless robotic materials \end{tabular} & \begin{tabular}{@{}@{}@{}l@{}}- energy consumption \\ - collision probability \\ - network throughput \end{tabular} \\
\hline

Shrestha \emph{et al.}~\cite{shrestha2016enhanced} &  Modulation & \begin{tabular}{@{}@{}l@{}}- TS-OOK with co-prime rates \\ - enhanced \& generalized co-prime generation \end{tabular} & \begin{tabular}{@{}@{}l@{}} - body-centric communication \\ - wireless robotic materials \end{tabular} & \begin{tabular}{@{}@{}l@{}}- collision probability \\ - aggregated inf. rate \end{tabular} \\
\hline

SRH-TSOOK~\cite{mabed2017enhanced}  & Modulation & \begin{tabular}{@{}l@{}} - based on TS-OOK \\ - random time between pulses \\- uniform average rate across users \end{tabular} &  \begin{tabular}{@{}@{}l@{}} - body-centric communication \\ - wireless robotic materials \end{tabular} & \begin{tabular}{@{}@{}l@{}}- collision probability \\- packet loss \\- network throughput\end{tabular} \\
\hline

ASRH-TSOOK~\cite{mabed2018flexible} &  Modulation & 
\begin{tabular}{@{}l@{}}- based on TS-OOK \\ - random time between pulses \\ - adaptive rate per user \end{tabular} & \begin{tabular}{@{}@{}l@{}} - body-centric communication \\ - wireless robotic materials \end{tabular} & \begin{tabular}{@{}@{}l@{}}- collision probability \\- packet loss \end{tabular} \\
\hline

Vavouris \emph{et al.}~\cite{vavouris2018energy}  &  Modulation & \begin{tabular}{@{}l@{}} - PPM with time-spread pulses \\- extreme energy efficiency \end{tabular}
  & \begin{tabular}{@{}@{}l@{}} - body-centric communication \\ - wireless robotic materials \end{tabular}  & \begin{tabular}{@{}l@{}} - energy consumption \\ - information rate \end{tabular} \\
\hline

Zarepour \emph{et al.}~\cite{zarepour2014frequency}   &  Modulation & \begin{tabular}{@{}l@{}} - frequency hopping to \\ avoid absorption peaks \end{tabular}  & \begin{tabular}{@{}@{}l@{}} - body-centric communication \\ - wireless robotic materials \end{tabular}  &\begin{tabular}{@{}@{}l@{}} - signal-to-noise ratio \\ - bit error rate \\ - capacity \end{tabular} \\
\hline

Multi-band OOK~\cite{Yu2014}  &  Modulation  & \begin{tabular}{@{}l@{}} - continuous-wave OOK in multiple bands \\ - high rates with simple transceivers \end{tabular}  &  - on-chip communication & \begin{tabular}{@{}l@{}} - energy consumption \\ - information rate \\ - silicon area \end{tabular} \\
\hline

DAMC~\cite{Han2014a}  &  Modulation  & \begin{tabular}{@{}l@{}} - continuous-wave multi-carrier \\ - bands chosen depending on distance \end{tabular}  &  - on-chip communication  & - information rate \\
\hline

Han \emph{et al.}~\cite{han2015multi} &  Modulation  & \begin{tabular}{@{}l@{}} - pulse-based version of DAMC \\ - waveforms chosen based on distance \end{tabular}  & \begin{tabular}{@{}@{}l@{}} - on-chip communication \\ - software-defined metamaterials \end{tabular} & \begin{tabular}{@{}@{}l@{}} - SINR \\ - bit error rate \\ - throughput \end{tabular} \\
\hline

Jornet \emph{et al.}~\cite{jornet2014low} &  Coding & \begin{tabular}{@{}l@{}} - low-weight channel coding \\
- minimizes energy in TS-OOK \end{tabular}  & \begin{tabular}{@{}@{}l@{}} - body-centric communication \\ -wireless robotic materials \end{tabular}  & \begin{tabular}{@{}l@{}} - information rate \\- codeword error rate \end{tabular} \\
\hline

MEC~\cite{kocaoglu2013minimum} & Coding & \begin{tabular}{@{}l@{}} - minimum energy channel coding \\ - assumes multi-carrier OOK \end{tabular}  & \begin{tabular}{@{}@{}l@{}} - on-chip communication \\ - software-defined metamaterials \end{tabular} & \begin{tabular}{@{}@{}l@{}} - energy consumption \\- transmission rate \\- error probability \end{tabular} \\
\hline

Chi \emph{et al.}~\cite{chi2014energy} & Coding & \begin{tabular}{@{}l@{}} - minimum energy coding \\ - prefix-free codes \end{tabular}  & \begin{tabular}{@{}@{}l@{}} - body-centric communication \\ - wireless robotic materials \end{tabular}  & - energy consumption \\
\hline

SBN~\cite{zainuddin2016sbn} & Coding & \begin{tabular}{@{}l@{}} - simple block codes \\ - efficiency-reliability trade-off \end{tabular}  & \begin{tabular}{@{}@{}l@{}} - body-centric communication \\ - wireless robotic materials \end{tabular}  & \begin{tabular}{@{}@{}l@{}} - bit error rate \\- energy efficiency \end{tabular} \\
\hline

DS-OOK~\cite{singh2020ds} &  Coding & - direct sequence + OOK  & \begin{tabular}{@{}@{}l@{}} - on-chip communication \\ - software-defined metamaterials \end{tabular} & \begin{tabular}{@{}@{}l@{}} - multi-user interference \\ - bit error rate \end{tabular}  \\
\hline

\end{tabular}
\end{center}
\vspace{-5mm}
\end{table*}

Finally, we highlight the recent work by Yao \emph{et al.}~\cite{yao2018ecpJournal}, which goes beyond existing forward error correction strategies and adopts a hybrid mechanism suitable for energy harvesting. Their proposed error control strategy is compatible with low-error codes, but incorporates probing packets. Before starting data transmission, the source sends probing packets. The receivers then acknowledge the probe only if they anticipate to have enough energy to receive the data packets; otherwise, they remain silent. This way, data packets are not sent if receivers are in a state of low energy, which is PHY-layer dependent.

\subsection{Beaming and Detection}
\label{sec:detection}
Beaming at the transmitter side (both beam switching and beam forming) and detection at the receiver side are two functions that deserve attention due to the highly limited resources and peculiar requirements of THz nanocommunication. 

While THz band communication is moving towards directive antennas and thus will likely require efficient beaming methods, the nanocommunication scenario discourages its use unless simple and effective methods are conceived. In this direction, recent works have discussed the role of graphene antennas. Graphene not only allows to miniaturize antennas, but also confers them with ultrafast reconfigurability achievable by simply changing the electrostatic voltage applied to the antenna. This has led to proposals where both the beam direction and frequency of resonance can be controlled with very simple approaches~\cite{Correas2017}. Leveraging these features, Hosseininejad \emph{et al.}~\cite{Hosseininejad2018WCNC} propose a programmable PHY interface to graphene antennas to expose such beam-switching and frequency tunability to upper layers. As an example, such a controller could easily implement the bit-level beam-switching~\cite{Lin2018} to implement beam multiplexing methods compatible with TS-OOK and the interference mitigation techniques discussed above.

At the receiver side, simple means of detection are crucial to ensure the viability of nanocommunication. 
Cid-Fuentes \emph{et al.} implement a low-complexity Continuous-Time Moving Average (CTMA) with a single low-pass filter and a peak detector~\cite{Cid-fuentes2012}. 
The evaluations contained therein demonstrate the potential for Tbps detection with relaxed synchronization requirements. 
In~\cite{gupta2015joint}, an iterative process employing an array of time-delayed CTMA detectors is proposed to achieve joint detection and synchronization for TS-OOK communication. A similar architecture is proposed in~\cite{Singh2019} that provides an estimation of time-of-arrival for time-hopping PPM modulation. Finally, the work by Iqbal \emph{et al.} is worth highlighting as it proposes a simple modulation mode detection and classification for intelligent nanonetworks where transmitters may switch between modulations type and order~\cite{iqbal2018modulation}.
Note that the approaches discussed above have not been included in Tables~\ref{tab:phy_layer_protocols} and~\ref{tab:mapping_to_requirements}, as they are not physical layer protocols, but proposals for hardware to support such protocols. $\blacksquare$

Table~\ref{tab:phy_layer_protocols} provides a summary of the above-discussed protocols for the physical layer in THz nanonetworks. It can be observed how the physical layer is well-researched from a theoretical and simulated perspective. However, the main hurdle for their realization is the actual circuit implementation of the analog front-end of the transceiver. THz signal generation with compact and efficient means remains as a huge open challenge, especially in the case of the hundred-femtosecond-long pulses assumed in most of the works in the field \cite{Aggrawal2016}. 

{\color{red}The literature mentions three main signal source technologies for THz nanocommunications. One of them is photoconductive technology. For instance, pulse-based photoconductive sources might provide the required signal with high power and, as such, have been proposed for the experimental testbeds of graphene antennas~\cite{Cabellos2014}. However, these sources depend on a bulky laser to excite the photo-carriers that turn into the THz signal and, therefore, are impractical for nanocommunications. A second alternative are electronic sources~\cite{lewis2014review} that generate THz signals from the upconversion of lower frequency ones. However, these sources generally cannot provide sufficient power with a compact form factor, although recent advances in nanoplasma switches may disprove that tenet~\cite{SamizadehNikoo2020}.}

A very promising technology in this field is, again, graphene plasmonics. Graphene supports the propagation of tunable plasmons in the THz band, leading to unprecedented miniaturization and reconfigurability opportunities when operating in this frequency band. These properties have been studied when using graphene transistors as very compact THz signal sources~\cite{Jornet2014TRANSCEIVER} exploiting the Dyakonov-Shur instability or also as direct modulators, translating changes in electrostatic biasing voltage into modulated plasmons~\cite{sensale2012broadband}. Graphene antennas, as discussed above, can not only be miniaturized down to a few micrometers and still resonate in the THz band, but also deliver joint frequency-beam reconfigurability with unprecedented simplicity~\cite{Correas2017, Hosseininejad2018WCNC}.

From the perspective of the receiver, the use of non-coherent detectors and CTMA approaches relax the synchronization requirements. However, synchronization keeps being the main challenge in impulse radio in general, and in THz nanocommunication in particular. To reach the promised Tbps barrier, sampling needs to occur at potentially very high speeds anyway, which goes against the simplicity and efficiency demands of most envisaged applications.


\section{Channel Modeling}
\label{sec:channel}

Channel characterization and modeling captures the changes that the electromagnetic waves suffer as they propagate through a medium until reaching the receiver. In general, comprehensive models incorporate all possible sources of losses (e.g., spreading, blocking, dielectric losses), dispersion (e.g., due to multipath), and noise (e.g., thermal noise and interferences). By accounting for all these effects, channel modeling provides the physical layer with the necessary considerations for the design of appropriate modulations and coding schemes that fulfill the application requirements.

Channel modeling is critical for the development of THz band nanocommunication due to the important differences of THz propagation with respect to microwave or optical propagation. The most striking peculiarity of THz channels is the appearance of molecular absorption effects, which create peaks of attenuation whose depth and frequency depend on the transmission distance and molecular composition of the medium, respectively. These effects limit the practicable bandwidth and, as we have seen in the previous section, may lead to the use of multi-carrier modulations for high-throughput applications. Another peculiarity of THz propagation is that materials that were effectively transparent and flat at the microwave or even mmWave regime start becoming lossy and producing rough scattering upon reflection. More details on these impairments are given in Section \ref{sec:chanMod}.

Obviously, the propagation channel is highly dependent on the actual application scenario. In the following subsections, we analyze the existing channel modeling efforts in two of the promising directions for  THz nanocommunication. In Section~\ref{sec:chanBody}, we review the works that characterize the channel within the body-centric applications. In Section \ref{sec:chanOnchip}, we discuss the attempts to model the THz channel within computing packages for on-chip communications. A summary of the papers discussed in the section is given in Table \ref{tab:channels}.

\subsection{THz Propagation Models}
\label{sec:chanMod}
In the THz band, phenomena that are generally neglected become significant as the wavelength reaches dimensions commensurate to the molecules found in the medium or the tiny irregularities of the surfaces upon which the waves may reflect. For instance, the pioneering work by Piesiewicz \emph{et al.} discusses how the resulting molecular absorption \cite{Piesiewicz2007} could impair communication in these frequencies. Later, scattering produced by certain particles suspended on the environment was factored in \cite{Kokkoniemi2015}. Thus, propagation models in the THz band need to account for these effects. In \cite{Kokkoniemi2015}, the path loss for \ac{LoS} propagation is given by:
\begin{equation}
    A_{LoS}(f) = A_{spr}(f) A_{mol}(f) A_{sca}(f),
    \label{eq:pathloss}
\end{equation}
where, on top of the typical spreading loss $A_{spr}(f)$, we have the molecular absoprtion $A_{mol}(f)$ and the particle scattering loss $A_{sca}(f)$. If waves reflect on rough surfaces, additional scattering would need to be accounted for \cite{Piesiewicz2007a}. Moreover, in \cite{jornet2011channel}, the noise temperature $T_{noise}$ is modeled as:
\begin{equation}
    T_{noise} = T_{sys} + T_{mol} + T_{other},
    \label{eq:noise}
\end{equation}
where, on top of the typical system electronic noise $T_{sys}$, noise caused by molecular absorption $T_{mol}$ is considered. $T_{other}$ refers to any additional noise source.

\begin{figure}
\centering
\includegraphics[width=0.85\columnwidth]{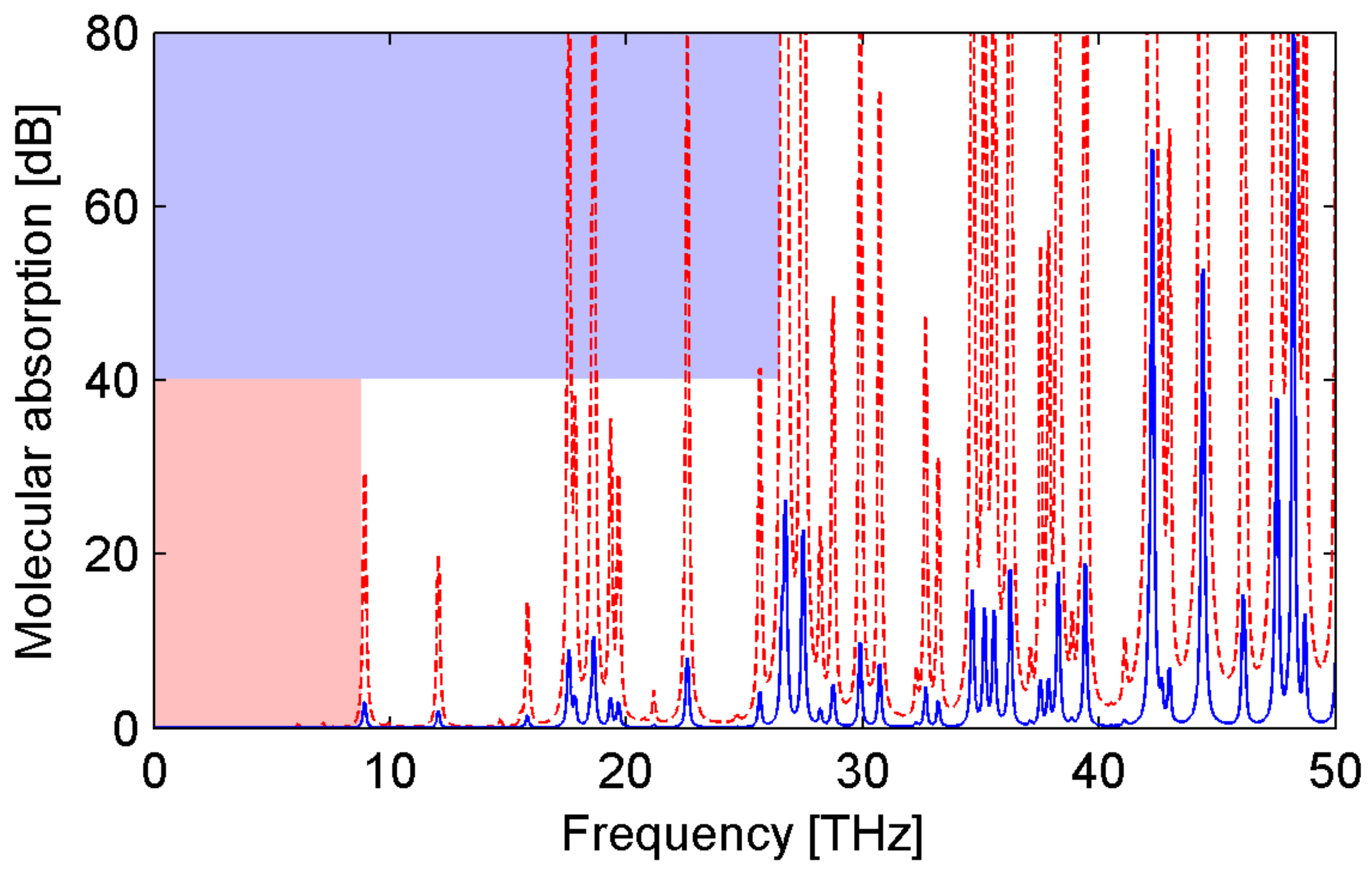}
\vspace{-1mm}
\caption{Molecular absorption for 1~cm (blue) and 10~cm (red)
 in a standard atmosphere as defined by the International Organization for Standardization (ISO) in the standard ISO 2533:1975. The blue and red backgrounds indicate the available bandwidth for distances of 1~cm and 10~cm, respectively, where available bandwidth is defined as the frequency band where the attenuation caused by molecular absorption is below 10 dB.}
\label{fig:absorption} \label{fig:available_channel}\label{absorption2}\label{absorption1}\vspace{-2mm}
\end{figure}

Molecular absorption is the process by which part of the wave energy excites molecules found along its path. This phenomenon occurs often in the atmosphere as many of the molecules comprised therein resonate in the THz regime. From the communications channel perspective, absorption manifests as (i) a frequency-selective attenuation and (ii) added noise from the molecules residual radiation. Jornet \emph{et al.}~\cite{jornet2011channel} were among the first to incorporate a model of the molecular absorption and noise into a complete THz channel model. On the one hand, the attenuation $A_{mol}(f)$ caused by this molecular absorption is modeled as \cite{jornet2011channel}:
\begin{equation}
  A_{mol}(f,d) = e^{k_{A}(f)d},
  \label{molAbs}
\end{equation} 
where $k_{A}(f)$ is the medium absorption coefficient that depends on the molecular composition of the medium, $f$ is frequency, and $d$ is the transmission distance. The result, illustrated in Figure~\ref{absorption1}, is a frequency-selective attenuation that scales with the distance and can effectively limit the channel bandwidth. On the other hand, the molecular noise is modeled as \cite{jornet2011channel}:
\begin{equation}
  T_{mol}(f,d) = T_{0}(1-e^{-k_{A}(f)d}),
  \label{molNoise}
\end{equation} 
where $T_{0}$ is the reference temperature. Based on these models, Jornet \emph{et al.}~\cite{jornet2011channel} studied the impact of absorption in terms of channel capacity for different medium compositions and distances.
Later, Javed \emph{et al.}~\cite{javed2013frequency} modeled the molecular absorption as a log-normally distributed attenuation (similar to how shadowing is generally accounted for) within a conventional log-distance path loss model. In~\cite{boronin2014capacity}, the authors focused on the most prominent transparency windows, i.e., bands where molecular attenuation is low, and performed a thorough capacity and throughput analysis both with and without energy constraints. Finally, Llatser \emph{et al.} analyzed the effect of molecular absorption in the time domain and targeting short distances~\cite{llatser2015time}. There, it was confirmed that molecular absorption and its dispersive effects are generally negligible up to a few centimeters, as exemplified in Figure~\ref{absorption2}.

Besides molecular absorption, diffuse scattering caused by particles or rough surfaces commensurate to the THz wavelength are also potential impairments in the THz band. These effects are also frequency-selective and, therefore, have an impact upon the response of the channel. Kokkoniemi \emph{et al.} provide a comprehensive channel model in both the time and frequency domains which accounts for the aforementioned effects, proving that they might not be negligible at certain distances~\cite{Kokkoniemi2015}. In the case of particle scattering, the attenuation $A_{sca}(f)$ is given by
\begin{equation}
\label{eq:partScat}
A_{sca}(f,d) = e^{k_{S}(f)d},
\end{equation} 
where the frequency dependence is modeled through the particle scattering coefficient $k_{S}(f) = \sum_{j}{N^{j}_{s}\sigma^{j}_{s}}$ \cite{Kokkoniemi2015}. This coefficient requires knowledge on the density of particles $N$ and of the scattering cross section $\sigma$ of each type of particle $j$. In the case of rough surface scattering, several measurement campaigns have been carried out to analyze the response of materials such as wood, plaster, concrete, plastic, glass, or metal~\cite{Jansen2011, Kokkoniemi2016, Fricke2017} and incorporate it into the channel model.


The relatively short wavelength of THz waves, besides leading to the impairments discussed above, also suggests the use of ray tracing techniques to develop comprehensive channel models even in nanocommunication scenarios. The work in~\cite{Han2014b} also argues that ray tracing could be particularly appropriate due to the high-gain antennas expected in THz applications. Moreover, it provides a comprehensive ray-based modeling methodology and exemplifies its use in indoor channel characterization. The methodology has been later extended to the particular case of THz wireless communications with graphene reflectarray antennas~\cite{Han2017b}. The same authors also discuss hybrid methodologies combining ray tracing and full-wave simulations to account for all effects accurately while being computationally affordable~\cite{Han2018a}. Furthermore, the authors of \cite{AbadalNANOCOM2015} propose to expose the design parameters of graphene antennas in order to facilitate the design space exploration of graphene-enabled wireless channels. Finally, and since most works in the field assume the co-existence of multiplexed links either spatially (via beaming) or temporally (via pulse-based modulations), Petrov \emph{et al.} also employ ray-based methods to evaluate interference and therefore derive SINR and spectral efficiency metrics~\cite{petrov2017a}. With the requirement of multiple transmissions, the authors also estimate the optimal distance between receiving nodes to maximize the area capacity and conclude that interference becomes more critical than molecular absorption in these multi-user scenarios.

\subsection{Intra-body Channels}
\label{sec:chanBody}
Wireless communication within the human body presents many exciting applications in the body-centric communication domain in as much as it poses a significant challenge from the perspective of the propagation of THz signals. From the modeling perspective, Equation \eqref{eq:pathloss} still holds. However, the terms related to spreading loss, molecular absorption loss, and scattering loss will vary. For instance, the human body is composed of multiple tissues such as skin, fat, or blood, each with their own response to THz radiation. Wireless propagation is impaired not only by the dielectric loss of each tissue, but also the transition between tissues, thereby affecting $A_{spr}$. Moreover, the composition very much depends on the position of transmitter and receiver within the body and of the patient itself, affecting $A_{mol}$. Finally, the presence of cells and other particles impact on the scattering term $A_{sca}$.

Among the first works on this regard, the authors in~\cite{yang2013numerical} studied the attenuation of fat in the THz band. The characteristics of the fat layer were extracted from characterization works in the optics domain, yielding an attenuation factor of around 30 dB/mm. Later, the authors extended the work and published a complete model in~\cite{yang2015numerical} containing blood, skin, and fat. The molecular absorption of those tissues is used to determine the system noise and, then, the channel capacity. Further, Elayan \emph{et al.} also considered a multi-layered model in the frequency and time domains and studied the response when on-body and intra-body devices communicate~\cite{elayan2017multi}. They concluded that up to 30\% of the incident power from outside the body may be reflected back and that the result is symmetrical. Finally, Zhang \emph{et al.} provide a detailed model of artificial skin and perform a capacity analysis in the THz band \cite{zhang2019power}.


\begin{table*}[!ht]
\vspace{-1mm}
\begin{center}
\caption{Summary of channel modelling works}
\label{tab:channels}
\begin{tabular}{m{1.6cm} m{1.3cm} m{1.2cm} m{1.5cm} m{4.4cm} m{5.2cm}} 
\hline
\textbf{Reference} & \textbf{Scope} & \textbf{Domain} & \textbf{Method} & \textbf{Analyzed Features} & \textbf{Evaluation Metrics} \\ 
\hline

\cite{jornet2011channel} & General & Frequency & Analytical & Molecular absorption, noise   & Path loss, capacity  \\ \hline

\cite{llatser2015time} & General & Frequency, time & Analytical & Molecular absorption   & Practicable bandwidth, pulse width  \\ \hline

\cite{Kokkoniemi2015} & General & Frequency, time & Analytical & Molecular absorption, particle scattering, rough surface scattering   & Path loss, delay spread, coherence bandwidth  \\ \hline

\cite{Kokkoniemi2016, Kokkoniemi2016a}  & General & Frequency & Experimental & Rough surface scattering, diffraction & Path loss (scattering power, diffraction angle) \\ \hline

\cite{Jansen2011} & Indoor & Frequency & Experimental & Rough surface scattering & Scattering power, path loss \\ \hline

\cite{Han2014b, Han2017b}  & Indoor & Frequency, time & Analytical & LoS and NLoS propagation, graphene reflectarray impact & Path loss, delay spread, coherence bandwidth, capacity   \\ \hline

\cite{petrov2017a} & Indoor & Frequency & Numerical & Blocking, Interference  & SIR, SINR \\ \hline

\cite{javed2013frequency} & Indoor, intra-body & Frequency & Analytical & Molecular absorption, noise (fat, blood, bone)   & Path loss, capacity  \\ \hline

\cite{yang2013numerical,yang2015numerical} & Intra-body & Frequency & Numerical & Absorption in blood, skin, fat  & Path loss, noise temperature, capacity   \\ \hline

\cite{elayan2017multi} & Intra-body & Frequency & Analytical, numerical & Discontinuities on multi-layer medium  & Reflected power, reflectance   \\ \hline

\cite{elayan2018stochastic} & Intra-body & Frequency & Analytical & Intra-body noise sources   & Noise spectral density  \\ \hline

\cite{zarepour2015reliability} & Intra-body & Frequency, time & Numerical & Time variation of medium composition & SNR, BER \\ \hline

\cite{Zajic2019} & Chip-to-chip & Frequency, time & Experimental & Reverberation &  Path loss, coherence bandwidth \\ \hline
 
\cite{OpokuAgyeman2016} & On-chip & Frequency & Numerical & Molecular absorption, dielectric losses  & Coupling (S$_{21}$)   \\ \hline

\cite{el2019analysis} & On-chip & Frequency & Numerical & Losses  &  Coupling (S$_{21}$) \\ \hline

\cite{Chen2019,chen2019channel} & On-chip & Frequency & Analytical, numerical & Losses, interference  &  Path loss, capacity \\ \hline

\cite{Zhang2007, Narde2019} & On-chip & Frequency & Experimental & Antenna orientation, position  & Insertion loss (S$_{11}$), Coupling (S$_{21}$)  \\ \hline

\cite{Tasolamprou2019} & Off-chip (SDM) & Frequency, time & Numerical & Propagation path, SDM geometry  & Path loss, delay spread   \\ \hline

\end{tabular}
\end{center}
\vspace{-4mm}
\end{table*}

The impact of the multiple phenomena occurring in intra-body channels on the noise temperature has been also studied in depth. Existing models evolved from the initial model from~\cite{yang2015numerical} where it was assumed that noise caused by molecular absorption dominates and other potential sources can be neglected, thereby obtaining an expression similar to Equation \eqref{eq:noise}. A more comprehensive discussion was included in~\cite{zhang2016modelling}, where the authors modeled signal-independent body radiation noise using Planck's law, as well as the signal-dependent molecular absorption noise. A comparison between the two validated the assumption of dominance of molecular absorption noise. Later, Elayan \emph{et al.} revisited the model to include, besides the two sources mentioned above, other components such as thermal noise at the transceiver circuitry or Doppler-shift-induced noise~\cite{elayan2018stochastic}. Finally, in~\cite{zhang2017analytical}, noise models are combined with interference models to derive SINR metrics towards accurately determining the capacity and throughput achievable in intra-body networks. It is therefore suggested that nanomachine density can be a factor as important as the composition of the intra-body channel in assessing the viability of the communication. 

We finally describe papers that extend existing models to account for changes in the transmission medium. The work of Zarepour \emph{et al.} is worth mentioning as it considers time-varying channels in the THz band~\cite{zarepour2015reliability}. The key takeaway is that nanocommunication channels are not static: the temperature, pressure, or molecular composition of the medium may vary over time. They provide as example the composition of the blood when breathing in, which is clearly different than when breathing out. A similar example is analyzed in~\cite{zarepour2015design}, where a nanonetwork for lung monitoring is explored. In that case, respiration clearly changes the volume and composition of the lung, and the authors adapt their design to that circumstance. Even further, all the works described in this section can be used to determine the attenuation caused by vegetation in plant monitoring nanosensor networks, which can also vary over time due to the effects of photosynthesis~\cite{afsharinejad2016performance}.

\subsection{On-chip Communication Channels}
\label{sec:chanOnchip}
A channel model that takes into consideration the peculiarities of the chip-scale scenario is fundamental to evaluate the available bandwidth and to properly allocate power. The enclosed nature of the chip package suggests that propagation losses may be small, but also that multipath effects may be present. Additionally, the multiple metalization layers and the Through-Silicon Vias (TSV) present in today's chips may further challenge propagation~\cite{Matolak2013CHANNEL}. Fortunately, the scenario is unique in that all these elements are fixed and known beforehand. Thus, the channel model will be virtually time-invariant and quasi-deterministic. Moreover, molecular absorption and diffuse scattering are not problems in this controlled environment~\cite{OpokuAgyeman2016, abadal2019wave}. Therefore, only the spreading loss term of Equation \eqref{eq:pathloss} stands. The model, however, has to take into consideration dielectric losses caused by the chip package materials and by transitions between the different layers found within the chip structure. In terms of noise, the most prominent source is thermal noise and, thus, conventional models can be employed. It is worth noting that the processor circuitry does not introduce noise as communication and computation sub-systems operate at disjoint frequencies.
 
Thus far, few works have explored the chip-scale wireless channel down to the nanoscale and in the THz band. The theory is well laid out~\cite{Matolak2013CHANNEL} and a wide variety of works exist in larger environments. THz propagation has been investigated in small and enclosed environments such as across a computer motherboard~\cite{Kim2016mother}, or within the metallic encasement of a laptop computer~\cite{Zajic2019}. The experimental results, up to 300~GHz, have confirmed that such systems act as reverberation chambers due to the metallic enclosure, leading to rather low path loss but very high delay spreads. 

Down to the chip level, most characterization efforts have thus far been limited to mmWave frequencies \cite{abadal2019wave}. Analytical models~\cite{Yan2009}, simulation-based studies~\cite{Rayess2017}, and actual measurement campaigns~\cite{Zhang2007,Narde2019} have been conducted under different assumptions. In most cases, path loss is modeled by an empirical log-scale approximation resulting in a derived the path loss exponent, rather than using a bottom up approach with Equation \eqref{eq:pathloss}. More recently, a few authors have tried to study propagation in sub-THz frequencies, i.e., up to around 140 GHz in \cite{Timoneda2018b} and 200 GHz in \cite{el2019analysis}. At the time of writing, however, the only channel model at THz frequencies is that of Chen \emph{et al.}~\cite{Chen2019, chen2019channel}, which employs ray tracing within a stratified model of the chip structure to extract path loss and dispersion metrics. 

Regardless, the main issue of the works mentioned above is that free-space propagation in an unpackaged chip is assumed for simplicity. Such a model generally falls short of capturing the enclosed nature of realistic on-chip communication environments. Hence, recent efforts are starting to model realistic flip-chip packages with lossy bulk silicon in the mmWave and sub-THz spectrum~\cite{Timoneda2018b}. The main conclusion is that the losses introduced by the silicon substrate prevent the package to act as a reverberation chamber. However, the price to pay is an unwanted path loss in excess of several tens of dBs. When scaling to the THz band, the results will likely worsen due to the smaller effective aperture of the antennas. For potential solutions we refer the reader to the next subsection. $\blacksquare$

Channel modeling in THz nanocommunication has been well-researched in theory and simulations. Actual measurements are more complex to achieve due to the lack of mature measuring equipment and the difficulty of accessing the nanoscale with enough accuracy. First experimental results have been achieved that confirm molecular absorption and scattering effects and, in the case of on-chip wireless communications, results up to sub-THz frequencies start becoming available. However, there is still a large room for improving the current characterizations experimentally.

In the intra-body networks, researchers have identified dielectric losses and molecular absorption as huge sources of attenuation and noise. Moreover, these can vary over time even assuming fixed transmitter-receiver positions. It is yet unclear, hence, how nanomachines will overcome these issues and how protocols will adapt to these very adverse and changing conditions while still being bio-compatible.  

In the on-chip scenario, a channel model in the THz band with a realistic chip package is still largely missing. Moreover, the problem of relatively high attenuation due to the losses within the silicon remains unsolved. In this respect, Timoneda \emph{et al.} propose to exploit the controlled nature of the chip scenario to actually \emph{design} the wireless channel without affecting the reliability of the digital circuits~\cite{Timoneda2018ADAPT}. To that end, the thickness of silicon and the thermal interface material are introduced as design parameters and optimized to minimize path loss while maintaining an acceptable delay spread.

Due to the similarities between the on-chip scenario and the intra-SDM network scenario, some of the knowledge of the former can be reused in the latter. First explorations in this regard \cite{Tasolamprou2019} indeed show that, internally, the mmWave wireless propagation paths within the SDMs suffer similar effects than in on-chip channels: low loss, wave-guiding effects, and relatively high delay spread. The challenge here is to extend those models to the THz band and confirm the viability of wireless communication within the metamaterials.


\section{Simulation and Experimentation Tools}
\label{sec:testbeds}

There are several tools for simulating the behaviour of nanonetworks operating the THz frequencies.   
The pioneering simulator is NanoSim~\cite{piro2013simulating,piro2013nano}, an event-based ns-3 module for modeling nanonetworks based on electromagnetic communications in the THz band. 
In NanoSim, a nanonetwork can be comprised of nanonodes, nanorouters, and nanointerfaces.
Nanonodes are small and simple devices with very limited energy, computational, and storage capabilities.
Nanorouters have resources larger than the nanonodes and they are envisioned to processing data coming from nanonodes, as well as controlling their behavior through control messages.
Nanointerfaces act as gateways between the nano- and macro-scale world.
All of them can be either static or different mobility models can be employed according to the application requirements (i.e., constant acceleration, constant velocity, random walk, random direction, and random way-points).
Moreover, their basic functionalities can be customized to the demands of a given evaluation scenario. 
The nanonetwork in NanoSim consists of the network, link, and physical layers.
On the network layer, random and selective flooding routing strategies are supported. 
Link layer currently supports Transparent-MAC (i.e., simple forwarding from network layer to the physical interface) and Smart-MAC protocols.
Moreover, NanoSim provides a TS-OOK-based physical layer with several adjustable parameters such as pulse duration, pulse transmission interval, and transmission duration.
Finally, the radio channel can be modeled based on a cut-off transmission distances between nanonodes.
In addition, the selectivity of the THz channel in both frequency and time domains can optionally be introduced based on the spectrum-aware channel modeling from~\cite{baldo2009spectrum}.

Vouivre~\cite{boillot2015scalable} (\cite{boillot2014using,boillot2015large} in its preliminary version) is a C++ THz nano-wireless simulation library developed as both an extension for Dynamic Physical Rendering Simulator (DPRSim) and as a standalone discrete event simulator. 
DPRSim~\cite{rister2007integrated} has been developed in the scope of the Claytronics project for supporting simulations with a large number (up to millions) of Claytronics micro-robots (also known as \textit{catoms}).
Original catoms do not have wireless transmission capabilities, as they are envisioned to communicate only through physical contact.
Vouivre introduces wireless communication capability to the catoms.
In particular, it can be used for simulating the THz radio channel and its concurrent accesses by catoms.
The THz radio channel is modeled by a continuous distance-dependent attenuation contribution increased by a certain noise value caused by the concurrent transmissions, with the noise value taken from~\cite{jornet2011channel,jornet2011low}.
In addition, transmission delay in combination with total attenuation have been utilized for determining packet reception probability.
Moreover, Vouivre implements a TS-OOK-based physical layer, while the upper layers have not been implemented, apart from the standard CSMA/CA scheme combined with the Friss propagation model in 2.4~GHz frequency for ``allowing ulterior studies of hybrid systems''. 

BitSimulator~\cite{dhoutaut2018bit} is another simulator specifically targeting THz band nanonetworks. 
BitSimulator is implemented in C++ and utilizes a discrete event model.
At the physical layer, BitSimulator implements the TS-OOK scheme with 100~fs long pulses and per-frame configurable parameter $\beta$.
The link layer is not implemented, as it is assumed that multiple frames can be temporally multiplexed and the nodes have the capability of tracking the transmissions intended for them.
Network layer implementation supports no routing, flooding-based, and \ac{SLR}~\cite{tsioliaridou2017packet}.
Two discovery modes are available: i) static, where neighbors are stored for each node and calculated at the beginning of the simulation; ii) dynamic, where neighbors are not stored, but computed at periodic time instances.
Hence, the support for simple mobility exists in BitSimulator. 
In terms of channel modeling, a simple \textit{communicationRange} parameter is used for specifying achievable transmission range.
In addition, collisions between frames are determined based on propagation delay and TS-OOK-specific bit values of the packets concurrently received at each nanonode~\cite{jornet2014femtosecond}.
If the number of collisions is above a set threshold, the packet is discarded. 

TeraSim~\cite{hossain2018terasim} is a newer alternative to the NanoSim simulator and also implemented on top of ns-3 . 
The simulator supports simulations of both major types of application scenarios, i.e., THz nano- and macro-scale communication.
In TeraSim, the THz radio channel for a nanoscale scenario is modeled by applying frequency and distance dependent spreading and absorption loss, accounting for waveforms with realistic bandwidth.
The simulator consists of a common channel module, separate physical and link layers for each scenario, and two assisting modules, namely, THz antenna module and energy harvesting module, originally designed for the macro- and nanoscale scenario, respectively.
TeraSim allows the user to select several channel attributes such as bandwidth, number of samples of the frequency-selective channel, and detection threshold (i.e., noise floor).
Collisions between two packets occur if two pulses of different TS-OOK receptions overlap in time (packet dropping based on \ac{SINR}) or if the pulse in reception overlaps with the pulse of an ongoing transmission (packet in reception is dropped). 
On the physical layer, TeraSim implements the TS-OOK modulation and channel coding scheme with user adjustable pulse and symbol durations. 
On the link layer in the nanoscenario, TeraSim implements the ALOHA protocol in which the data is sent if there is enough energy for the transmission.
The receiver receives the data if there is enough energy for reception and replies with an acknowledgment (ACK) packet.
The packet is dropped upon exceeding the maximum number of retransmissions.
In addition, TeraSim implements the \ac{CSMA} handshake protocol, in which the usual \ac{RTS}~/~\ac{CTS} packet exchange occurs. 
A CTS packet is transmitted if the receiver has enough energy to complete the reception.
Same as before, the packet is dropped upon exceeding the maximum number of retransmissions.
Upper layer protocols are not specifically tuned to THz nanocommunication, but utilize available ns-3 modules (e.g., \ac{UDP} client/server, IPv4 addressing). 
Similarly, node mobility support is based on the available ns-3 modules. 
An interesting feature of TeraSim lies in the fact that it implements energy harvesting capability of the nanonodes.
Hence, the current energy levels of the nodes play a role in the network performance simulations, arguably making TeraSim the most suitable simulation tool for a variety of low-power THz nanoscale applications. $\blacksquare$

As outlined above, there are several tools currently available for THz nanocommunication and nanonetworking simulations.
Being one of the most recently proposed simulators, TeraSim seemingly provides the most extensive capabilities, including the energy-harvesting module for nanonodes. 
BitSimulator, another recently proposed tool, trades-off the complexity for scalability, arguing that for many of the envisioned nanocommunication applications scalability will be the primary requirement.
However, at this point the scalability vs. realism trade-off is merely a speculation, as it is not clear how scalable the outlined simulators are.
In addition, it would be interesting to investigate the difference in simulation results of the different simulators in order to evaluate their usability for different scenarios, as well as the reliability of the simulation results. 
In other words, large discrepancies in the results derived using different simulators could put into question the reliability of findings obtained using these simulators.

In many  of the envisioned applications, the only feasible powering option for the nanonodes will be through energy harvesting. 
This fact is only reflected in the TeraSim simulator.
Even there, the energy storage capacity of the nanonodes is unlimited and the energy harvester is implemented as a simple process in which energy is harvested at a constant rate.
However, the majority of nanoscale harvesters (e.g., exploiting piezo-electric effect of \ac{ZnO} nanowires~\cite{wang2008towards} or ultrasound-based power transfer~\cite{canovas2018nature}) charge the node in a non-linear way, with their harvesting rates being highly dependent on their current energy levels and maximum storage capacity~\cite{jornet2012joint}.
To address these limitations, the insights from~\cite{llatser2014n3sim} could be utilized.
\cite{llatser2014n3sim} provides N3Sim, an ns-3-based simulator for molecular nanocommunication.  
It provides several harvesting options (with harvesting being a synonym for collecting molecules from nanonodes' local neighborhood), some of them realistically assuming that the nanonode's energy storage capacity is limited.
In addition, the charging operation due to harvesting is in N3Sim a non-linear process. 
Moreover, in the currently available simulators the energy consumption of a nanonode is attributed to transmission and reception only.
However, idling energy should be accounted for if the aim is accurate energy consumption modeling, as well as energy consumed due to for example information processing or data storage.
Due to nanonode's highly constrained energy resources, accurate energy modeling should be of paramount importance, as will be discussed in the next section in more details.

Furthermore, novel mobility models are needed for accurate simulations of nanonetworks for several application scenarios.
For example, as body-centric applications obviously assume in-body communication, hence fine-grained models of human mobility are needed, as well as models for blood stream and various other in-body mobility effects (e.g., heart-beats).
A very good initial step in this direction is BloodVoyagerS~\cite{geyer2018bloodvoyagers}, a model of a human body’s cardiovascular system, developed with the idea of simulating nanonodes’ mobility.
Similar tools are needed for other aspects of a human body.
In addition, the integration of such mobility models with the current simulators will be needed for maximizing the benefits and realism of the THz nanocommunication and nanonetworking simulations.   

Finally, in terms of experimentation facilities or experimental datasets, the areas of THz nanocommunication and nanonetworking are still uncharted.
Publicly available experimental results of experimentation infrastructure would presumably give a strong boost to the research in these domains, which has already been recognized in the community. 
One good example going in this direction is VISORSURF, a \ac{EU}-funded research project whose aim is to develop a full stack of hardware and software tools for THz-based control of metamaterials~\cite{liu2019intelligent,liaskos2019novel}.  
More initiatives targeting development of hardware tools, integration into full prototypes, or generation of public datasets are needed for other nanocommunication scenarios. 

\vspace{-1mm}
\section{Additional Challenges, Open Issues, and Future Research Directions}
\label{sec:challenges}

The optimization objectives for the above-discussed protocols are indicated in Table~\ref{tab:mapping_to_requirements}.
These objectives represent the design aims of the protocols and should not be mixed with the performance metrics used in the evaluation of the protocols and listed in Tables~\ref{tab:network_layer_protocols},~\ref{tab:link_layer_protocols}, and~\ref{tab:phy_layer_protocols}. 
The aim of Table~\ref{tab:mapping_to_requirements} is to help the reader in the selection of suitable protocols for a given application with specific requirements. 
In addition, the aim is to indicate the ``missing pieces'' in the existing protocols, i.e., the potential improvement directions. 
For example and as already mentioned, in terms of link-later protocols Akkari~\emph{et al.}~\cite{akkari2016distributed} is the only one explicitly aiming at latency optimization.
Hence, if an application requires certain bounds on latency, Akkari~\emph{et al.}~\cite{akkari2016distributed} would naturally be the first choice for the link layer.
Moreover, given that there is only one proposal targeting latency optimization, new link layer protocols could be developed for its further optimization.  



\begin{table*}[!ht]
\vspace{-3mm}
\begin{center}
\caption{Optimization objectives for existing protocols in different layers of the protocol stack}
\label{tab:mapping_to_requirements}
\begin{tabular}{l m{1.1cm} m{1.0cm} m{1.0cm} m{1.3cm} m{1.45cm} m{1.2cm} m{1.4cm} m{1.0cm} m{1.1cm} m{1.1cm}} 
\hline
\textbf{Protocol} & \begin{tabular}{@{}l@{}}\textbf{Network}  \\ \textbf{scalability} \end{tabular} & \begin{tabular}{@{}l@{}}\textbf{Node} \\ \textbf{density} \end{tabular} & \textbf{Latency} & \textbf{Throughput} & \begin{tabular}{@{}l@{}} \textbf{Bidirectional} \\ \textbf{traffic} \end{tabular} & \textbf{Reliability} & \begin{tabular}{@{}l@{}} \textbf{Energy} \\ \textbf{consumption} \end{tabular} & \textbf{Mobility} & \textbf{Security} \\ \hline

\multicolumn{10}{c}{\textbf{Network layer protocols}} \\ 

Xia~\emph{et al.}~\cite{xia2017cross} & & & & \hfil \checkmark & \hfil \checkmark & & \hfil \checkmark & & \\ 

Rong~\emph{et al.}~\cite{rong2017relay} & & & & & \hfil \checkmark & \hfil \checkmark & & & \\ 

Yu~\emph{et al.}~\cite{yu2015forwarding} & \hfil (\checkmark) & \hfil \checkmark & \hfil \checkmark & \hfil \checkmark & \hfil \checkmark & & \hfil (\checkmark) & & \\ 

PESAWNSN~\cite{yen2017energy} & \hfil \checkmark & & & & & & \hfil \checkmark & & \\ 

{\color{red}C\'{a}novas-Carrasco \emph{et al.}~\cite{canovas2018nanoscale}} & & & &  {\color{red}\hfil (\checkmark)} & & &  {\color{red}\hfil \checkmark} &  {\color{red}\hfil \checkmark}  & \\

 {\color{red}C\'{a}novas-Carrasco \emph{et al.}~\cite{canovas2019optimal}} & & & &  {\color{red}\hfil \checkmark} & & &  {\color{red}\hfil \checkmark} &  {\color{red}\hfil \checkmark}  & \\

Liaskos~\emph{et al.}~\cite{liaskos2015promise} & \hfil \checkmark & & \hfil \checkmark & & & \hfil \checkmark & \hfil \checkmark & & \\ 

Tsioliaridou~\emph{et al.}~\cite{tsioliaridou2016lightweight} & \hfil \checkmark & & \hfil \checkmark & & & \hfil \checkmark & \hfil \checkmark & & \\ 

Afsana~\emph{et al.}~\cite{afsana2015outage} & & & & \hfil \checkmark &  \hfil \checkmark & \hfil \checkmark & \hfil \checkmark & & \\ 

Stelzner~\emph{et al.}~\cite{stelzner2018body} &  \hfil \checkmark & & & & &  \hfil \checkmark & & & \\ 

Buther~\emph{et al.}~\cite{buther2018hop} & \hfil \checkmark & & & & & & \hfil \checkmark & \hfil \checkmark & \\ 

E\textsuperscript{3}A~\cite{al2017cognitive} & & & \hfil \checkmark & \hfil \checkmark & & \hfil \checkmark & \hfil \checkmark & & \\ 

Pierobon~\emph{et al.}~\cite{pierobon2014routing} & & & \hfil \checkmark & \hfil \checkmark & & \hfil \checkmark & \hfil \checkmark & & \\ 

Tsioliaridou~\emph{et al.}~\cite{tsioliaridou2015corona} & \hfil \checkmark & & & & \hfil \checkmark & \hfil \checkmark & \hfil \checkmark &  & \\ 

Tsioliaridou~\emph{et al.}~\cite{tsioliaridou2017packet}  & \hfil \checkmark & & & & \hfil \checkmark & \hfil \checkmark & \hfil \checkmark &  & \\

\hline

\multicolumn{10}{c}{\textbf{Link layer protocols}} \\

Akkari~\emph{et al.}~\cite{akkari2016distributed} & & & \hfil \checkmark & \hfil \checkmark & & & & & \\ 

Alsheikh~\emph{et al.}~\cite{alsheikh2016grid} & & & & & \hfil \checkmark & \hfil \checkmark & & & \\

PHLAME~\cite{jornet2012phlame} & \hfil \checkmark & \hfil \checkmark & & & \hfil \checkmark  & \hfil \checkmark & \hfil \checkmark & & \\ 

DRIH-MAC~\cite{mohrehkesh2015drih} & \hfil \checkmark & & & & \hfil \checkmark & \hfil \checkmark & \hfil \checkmark & & \\ 

TCN~\cite{d2015timing} & & & & & & \hfil \checkmark & \hfil \checkmark & & \\

Xia~\emph{et al.}~\cite{xia2015link}  & & & & \hfil \checkmark & & \hfil \checkmark & & & \\

Smart-MAC~\cite{piro2013nano} & & & & & & \hfil \checkmark & & & \\

APIS~\cite{yu2017pulse} & & & & \hfil \checkmark & & \hfil \checkmark & \hfil \checkmark & & \\

Wang~\emph{et al.}~\cite{wang2013energy} & \hfil \checkmark & \hfil \checkmark & & \hfil \checkmark & & & & & \\

EEWNSN-MAC~\cite{rikhtegar2017eewnsn} & \hfil \checkmark & \hfil \checkmark & & & & \hfil \checkmark & & \hfil (\checkmark) & \\ 

CSMA-MAC~\cite{lee2018slotted}  & & & & \hfil \checkmark & & & & & \\ 

Mansoor~\emph{et al.}~\cite{Mansoor2016} & & & & \hfil \checkmark & \hfil \checkmark & \hfil \checkmark & \hfil \checkmark & & \\ 

BRS-MAC~\cite{Mestres2016} & \hfil \checkmark & & & & \hfil \checkmark & \hfil \checkmark & & \\

Dynamic \ac{MAC}~\cite{Mansoor2015} & & & & \hfil \checkmark & \hfil \checkmark & \hfil \checkmark & \hfil \checkmark & \\ \hline

\multicolumn{10}{c}{\textbf{Physical layer protocols}} \\

TS-OOK~\cite{jornet2014femtosecond} & \hfil \checkmark & \hfil \checkmark & & &  &   & \hfil \checkmark  &   & \\ 

PHLAME~\cite{jornet2012phlame} & \hfil \checkmark & \hfil \checkmark & & & \hfil \checkmark &  \hfil \checkmark & \hfil \checkmark & & \\ 

Shrestha~\emph{et al.}~\cite{shrestha2016enhanced} & \hfil \checkmark & \hfil \checkmark & & & \hfil \checkmark &  \hfil \checkmark & \hfil \checkmark & & \\ 

SRH-TSOOK~\cite{mabed2017enhanced} & \hfil \checkmark & \hfil \checkmark & & \hfil \checkmark & \hfil \checkmark &  \hfil \checkmark & & & \\ 

ASRH-TSOOK~\cite{mabed2018flexible} & \hfil \checkmark & \hfil \checkmark & & \hfil \checkmark & \hfil \checkmark &  \hfil \checkmark & & & \\ 

Vavouris~\emph{et al.}~\cite{vavouris2018energy} & \hfil \checkmark & & & & & & \hfil \checkmark & & \\   

Zarepour~\emph{et al.}~\cite{zarepour2014frequency} & & & & & & \hfil \checkmark & & \hfil \checkmark & \\  

Multi-band OOK~\cite{Yu2014} & & & \hfil \checkmark & \hfil \checkmark & & \hfil \checkmark & \hfil \checkmark & & \\

DAMC~\cite{Han2014a} & & & & \hfil \checkmark & & \hfil \checkmark & & \hfil \checkmark & \\  

Han~\emph{et al.}~\cite{han2015multi} & & & & \hfil \checkmark & & \hfil \checkmark & & \hfil \checkmark & \\ 

Jornet~\emph{et al.}~\cite{jornet2014low} & \hfil \checkmark & \hfil \checkmark & & & &\hfil \checkmark & \hfil \checkmark & & \\ 

MEC~\cite{kocaoglu2013minimum} & & & & & & \hfil \checkmark & \hfil \checkmark & & \\ 

Chi~\emph{et al.}~\cite{chi2014energy} & & & \hfil \checkmark & & & \hfil \checkmark & \hfil \checkmark & & \\ 

SBN~\cite{zainuddin2016sbn} & & & \hfil \checkmark & & & & \hfil \checkmark & & \\  

DS-OOK~\cite{singh2020ds} & & \hfil \checkmark & & \hfil \checkmark & & & & & \\

\hline

\end{tabular}
\end{center}
\vspace{-4mm}
\end{table*}

In addition, there are several overarching challenges not directly related to the ones outlined in previous sections, which we discuss in the reminder of this section. 

{\color{red}{\subsection{Common Design Frameworks}

There is a need for common design frameworks in order to simplify the development of the THz nanonetworks for supporting different types of the envisioned applications. 
We believe this issue should be approached from the perspectives of both the applications and the nanodevices. 
In other words, in order to support a certain application in one of the specified application domains, one should be aware of the requirements that this application poses on the underlying nanonetwork. 
Simultaneously, one should consider the capabilities of the nanodevices expected to be used in that particular context, as it is hardly imaginable that the same types of devices will be used for enabling e.g., the on-chip, in-body, or software-defined metamaterial-based applications.  

In Section~\ref{sec:applications}, we aimed at deriving the application requirements on the level of different application domains.
Nevertheless, we believe such derivations will be needed on the level of particular applications.
The current literature on this issue is sparse, with only a few works mostly targeting applications in the domain of in-body communication~\cite{ali2015internet,asorey2020analytical,asorey2020throughput,asorey2020flow}.
Further research is needed and should, in contrast to the existing works, aim at \textit{qualitatively specifying exhaustive sets} of application requirements for all applications in all of the promising application domains.  

From the perspective of the nanodevices, we argue the full specifications of their features are needed, with primarily concerns being their size, available energy, processing power, and memory. 
In this direction, it is worth emphasizing~\cite{canovas2016conceptual}, where by following the guidelines originally outlined in~\cite{akyildiz2010electromagnetic} the authors discuss a conceptual design of a nanodevice, consisting of a nanoprocessor, nanomemory, nanoantenna, and nanogenerator.
Seemingly, the nanodevice will not be suitable for all the applications, for example because it assumes a nanogenerator based on energy harvesting or wireless power transfer. 
As such, it can be hardly characterized as feasible e.g., on-chip communication in which the energy is abundant.

We believe the derivation of the mentioned perspectives will not only substantially simplify the development of the supporting nanonetworks along well-specified design requirements and constraints, but also highly contribute to the intuition and reasoning on the feasibility of different approaches for a particular set of applications.}}

\subsection{Transport Layer Protocols}

In~\cite{akyildiz2014teranets,akyildiz2014terahertz}, the authors state that, as Gbps and Tbps links become a reality, the network throughput will increase dramatically. 
It will, therefore, be necessary to develop new transport layer solutions for mitigating network congestion problems. 
The authors in~\cite{akyildiz2014teranets,akyildiz2014terahertz} also state that the overhead of existing transport layer protocols requires minimization in order to reduce the performance constraints.
This has been recognized by only a few early works on the topic~\cite{tsioliaridou2017novel,georgiadis2006resource,alam2017energy}.

In the scope of the VISORSURF project, Tsioliaridou \emph{et al.}~\cite{tsioliaridou2017novel} consider a software-defined metamaterial named the HyperSurface. 
The HyperSurface encompasses a hardware layer that can change its internal structure by tuning the state of its active elements, as well as a nanonetwork in which each nanonode controls a single active element.
For carrying sensed information to the external world for processing, as well as configuration commands from the external world to the HyperSurface, the authors propose HyperSurface Control Protocol (HyperCP). 
The Hyper-CP uses Lyapunov drift analysis for avoiding congested or out-of-power nanonetwork areas~\cite{georgiadis2006resource}.
Specifically, the protocol aims at optimizing the network throughput by accounting for outdated nanonodes' status information, as well as battery and latency constraints.

Alam \emph{et al.}~\cite{alam2017energy} state that the existing studies on nanocommunication and nanonetworking in body area nanonetworks focus on lower layers of the protocol stack, resulting in the upper layers (e.g., transport layer) remaining unexplored. 
They further argue that electromagnetic waves will potentially be harmful to sensitive body areas.
Moreover, they claim that molecular communication is slower and error-prone compared to electromagnetic one.
Motivated by the above arguments, they propose an energy-efficient transport layer protocol for hybrid body area nanonetworks (i.e., both electromagnetic and molecular communication supported). 
The protocol is based on a congestion control mechanism with very limited overhead, in which the sender upon receiving a ``halt'' signal suspends the packet transmission for a predefined timeout period.

These initial contributions are constrained to narrow application domains, specifically to software-defined metamaterials and body-centric communication. 
Transport layer protocols for other application domains with differing requirements are currently lacking. 
Even the two outlined protocols are lacking relevant performance details, pertaining primarily to their scalability and protocol overhead. 
For example, for Hyper-CP it is unclear if the distribution of battery states and latency constraints among nanonodes is at all feasible, given that it unavoidably causes signalling-related energy dissipation at the nanonodes.   
Similarly, for the protocol proposed in~\cite{alam2017energy}, halt signals can potentially be infeasible for low-energy nanonodes.
In such a case, the transmission of data packets could unnecessarily continue until the depletion of the transmitter's energy.  
In summary, transport layer protocols for nanonetworks are currently to a large extent unexplored.
Even more, it is unclear if such protocols will at all be needed for many scenarios, as they will inevitably increase the protocol overhead, which for energy-constrained nanonodes could be infeasible. 

\subsection{Reduced~/~Integrated Protocol Stack}

Along the above conclusions, in case of energy-constrained nanonodes, the protocol stack will have to be substantially condensed and integrated for maintaining feasible nanocommunication and nanonetworking.  
This it predominantly due to the fact that more traditional networks trade-off a high overhead of the utilized protocols with the support for heterogeneous application requirements.
In nanocommunication and nanonetworking, this paradigm will potentially be shifted, which is currently largely unexplored with the only two examples being the ones outlined below.
Certainly, further insights are needed in terms of the design choices for nanonetworks for enabling different applications, pertaining to either reducing the complexity of the protocol stack by making it condensed and application specific, or  providing a full stack at the cost of increased energy consumption, latency, and complexity.  

The first example where the authors argue that the paradigm shift is needed is~\cite{hassan2017event} {\color{red}{(extended in~\cite{hassan2019eneutral})}}, in which a framework is proposed for enabling energy-harvesting nanonodes to communicate their locations (i.e., addresses) and sensed events using only one wireless pulse. 
These wireless pulses feature two degrees of freedom pertaining to any two of the following parameters: amplitude, pulse width, and transmitted energy (equaling amplitude multiplied by pulse width). 
The framework requires each nanonode to use a particular pulse width for their identification, while the event types are identified by the amplitude/energy emitted by each nanonode. 

Similarly, the authors in~\cite{tsioliaridou2018bitsurfing} argue that \textit{even exotic (i.e., hard to integrate) power supplies relying on energy harvesting can only scavenge energy for 1 packet transmission per approximately 10 sec~\cite{jornet2012joint}.
 This makes the development of even basic protocols – such as addressing and routing – highly challenging.}
To mitigate this effect, the authors propose BitSurfing, a network adapter that does not generate physical data packets when transmitting information, but assigns meaning to symbols created by an external source.
It does that by reading incoming symbols and waiting for desired messages to appear in the stream. 
Once they appear, short low-energy pulse is then emitted to notify neighboring nodes. 
The authors demonstrate BitSurfing's perpetual operation, as well as the ability to operate without link and transport layer protocols.

Another example of a shallow protocol stack is given in the on-chip communication context, where packets need to be served to guarantee forward progress in the computation. 
This means that cores are \emph{self-throttling}, hence their injection speed will be reduced if the network becomes congested~\cite{Nychis2012}. 
As a result, the responsibilities of the transport layer are reduced and generally implemented at the architecture level. For instance, cache coherence protocols (which generate most of the traffic in multiprocessors) typically implement end-to-end acknowledgment and timeouts to confirm the reads and writes on shared data. 
In fact, on-chip communication take advantage of the monolithic nature of the multiprocessor system to reduce the depth of the protocol stack.

\subsection{Energy Lifecycle Modeling}

From the above discussions, it is clear that the energy consumption is one of the most stringent constraints that will potentially impede on the feasibility of nanocommunication and nanonetworking in the THz frequencies.
In order to develop feasible nanocommunication and nanonetworking protocols, there is a need for accurate energy lifecycle modeling, especially for the nanonodes whose only powering option is through energy harvesting. 
As the practical implementations of the nanonodes are currently lacking, the development of accurate analytical energy lifecycle models can be seen as a fundamental step towards the design of feasible nanonetwork architectures and protocols~\cite{jornet2012joint2}.
This has to an extent been recognized in the research community.

In the pioneering works on the topic~\cite{jornet2012joint2,jornet2012joint}, an energy model for self-powered nanonodes is developed with the aim of capturing the correlation between the nanonodes' energy harvesting and the energy consumption processes. 
The energy harvesting process is realized by means of a piezoelectric nanogenerator, while the nanonode's energy consumption is quantified by assigning certain amounts of energy for the transmission and reception of ``1'' bits, under the assumption of the TS-OOK communication scheme being employed.
A mathematical framework is then developed for deriving packet delivery probability, end-to-end delay, and achievable network throughput.
A similar approach for energy lifecycle modeling has been taken in~\cite{mohrehkesh2013optimizing,mohrehkesh2017energy}, in which the authors additionally evaluate the contributions of packet sizes, repetitions, and code weights on the nanonode's energy consumption. 

The authors in~\cite{canovas2018nature} extended the work from~\cite{jornet2012joint2,jornet2012joint} by highlighting the effects of the employed network topology, as well as of different energy harvesting approaches and rates (i.e., blood currents and ultrasound-based power transfer). 
Their results show that a micrometer-sized piezoelectric system in lossy environments becomes inoperative for transmission distances over 1.5~mm.
Similarly, the authors in~\cite{lemic2019modeling,lemic2020idling} reason that, as any other practical transceiver, a nanoscale transceiver will consume certain amounts of energy during its idling periods, in addition to the consumption due to transmission and reception.  
Once the idling energy is accounted for in the overall energy consumption modeling, the authors show that for feasible communication this energy consumption has to be nine orders of magnitude smaller than the energy consumed for reception during the same period.  
This is certainly a challenging requirement, as in current nanoscale systems the idling energy is in the best case scenario up to three orders of magnitude smaller than the corresponding energy in reception.

The above mentioned contributions are only the initial steps in a promising direction. 
As discussed in~\cite{lemic2019modeling,lemic2020idling}, certain energy will inevitably be consumed when a nanonode is idling. 
By the same token, the nanodevice's energy will be distributed to different functions such as sensing, processing, etc., and not all of it can be used for transmission or reception.
That being said, seemingly many current works on the design of nanocommunication and nanonetworking protocols could be infeasible under more realistic energy consumption models, as they indeed make an assumption that the overall energy of a nanodevice can be utilized for transmission and reception of information.    
Example-wise, as the nanonodes are expected to experience intermittent on-off behavior, they will often have to wake-up after harvesting sufficient energy~\cite{lemic2019assessing}.
This wake-up process per-se will consume a certain amount of energy, which the current energy models and consequently the protocols utilizing such models do not account for. 

{\color{red}{As mentioned, in many applications the nanonodes will be powered solely through harvesting environmental energy.
Hence, it is first worth pointing out that there is a variety of potentially feasible candidates for energy harvesting, for example blood currents~\cite{canovas2018nature} and ultrasound-based power transfer~\cite{donohoe2017nanodevice} in in-body communication applications, and \ac{RF}-based power transfer in the domain of SDMs~\cite{rong2017simultaneous}.  
Similar to energy consumption modelling, there is a need for an accurate modelling of the charging of an energy-harvesting nanonode.
Here, it is worth pointing out the effort in~\cite{donohoe2017nanodevice}, where the authors outline an approach for ultrasound-based powering of piezoelectric \ac{ZnO} nanowires-based nanonodes placed on peripheral nerves in the human body. 
In their characterization of the harvested energy (i.e., intensity of ultrasound), the authors consider a variety of relevant parameters such as the stimulus depth, size of the energy-harvesting array, and duration of ultrasound pulses.
Such a comprehensive characterization makes it easy for the proposed approach to be utilized in future works, among others in the domain of THz nanonetworks.   
Such efforts are needed for other types of energy harvesters and harvesting sources.
In more general terms, accurate energy harvesting and consumption models would be highly beneficial.}}

\subsection{End-to-End Architectures}

To truly support many of the envisioned applications, seamless integration of the nanonetworks with existing networking infrastructures will be needed~\cite{petrov2015feasibility}. 
Addressing this issue is not straightforward, as existing networks predominantly utilize carrier-based electromagnetic communication, while nanonetworks will seemingly have to rely on energy-constrained pulse-based communication. 
Thus, special gateway nodes between the macro- and nano-worlds will be required, which has been only sporadically addressed in the literature to date.

The authors in~\cite{yu2017demand} argue that data acquisition from nanonetworks faces two challenges.
First, the mismatch between the demands of the nanonetworks and the available bandwidth of the backhaul link with the macro-world reduces the bandwidth efficiency of the backhaul, as well as the energy efficiency of THz band nanonetworks. 
To address this issue, the authors propose a polling mechanism for the backhaul tier which is composed of nanosinks that aggregate and transport data from nanonodes to the gateway. 
The mechanism is based on on-demand polling and accounts for the dynamic backhaul bandwidth and THz channel conditions.
The mechanism is developed with the goal of supporting applications in the domain of wireless robotic materials. 

Several works (i.e.,~\cite{akyildiz2015internet,dressler2015connecting,piro2015design,jarmakiewicz2016internet}) propose hierarchical network architectures for enabling body-centric communication and sensing only-based applications.
Specifically, the aim in these works is to enable uplink communication between a nanonetwork deployed inside of a human body and external monitoring devices through nanointerfaces (i.e., gateways). 
Moreover, in~\cite{akyildiz2010internet,jornet2012internet} the authors aim at enabling interconnection of sensing nanodevices with existing communication networks, eventually forming the \ac{IoNT}.

However, the current results mostly aim at enabling uplink communication from the nanonetworks toward the macro-world. 
In order to unlock the full potential of the envisioned applications, downlink (predominantly control) communication will also be required.
This will enable applications ranging from software-controlled metamaterials to control of and actuation using wireless robotic materials.  
Future investigations should also aim at minimizing the latency of such communication, especially for control-related applications. 

\subsection{Security, Integrity, and Privacy}

One question that did not receive substantial research attention is the security of THz band nanocommunication. 
The authors in both~\cite{dressler2012towards} and~\cite{atlam2018internet} argue that the security-related goals in THz band nanocommunication should be confidentiality (protection against malicious or unauthenticated users), integrity (protection against modification), and availability (protection against disruption by a malicious user). 
There are several challenges pertaining to achieving those goals, as in more details discussed in~\cite{dressler2012towards}.
First, it will be necessary to develop methods for the establishment of shared encryption keys, as well as for revoking them when needed.
Second, the overhead of secure communication protocols, cryptographic algorithms, and access control and authentication methods will have to be minimized and potentially reconsidered for ultra-low power nanonetworks. 
Finally, as the prevention of all malicious attacks can hardly be guaranteed, it will be important to at least develop means for their detection, as well as strategies for reacting to them. 
However, the above-listed challenges are all but explored  in the context of THz band nanocommunication and nanonetworking.

\vspace{-1mm}
\subsection{Localization and Tracking}

Many of the envisioned applications supported by THz nanonetworks require localization or even tracking of the nanonodes. 
This is a highly challenging requirement, given that such localization and tracking capabilities will, due to the nature of the nanonodes, have to operate in a highly energy-constrained way, as well as provide very high accuracy (due to the small sizes of the nanonodes).  
Moreover, due to the low range of THz band nanocommunication, the localization capability will potentially have to be based on multi-hopping, which has a drawback in terms of propagation of localization errors with the number of hops~\cite{karl2007protocols}.

There are only a handful of attempts in localizing THz-operating nanonodes. 
In~\cite{tran2014localization}, the authors propose two ranging- and hop-counting-based localization algorithms. 
The algorithms are envisioned to be used for estimating the locations of all nanonodes deployed within a certain area. 
The first technique uses flooding-based forwarding to all nanonodes, where the locations of the two considered nanonodes are estimated by counting the number of hops between them. 
To reduce the overhead and energy dissipation, the second algorithm works under the assumption that all nanonodes are grouped into clusters. 
Cluster heads communicate together and count the number of hops in order to localize different nodes within a cluster. 
Similarly, Zhou \emph{et al.}~\cite{zhou2017pulse} propose a pulse-based distance accumulation (PBDA) localization algorithm. 
The PDBA algorithm adopts TS-OOK pulses for estimating the distance between clustered nanonodes.
The algorithms proposed in~\cite{tran2014localization,zhou2017pulse} can potentially operate within the nanonodes' energy constraints.
However, their accuracy is intrinsically going to be relatively low due to the propagation of localization errors as the number of hops increases.   

One potential direction in enhancing localization accuracy, while at the same time maintaining its low energy consumption profile, is to base the localization capabilities on backscattered signals. 
The feasibility and promising accuracy of such an approach has been demonstrated in~\cite{el2018high}.
The approach in~\cite{el2018high} utilizes a backscattered signal from a nanonode (i.e., DR-Lens tag) for extracting the round-trip time-of-flight (RToF) between the tag and the localization anchor. 
The RToF readings from multiple anchors are then used for estimating the corresponding distances to the nanonode, with the nanonode's location consequently being determined using linear least square algorithm. 
The yet unresolved challenges of developing such systems include angle and frequency-dependent response of the nanonodes, non-free space (e.g., in-body) propagation, and almost certainly a variety of hardware imperfections.   

\vspace{-1mm}
\subsection{Standardization} 

As  stated in~\cite{akyildiz2014terahertz}, the THz band is not yet regulated and it is up to the scientific community to jointly define the future of the paradigm. 
The IEEE~P1906.1 standard~\cite{7378262} is, to the best of our knowledge, the only attempt going in this direction for THz-based nanocommunication.
The standard provides a conceptual framework for future developments of nanoscale communication networks. 
C{\'a}novas-Carrasco \emph{et al.}~\cite{canovas2017ieee} provided a review of the latest IEEE~P1906.1 recommendations, in which they outlined the main features and identified several shortcomings of the standard, to which, they argue, further research efforts should be devoted. 
First, the characteristics and reference powering solutions of potential nanodevices have not been discussed in the standard.  
Second, the recommended values and ranges for respectively transmission power and SNR for reception have not been specified.  
Third, the standard does not specify the \ac{OSI} layers 2 and 3 techniques, which hampers protocol interoperability.  
In order words, standardization efforts aiming at media access control, addressing schemes, flow control, error detection, and routing procedures are needed. 
Finally, the standard currently lacks recommendations about the interconnection between nanonetworks and existing communication networks.
Substantial research efforts are certainly still needed for addressing the above-stated limitations of the standard. 



\section{Conclusion}
\label{sec:conclusion}


In this survey, we have outlined the most promising application domains that could be enabled by the nanonetworks operating in the THz frequencies. 
Moreover, we have derived the requirements that such applications pose on the supporting nanonetworks, which could be utilized as a rule-of-thumb guidelines in the development of the supporting nanocommunication and nanonetworking protocols.  
We have then outlined the current \acf{SotA} nanocommunication and nanonetworking protocols, as well as the applicable channel models and experimentation tools.
In addition, we have discussed their strengths and weaknesses, as well as summarized a set of potential directions for future research.
Future efforts could target the development of THz nanonetwork prototypes and experimental testing infrastructures, development of protocols for primarily higher layers of the protocol stack, mitigation of mobility effects, enabling additional features such as security or localization of nanodevices, to name some.

We believe that this survey demonstrates that THz band nanocommunication is a vivid and promising research domain, as Prof. Feynman suggested it will be more than 60 years ago.
We encourage the community to focus on resolving the indicated major challenges, so that the outlined set of exciting applications could become a reality in the near future.

\section*{Acknowledgments}

The author Filip Lemic was supported by the EU Marie Skłodowska-Curie Actions Individual Fellowships (MSCA-IF) project Scalable Localization-enabled In-body Terahertz Nanonetwork (SCaLeITN, grant nr. 893760). 
The author Pieter Stroobant was supported by a PhD grant of Ghent University's Special Research Fund (BOF).
This work was partly funded by the Ghent University's BOF/GOA project ``Autonomic Networked Multimedia Systems''. 
This work was also supported by the Spanish Ministry of Economy and Competitiveness under contract TEC2017-90034-C2-1-R (ALLIANCE project) that receives funding from FEDER, and by the European Commission under grants No. 736876 and No. 863337.

\renewcommand{\bibfont}{\footnotesize}
\printbibliography

\end{document}